\documentclass[11pt,a4paper,utf8]{article}

\usepackage[utf8]{inputenc}
\def\bea#1\eea{\begin{align}#1\end{align}}
\def\bea#1\eea{\begin{eqnarray}#1\end{eqnarray}}
\def\be#1\ee{\begin{equation}#1\end{equation}}
\def\ba#1\ea{\begin{align}#1\end{align}}
\def\nl{\nonumber\\}

\newcommand{\calP}{\mathcal{P}}
\newcommand{\dcalP}{{\mathcal{P}^\circ}}
\newcommand{\calA}{\mathcal{A}}

\newcommand{\calE}{\mathcal{E}}

\newcommand{\bms}{{\bm S}}

\def\yz#1\yz {{\color{blue} [[YZ: #1]] }}

\usepackage{amsthm}

\usepackage[utf8]{inputenc}
\usepackage{jheppub}
\usepackage{amsmath}
\usepackage{multicol}
\usepackage{bbm}
\usepackage{enumerate}
\usepackage[inline]{enumitem}
\usepackage{bibentry}
\usepackage{comment}
\usepackage{amsthm}
\usepackage{mathrsfs}
\usepackage{upgreek}
\usepackage{amssymb}
\usepackage{bm}
\usepackage{setspace}
\usepackage{array,multirow,arydshln}
\usepackage{bigdelim}
\usepackage{scalerel}
\usepackage{diagbox}
\usepackage{caption}
\usepackage{tabularx}
\usepackage{tikz}
\usepackage{tcolorbox}
\usepackage{subfig}

\usepackage{stackengine}

\usepackage{tikz}
\usetikzlibrary{calc,shapes.geometric,arrows,arrows.meta,decorations.pathmorphing,decorations.markings,patterns}

\def\<{\langle}
\def\>{\rangle}
\def\a{\alpha}
\def\b{\beta}
\def\c{\cdot}

\def\r{\ref}

\title{Notes on polytopes, amplitudes and boundary configurations for Grassmannian string integrals}

\author[a,b]{Song He,}

\author[c]{Lecheng Ren,}

\author[a,d,b]{and Yong Zhang}
\affiliation[a]{CAS Key Laboratory of Theoretical Physics, Institute of Theoretical Physics, Chinese Academy of Sciences, Beijing 100190, China}
\affiliation[b]{School of Physical Sciences, University of Chinese Academy of Sciences, No.19A Yuquan Road, Beijing 100049, China
}
\affiliation[c]{Department of Physics, Brown University, Providence, RI 02912, USA}
\affiliation[d]{Perimeter Institute for Theoretical Physics, Waterloo, ON N2L 2Y5, Canada}

\emailAdd{songhe@itp.ac.cn}
\emailAdd{lecheng$\_$ren@brown.edu}
\emailAdd{yongzhang@itp.ac.cn}

\date{\today}

\abstract{We continue the study of positive geometries underlying the {\it Grassmannian string integrals}, which are a class of ``stringy canonical forms", or stringy integrals, over the positive Grassmannian mod torus action, $G_+(k,n)/T$. The leading order of any such stringy integral is given by the canonical function of a polytope, which can be obtained using the Minkowski sum of the Newton polytopes for the regulators of the integral, or equivalently given by the so-called scattering-equation map.  
The canonical function of the polytopes for Grassmannian string integrals, or the volume of their dual polytopes, is also known as the generalized bi-adjoint $\phi^3$ amplitudes.
We compute all the linear functions for the facets which cut out the polytope for all cases up to $n=9$, with up to $k=4$ and their parity conjugate cases. 
The main novelty of our computation is that we present these facets in a manifestly gauge-invariant and cyclic way, and identify the boundary configurations of $G_+(k,n)/T$ corresponding to these facets, which have nice geometric interpretations in terms of $n$ points in $(k{-}1)$-dimensional space. All the facets and configurations we discovered up to $n=9$ directly generalize to all $n$, although new types are still needed for higher $n$.}

\begin{document}
\maketitle
\flushbottom


\section{Introduction}

In~\cite{Arkani-Hamed:2019mrd}, N. Arkani-Hamed, T. Lam and one of the authors introduced {\it stringy canonical forms}, which are string-like integrals as $\alpha'$-deformation of canonical form~\cite{Arkani-Hamed:2017tmz} for general polytopes. Let's briefly review the idea: we consider an integral over $\mathbb{R}^d_{>0}$, which depends on a collection of polynomials, $p_I$, with positive coefficients:
\be\label{stringyint}
{\cal I}_{\{p\}} (X,c)=(\alpha')^d \int_{\mathbb{R}^d_{>0}} \prod_{i
=1}^d \frac{d x_i}{x_i} x_i^{\alpha' X_i} \prod_I p_I(x)^{-\alpha' c_I}\,.
\ee
and it is a meromorphic function of the exponents, $X_i>0$ for $x_i$ factor and $c_I>0$ for polynomial $p_I$, respectively. As explained in~\cite{Arkani-Hamed:2019mrd}, such integrals resemble open-string amplitudes in a number of ways, and there is a natural polytope associated with it. 

The most basic observation of~\cite{Arkani-Hamed:2019mrd} is that the integral converges if ${\bf X}:=(X_1, \cdots, X_d)$ is inside the polytope ${\cal P}$ that is defined as the {\it Minkowski sum} of {\it Newton polytopes} of $p_I$, ${\cal P}=\sum_I c_I {\bf N} (p_I)$. Exactly this polytope controls the leading order of ${\cal I}$ as $\alpha'\to 0$, which is given by the canonical function of ${\cal P}$ (or the volume of the dual polytope) computed at ${\bf X}$:
\be\label{fahd}
\lim_{\alpha'\to 0} {\cal I}=\underline{\Omega}({\cal P}; {\bf X})\,.
\ee
Another remarkable property is that the same polytope can be obtained as the image of the so-called {\it scattering-equation (SE) map} from the integration domain. The map is given by a rewriting of the saddle-point equations, which in turn are given by taking $\alpha'\to \infty$ limit of the regulator $R:=\prod_{i=1}^d x_i^{\alpha' X_i} \prod_I p_I(x)^{-\alpha' c}$:
\be\label{sesss}
d\log R=0 \,\iff \, X_i=x_i \sum_I \frac{c_I }{p_I }\frac{ \partial p_I}{ \partial x_i}\,, \quad {\rm for}~i=1,\cdots, d\,.
\ee
As shown in~\cite{Arkani-Hamed:2019mrd}, this provides a diffeomorphism from ${\bf x}\in {\mathbb{R}}_{>0}^d$ to the interior of the polytope ${\cal P}$, thus one can obtain the canonical form of the latter, $\Omega({\cal P}; {\bf X}):=d^d {\bf X} \underline{\Omega}({\cal P}; {\bf X})$, as the pushforward of that of the former by summing over all solutions of \eqref{sesss}:
\be\label{CHYgen}
\Omega({\cal P}; {\bf X})=\sum_{\rm sol.} \prod_{i=1}^d \frac{d x_i}{x_i} \quad \iff \quad \underline{\Omega}({\cal P}; {\bf X})=\int \prod_{i=1}^d \frac{d x_i}{x_i} \delta ( X_i -x_i \sum_I \frac{c_I}{p_I}\frac{\partial p_I}{\partial x_i})\,. 
\ee
where equivalently the canonical function can be written as a generalization~\cite{Arkani-Hamed:2017mur} of the CHY formula for bi-adjoint $\phi^3$ amplitudes~\cite{Cachazo:2013hca,Cachazo:2013iea} (see also \cite{Gao:2017dek, He:2018pue}).

As discussed in~\cite{Arkani-Hamed:2019mrd}, open-string amplitudes can be thought of as the stringy canonical form for the so-called ABHY associahedron~\cite{Arkani-Hamed:2017mur}; in the $\alpha'\to 0$ limit we obtain the field-theory, bi-adjoint $\phi^3$ amplitudes as its canonical function, which can also be obtained using the pushforward/CHY formula from the scattering-equation map for the associahedron~\cite{Arkani-Hamed:2017mur}. As shown in~\cite{Arkani-Hamed:2019mrd},  the two most natural extensions of string amplitudes are the so-called cluster string integrals and the Grassmannian string integrals. The former are stringy integrals for generalized associahedra of finite-type cluster algebra (usual string integrals correspond to type ${\cal A}$), which exhibit various remarkable properties relevant for scattering of generalized particles~\cite{baziermatte2018abhy,arkanihamed2019causal} and strings~\cite{Arkani-Hamed:2019plo,Arkani-Hamed:2019clu}. The latter are natural integrals over positive Grassmannian mod torus action, $G_+(k, n)/T$ (string integrals correspond to $k=2$), which were first studied very recently in~\cite{Arkani-Hamed:2019rds}. At leading order, the Grassmannian string integral is computed by the canonical function of a class of polytopes, which are closely related to the positive tropical Grassmannian~\cite{speyer2005tropical}, as shown in~\cite{Cachazo:2019ngv, Drummond:2019qjk, Drummond:2019cxm,Henke:2019hve}). It has also been shown that the canonical function can also be obtained using a CHY formula which are higher-$k$ generalization of the bi-adjoint $\phi^3$ amplitudes (see also \cite{Cachazo:2019apa,Cachazo:2019ble,Sepulveda:2019vrz}). 

On the other hand, just as the momentum twistor space~\cite{hodges2009eliminating}, $G(4,n)$, is crucial for the amplituhedron reformulation~\cite{Arkani-Hamed:2013jha,
Arkani-Hamed:2017vfh} of all-loop integrand in planar ${\cal N}=4$ SYM~\cite{ArkaniHamed:2012nw}, the space $G_+(4,n)/T$ can play a crucial role for all-loop {\it integrated} amplitudes (or even non-perturbatively). Recall that for $n=6,7$, all evidence so far~\cite{Caron-Huot:2019bsq, Goncharov:2010jf, Dixon:2016nkn, Prlina:2017azl, Prlina:2017tvx, Drummond:2017ssj, Zhang:2019vnm} is consistent with the conjecture that the alphabet for symbol letters consists of the $9$ and $42$ cluster variables associated with $G_+(4,n)/T$, respectively; it is thus highly desirable to study the structure of $G_+(4,n)/T$ for $n\geq 8$. In~\cite{Arkani-Hamed:2019rds, Drummond:2019cxm,Henke:2019hve}, it has been shown that polytopes ${\cal P}(4,n)$ and the associated tropical positive Grassmannian indeed have intriguing applications to the mathematical structures of multi-loop scattering amplitudes in ${\cal N}=4$ SYM. 

In this paper, we continue to investigate the leading order of Grassmannian string integral for $G_+(k,n)/T$ as $\alpha'\to 0$, which is the canonical function of polytope denoted as ${\cal P}(k,n)$, and we call it the generalized amplitudes ${\cal A}_{k,n}=\underline{\Omega}({\cal P}(k,n))$ since these functions generalize bi-adjoint $\phi^3$ amplitudes for $k=2$ case~\cite{Cachazo:2019ngv}:
\ba\label{lim}
 {\cal A}_{k,n}:=\lim_{\a' \to 0}  {(\a')^{d}}\int_{G_+(k,n)/T }  \Omega(G_+(k,n)/T)\,
 R_{k,n} 
\ea
where $d=(k-1)(n-k-1)$, and $\Omega(G_+(k,n)/T)$ denotes the canonical form of the domain $G_+(k,n)/T$. The regulator $R_{k,n}$ is defined as 
\be 
 R_{k,n} :=
\prod_{1\leq i_1<i_2<\cdots<i_k\leq n }
(i_1,i_2,\cdots,i_k)^{\a' \, s_{i_1,i_2,\cdots,i_k}}\,,
\ee 
where the ${\mathrm {SL}}(k,{\mathbb C})$ invariant $(i_1,i_2,\cdots,i_k)$ 
is the minor of the $k$ vectors labelled by $i_1, \cdots, i_k$. 
The generalized Mandelstam variables are symmetric tensors  satisfying ``on-shell'' condition  $s_{i, i, \cdots }=0$ and  ``momentum conservation'' condition
\ba\label{fqfeq}
\sum_{\{ i_2,\cdots,i_k\} \subset \{1,\cdots,n\}\backslash \{i_1\} } s_{i_1,i_2,\cdots,i_k}=0 \quad \forall i_1\,.
\ea
Note that the dimension of kinematic space for $(k,n)$ is thus ${n\choose k}-n$. 

As we will review shortly, once we recast the Grassmannian string integral in a positive parametrization \eqref{stringyint},  we can directly apply the two general methods mentioned above to compute the polytope ${\cal P}(k,n)$: (1). by taking the Minkowski sum of the Newton polytopes, and (2). by working out the image of the scattering-equation (SE) map. These two methods are usually complementary: while the Minkowski sum naturally gives the vertices of the polytope, the SE map gives the facets which cut out the polytope. Once the polytope ${\cal P}(k, n)$ is obtained, it is straightforward to compute its canonical function, ${\cal A}_{k, n}$ {\it e.g.} by any triangulation of the dual polytope. However, we emphasize that a priori our computation for ${\cal P}(k,n)$ and ${\cal A}_{k,n}$ are independent of the positive tropical Grassmannian or higher-$k$ generalization of CHY formulas. Note that in the form of \eqref{stringyint}, we automatically have the subspace where the polytope lives: the $d$-dim subspace of the kinematic space can be specified by choosing all the $c$'s to be constant, and a basis for it is given by the ${\bf X}$. Thus it suffices to give all the linear functions for the facets of ${\cal P}(k,n)$ (or equivalently all the poles of ${\cal A}_{k,n}$), which we denote as $F_a$ for $a=1,2, \cdots, N$, and requiring all $F_a\geq 0$ immediately cut out ${\cal P}(k,n)$. The canonical function ${\cal A}_{k,n}$ follows from any triangulation of the dual polytope: for a simple vertex of ${\cal P}(k,n)$, {\it i.e.} with $d$ facets touching it, so its contribution to ${\cal A}_{k,n}$ (the volume) is given by the product of the $d$ factors $1/F$; for non-simple vertex we will need to triangulate its dual which we will discuss. We emphasize that having the polytope is the most important part of the calculation. All the facets (or equivalently the poles) up to $k=3$, $n=10$ and $k=4$, $n=9$, and the explicit results for generalized amplitudes up to $n=8$, are presented in the supplementary material.

In addition to providing more data and an independent method for computing the polytope and its canonical function, the main novelty of our calculation is that, up to $n=9$ (for $k=3,4$ and their parity conjugate $k \to n{-}k$), we obtain interesting patterns for all the facets of the polytope, or the poles of the generalized amplitudes, which can be written in a completely invariant way using generalized Mandelstam variables (independent of any particular parametrization of $G_+(k,n)/T$). These facets/poles are direct generalizations of {\it planar variables} $X_{a,b}=\sum_{a\leq i<j<b} s_{i,j}$ of the $k=2$ case to higher $k$. As we will see, already for $k=3$, as $n$ increases there are more new types of facets, and even more so for $k=4$ case. An interesting phenomenon is that there are usually more than one way to represent a facet, and their equivalence is guaranteed by momentum conservation. 

More importantly, we will see that all the facets of the polytope are associated with certain ``boundary" configurations of $G_+(k,n)/T$; each configuration can be naturally mapped to the corresponding facet via the SE map. These boundary configurations and more generally the stratification of $G_+(k,n)/T$ have been studied in~\cite{Arkani-Hamed:2020cig} in terms of ``hypersimplex decomposition". Here we will not try to systematically study the boundary configurations in general, but instead we find the configurations by finding the pre-image of the facets. Such a  boundary is controlled by a single parameter $\epsilon \to 0$ and it is specified by to which order in $\epsilon$ some of all the minors vanish; the facet is then given by the residue at $\epsilon \to 0$ of the $d\log R_{k,n}$ at the boundary. In this way, we find all facets and boundary configurations for up to $n=9$, written in a manifestly gauge-invariant and cyclic manner. 

The boundary configurations have nice geometric interpretations which can be described in terms of $n$ points in $(k{-}1)$-dim space. The identities between different representations of a given facet reflect equivalence relations between different configurations of $G_+(k,n)/T$, which can be understood systematically~\cite{Arkani-Hamed:2020cig}. All the facets and associated boundary configurations we found directly generalize to all $n$, which represent a subset of the facets and configurations, although new types will appear for higher $n$.

Before proceeding, let's briefly spell out the counting of the facets/configurations. It is convenient to organize them into cyclic classes: when $k$ and $n$ are coprime, the length for each cyclic class is always $n$, but if there is a common factor of $k$ and $n$, the length of some classes may be smaller. For $(k, n)=(3,7), (3,8), (3,10)$ and $(4,9)$, the number of facets/configurations is $7\times 6$, $8 \times 15$, $10 \times 314$ and $9 \times 2155$, respectively. For $(3,6), (3,9)$ and $(4,8)$, we have $6\times 2+2\times 2$, $9\times 50+3 \times 7$ and $8\times 42+ 4\times 3+2\times 6$, respectively.

\section{${\cal P}(k,n)$ from Minkowski sum and scattering-equation map}

\begin{figure}[!htb]
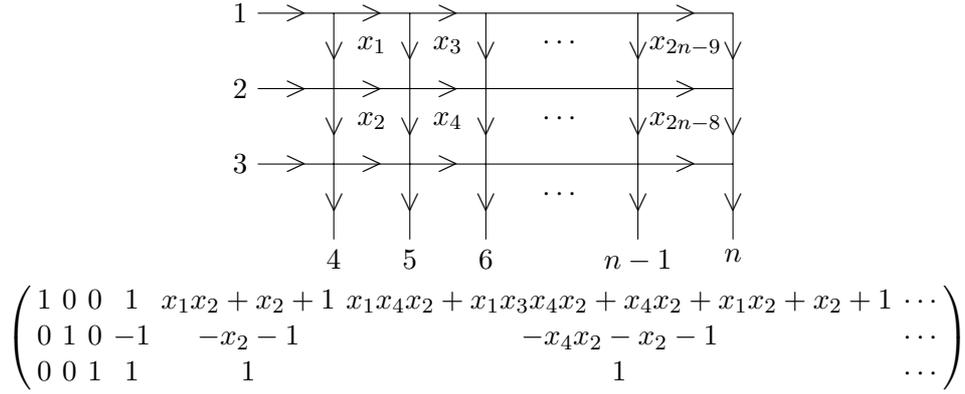

\centering
   \tikz[scale=1]{
   \draw (0,0) node[left] {1}
   --(1,0) node[pos=.5]{$>$}node[pos=.5,below=5pt]{}
   --(2,0) node[pos=.5]{$>$}node[pos=.5,below=5pt]{$x_1$}
    --(3,0) node[pos=.5]{$>$}node[pos=.5,below=5pt]{$x_3$}
    --(5,0)node[pos=.5,below=5pt]{$\cdots$}
     --(6.25,0) node[pos=.5]{$>$}  node[pos=.5,below=5pt]{$x_{2n-9}$}
    --(6.25,-1) node[pos=.5]{\rotatebox {-90} {$>$}} 
    --(6.25,-2) node[pos=.5]{\rotatebox {-90} {$>$}} 
    --(6.25,-3) node[pos=.5]{\rotatebox {-90} {$>$}} node[below] {$n$}
    (0,-1) node[left] {2}
       --(1,-1) node[pos=.5]{$>$}node[pos=.5,below=5pt]{}
   --(2,-1) node[pos=.5]{$>$}node[pos=.5,below=5pt]{$x_2$}
    --(3,-1) node[pos=.5]{$>$}node[pos=.5,below=5pt]{$x_4$}
    --(5,-1)node[pos=.5,below=5pt]{$\cdots$}
    --(6.25,-1) node[pos=.5]{$>$}  node[pos=.5,below=5pt]{$x_{2n-8}$}
    
        (0,-2) node[left] {3}
       --(1,-2) node[pos=.5]{$>$}node[pos=.5,below=5pt]{}
   --(2,-2) node[pos=.5]{$>$}node[pos=.5,below=5pt]{}
    --(3,-2) node[pos=.5]{$>$}node[pos=.5,below=5pt]{}
    --(5,-2)node[pos=.5,below=5pt]{$\cdots$}
    --(6.25,-2) node[pos=.5]{$>$}  node[pos=.5,below=5pt]{}
    (1,0)
     --(1,-1) node[pos=.5]{\rotatebox {-90} {$>$}} 
    --(1,-2) node[pos=.5]{\rotatebox {-90} {$>$}} 
    --(1,-3) node[pos=.5]{\rotatebox {-90} {$>$}} node[below] {4}
        (2,0)
     --(2,-1) node[pos=.5]{\rotatebox {-90} {$>$}} 
    --(2,-2) node[pos=.5]{\rotatebox {-90} {$>$}} 
    --(2,-3) node[pos=.5]{\rotatebox {-90} {$>$}} node[below] {5}
            (3,0)
     --(3,-1) node[pos=.5]{\rotatebox {-90} {$>$}} 
    --(3,-2) node[pos=.5]{\rotatebox {-90} {$>$}} 
    --(3,-3) node[pos=.5]{\rotatebox {-90} {$>$}} node[below] {6}
                (5,0)
     --(5,-1) node[pos=.5]{\rotatebox {-90} {$>$}} 
    --(5,-2) node[pos=.5]{\rotatebox {-90} {$>$}} 
    --(5,-3) node[pos=.5]{\rotatebox {-90} {$>$}} node[below] {$n-1$}
    ; }
$\left(
\begin{array}{ccccccc}
 1 & 0 & 0 & 1 & x_1 x_2+x_2+1 & x_1 x_4 x_2+x_1 x_3
   x_4 x_2+x_4 x_2+x_1 x_2+x_2+1&\cdots \\
 0 & 1 & 0 & -1 & -x_2-1 & -x_4 x_2-x_2-1 &\cdots\\
 0 & 0 & 1 & 1 & 1 & 1 &\cdots \\
\end{array} 
\right)$
\caption{\label{threefafdaafd}  A  parameterization of $G_+(k,n)/T$ }
\end{figure}

Let us first review a canonical way to parametrize $G_+(k,n)/T$  using the so-called ``web variables"~\cite{speyer2005tropical}. We illustrate it by showing an example of $k=3$ in figure \ref{threefafdaafd}.  

We first gauge fix the $k\times n$ matrix to have an identity matrix $I_{k \times k}$ in its first $k$ columns. The remaining entries are determined as follows. There are $(k-1)(n-k-1)$ web variables, one in each box (while all the boundary slots are assigned with $x=1$). Following the decorated directions, we find all paths from $i$ with $1\leq i\leq k$ to $j$ with $k+1\leq j\leq n$, and for each path we assign the product of all web variables below it, which we denote as ${\rm Prod}_{\rm path} (x)$. The sum over all paths gives the entry $(i, j)$ of the matrix:
 \be\label{eqprodpath}
(-)^{i+1} \sum_{\rm paths}{\rm Prod}_{\rm path} (x) \quad {\rm for}~1\leq i\leq k,~k{+}1\leq j\leq n\,.
  \ee   
With this parameterization, note that entries of the last row are trivial, and we have $n$ trivial minors that equal to $1$: $k{+}1$ of them from the first $k{+}1$ columns and $n{-}k{-}1$ ones, $(1,2,\cdots, k{-}1, i)$ with $k{+}2\leq i\leq n$. The following $d=(k-1)(n-k-1)$ minors are monomials of $x$'s, $(1,2,\cdots,j, l, l{+}1,\cdots, l{+}k{-}j{-}1)$ with $0\leq j\leq k{-}2$  and $j{+}3\leq l \leq n{-}(k{-}j{-}1)$. For example,  for $(k,n)=(3,6)$ we have
 \be 
 (3,4,5) = x_1 x_3, \quad
 (4,5,6) = x_1 x_2 x_3^2 x_4 , \quad
  (1,4,5) = x_3 , \quad
 (1,5,6) = x_3 x_4 .
 \ee  
The remaining ${n \choose k}{-}n{-}d$ minors are non-trivial polynomials with positive coefficients, and for each of them one can always factor out a monomial of $x$'s which gives a polynomial $p_I$ with constant term $1$. We can reorganize the regulator into $d$ monomials and ${n \choose k}-n-d$ polynomials, so the Grassmannian string integral is written in the standard form:
 \ba\label{fqe3fq}
{(\a')^{d}}~\int_{\mathbb{R}_{>0}^d} \,  \prod _{a=1}^{d} d\log x_a \, R_{k,n}\quad  ~{\rm with}~   R_{k,n}=
\prod_{a=1}^d x_a^{\a' X_a}  \prod_{I} p_I(\{x\})^{-\a' c_I}\,,
\ea
where $c_I=-s_I$ for each polynomial (non-trivial minor) $p_I$, and we collect all contributions to $x_a$ thus its exponent $X_a$ is a sum of the $s_I$. Note that this simple rewriting already gives a family of $d$-dim subspaces of the kinematic space for $(k,n)$, where the polytope ${\cal P}(k,n)$ lives in. Such a subspace is defined by setting all $c_I$ to be {\it positive} constants, and it can be spanned using the basis ${\bm X}=\{X_1,\cdots,X_d\} $. 
Now the polytope ${\cal P}(k,n)$ can be cut out by all inequalities $F_a ({\bf X}, \{c\}) \geq 0$ for $a=1,2, \cdots, N$, where $F_a$ is a linear combination of $X$'s and $c$'s. In the rest of the paper, we will present these facets (which are poles of ${\cal A}_{k,n}$), and by \cite{Arkani-Hamed:2019mrd} they can be computed by taking the Minkowski sum of Newton polytopes for all $p_I$, 
${\cal P}(k,n)=\sum_I c_I {\bf N}(p_I)$. The same polytope ${\cal P}_{k,n}$ can be obtained as the image of the scattering-equation (SE) map from the integration domain ${\mathbb R}_{>0}^{d}$. 
 
\begin{figure}[!htb]
\centering
\subfloat[${\cal P}(3,5)$]{
\begin{tikzpicture}
\draw[->,blue] (-1,0) -- (4.5,0) node[right]{$X_1$};
\draw[->,blue] (0,-1) -- (0,4) node[above]{$X_2$};
\draw[thick] (0,0) -- node[left,brown]{$s_{345}$} (0,3) node[left,blue]{$c_{135}+c_{235}$} -- node[above,brown]{$s_{512}$} (3,3) -- node[right,brown]{$s_{234}$} (3,1) -- node[left,brown]{$s_{451}$} (2,0) node[below,blue]{$c_{245}\ $} -- node[above,brown]{$s_{123}$} (0,0);
\draw[thick,dashed] (3,1) -- (3,0) node[below,blue]{$\quad c_{235}+c_{245}$};
\end{tikzpicture}
}\subfloat[ ${\cal P}^\circ(3,5)$ and its triangulation]{
\begin{tikzpicture}
\draw[->,blue] (-2.5,0) -- (2.5,0) node[right]{};
\draw[->,blue] (0,-2.5) -- (0,2.5) node[above]{};
\draw[thick] (1.5,0) node[below,brown]{$\qquad s_{345}$} -- (0,-1.5) node[right,brown]{$\ s_{512}$} -- (-1.5,0) node[below,brown]{$s_{234}\quad$} -- (-1.5,1.5) node[above,brown]{$s_{451}$} -- (0,1.5) node[right,brown]{$\ s_{123}$} -- (1.5,0);
\draw[dashed,thick] (0,0) -- (-1.5,1.5) (0,0)--(1.5,0)(0,0)--(-1.5,0)(0,0)--(0,1.5)(0,0)--(0,-1.5);
\end{tikzpicture}
}
\caption{\label{qiupfhq9} Polytope ${\cal P}(3,5)$ and its dual ${\cal P}^\circ(3,5)$ for $G_+(3,5)/T$ }
\end{figure}
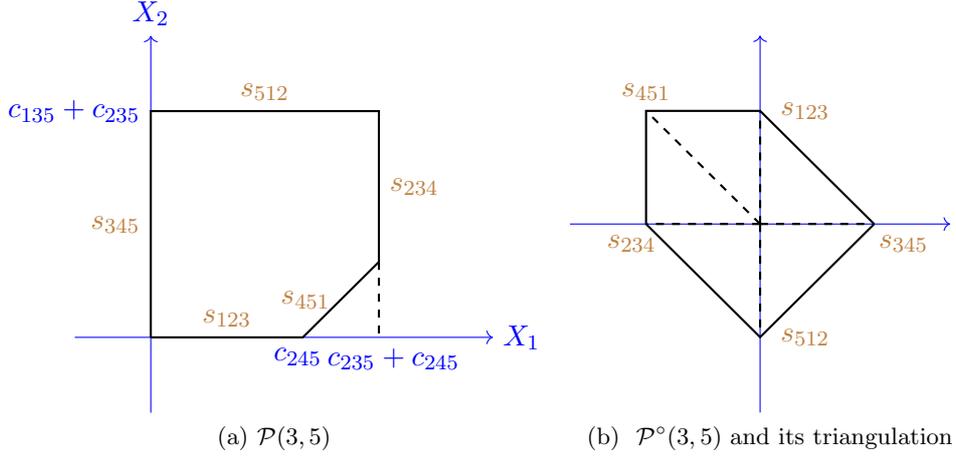
   
{\textbf {Example}  
$(k,n)=(3,5)$.}   Using web variables, the regulator reads
 \ba\label{qiohf}
 R_{3,5}=
x_1^{\a'X_1} x_2^{\a'X_2} \left(x_1+1\right){}^{-\a'c_{245}} \left(x_2+1\right){}^{-\a'c_{135}} \left(x_1
   x_2+x_2+1\right){}^{-\a'c_{2
   35}} \,,
   \ea
  with 
$  X_1= s_{345}, \,X_2= s_{123}$. The Newton polytopes from the regulator  $ R_{3,5}$ are two intervals ${\bf conv} (\{0,0\},\{1,0\})$, ${\bf conv} (\{0,0\},\{0,1\})$  and a triangle ${\bf conv}(\{0,0\},$ $\{0,1\},\{1,1\})$. The Minkowski sum of them (weighted by the $c$'s) gives a pentagon ${\cal P}(3,5)$ shown in Fig. \ref{qiupfhq9} (a). The facets of ${\cal P}(3,5)$, which correspond to the poles of ${\cal A}_{3,5}$, are
    \begin{alignat}{4}\label{facet35}
  X_1&=0 , \quad s_{345} ~~~~~~~& c_{235}+c_{245}-X_1&=0 , \quad s_{234}  \nl
      X_2&=0 ,  \quad s_{123}~~~~~~~&  c_{135}+c_{235}-X_2&=0 ,\quad s_{512}  \nl
&&c_{245}-X_1+X_2&=0, \quad s_{451} 
  \end{alignat}
where we have seen that each facet/pole can be written more invariantly as $s_{i, i{+}1, i{+}2}$ for $i=1,2,\cdots 5$. The pentagon ${\cal P}(3,5)$ can also be obtained as the image of SE map
\ba \label{se35}
X_1= x_1 \left(\frac{c_{245}}{1+x_1}+\frac{c_{235}\, x_2}{1+x_2+x_1x_2} \right)\,,\quad
X_2= x_2 \left(\frac{c_{135}}{1+x_2}+\frac{c_{235}\, (1+x_1)}{1+x_2+x_1x_2} \right)\,.
\ea 
Obviously the boundary $x_i=0$ of the domain, is mapped to the facet $X_i =0$ of ${\cal P}(3,5)$, for $i=1,2$.  With $x_1\to\infty$ and $x_2 \to \infty$, the map gives facets $c_{235}+c_{245}-X_1=0$ and $c_{135}+c_{235}-X_2=0$, respectively. The last facet is more non-trivial which can be seen as blowing up the corner with $x_1\to \infty$ and $x_2\to 0$. Let's use $x_1={\hat x}_1\, \epsilon^{-1}, x_2={\hat x}_2 \,\epsilon$  with $\epsilon\to 0$ to parametrize this corner, then eq. \eqref{se35} becomes
  \ba \label{se325}
X_1\to c_{245}+\frac{c_{235}\, \hat{x}_1 \hat{x}_2}{1+\hat{x}_1\hat{x}_2} \,,
\quad 
X_2\to \frac{c_{235}\, \hat{x}_1\hat{x}_2}{1+\hat{x}_1\hat{x}_2} \,,
\ea 
which implies that  $X_1-X_2\to c_{245}$. Thus under SE map it indeed gives the last facet in \eqref{facet35}, and we see $\hat{x}_1\hat{x}_2$ can be used as a new variable to parameterize this facet. 
  
Finally, let's comment on computing the canonical function ${\cal A}_{k,n}$ once the polytope is obtained, which is in fact the volume of the dual polytope ${\cal P}^\circ (k,n)$~\cite{Arkani-Hamed:2017tmz}. It is straightforward to triangulate ${\cal P}^\circ (k,n)$ by using the reference point $(1,0,\cdots,0)$ inside it; the total volume is given by the sum of the volume of ``cells" formed by connecting the point to each facet of  ${\cal P}^\circ (k,n)$, which corresponds to a vertex of ${\cal P}(k,n)$. The facet is a $(d{-}1)$-dim simplex if and only if the corresponding vertex in ${\cal P}(k,n)$ is simple, in which case the cell is a $d$-dim simplex, with volume given by the  product of $1/F$'s for the $d$ facets adjacent to that vertex in ${\cal P}(k,n)$. Otherwise, we have to triangulate that facet into simplices. We emphasize that there are numerous other triangulations and ways for evaluating the canonical function, and once the (invariant) polytope is given, it is not very important which particular triangulation we use. For example, the dual polytope ${\cal P}^\circ(3,5)$ and its triangulation are shown in figure \r{qiupfhq9} (b).  In this case, all facets of the dual polytope are simplices (intervals) and the contribution to the volume from the $i$-th facet is $1/(s_{i{-}2, i{-}1, i} s_{i, i{+}1, i{+}2})$, and the sum obviously gives the amplitudes $m_5^{(3)}(\mathbb{I}|\mathbb{I})$.  

Next, let us study the first non-trivial example with $(k,n)=(3,6)$. 

\subsection{Example: ${\cal P}(3,6)$ and its generalized amplitudes}

Using web variables,  the regular for $G_+(3,6)/T$ becomes 
   \ba\label{cqeh;lk}
  R_{3,6}=\,
&x_1^{\a' X_1}  x_2^{\a' X_2}   x_3^{\a' X_3} x_4^{\a' X_4}  \left(x_1+1\right){}^{-\a' c_{245}} \left(x_2+1\right){}^{-\a' c_{135}}    \left(x_4+1\right){}^{-\a' c_{146}}
\nl
   & 
     \left(x_1 x_2+x_2+1\right){}^{-\a' c_{235}} 
\left(x_2 x_3+x_3+1\right){}^{-\a' c_{356}}
   \left(x_3 x_4+x_4+1\right){}^{-\a' c_{346}}  \nl
   &\left(x_1x_3 +x_1+1\right){}^{-\a' c_{256}}
   \left(x_2 x_4+x_2+1\right){}^{-\a' c_{136}}   \left(x_1x_3 x_4 + x_1x_4+x_1+x_4+1\right){}^{-\a' c_{246}}
   \nl& \left(x_1 x_2x_3 x_4 +x_1  x_2x_4+x_1 x_2+x_2x_4
   +x_2+1\right){}^{-\a' c_{236}}\,,
   \ea
where we have $X_2=s_{123}$, $X_3=s_{456}$, and $X_1=S_{3456}$, $X_4=S_{1234}$. For the latter we have defined $S_{A}$ for any subset $A$ with $|A|\geq 3$  (for $|A|=3$, $S_A:=s_A$):
   \be \label{fhuaihfjd}
    S_{A}:= \sum_{a_1<a_2<a_3 \in A }s_{a_1,a_2,a_3}\,.
    \ee 

It is straightforward to compute the polytope ${\cal P}(3,6)$ by taking the Minkowski sum of Newton polytopes from $R_{3,6}$. Let's list the result first, namely the $16$ linear functions for the facets which cut out the polytope (or the $16$ poles of the canonical function). Clearly $X_1$ to $X_4$ are $4$ of the facets, and by cyclic permutations we expect the following two cyclic classes:
\be\label{fasiduhf}
S_{123}, \cdots, S_{612};\quad S_{1234}, \cdots, S_{6123}\,.
\ee
Indeed we find these $12$ facets, but there are $4$ additional ones, which are of a distinct type. Let's first define $S_{(B)}=S_{\bar{B}}$ where $\bar{B}$ denotes the complement set, for any subset $B$ with $2\leq |B|\leq n-3$   (this notation will become clear in the next section), thus {\it e.g.} $S_{1234}=S_{(56)}$. With this definition it turns out it is convenient to define a new type of facets
\be \label{fjadsfi}
   S_{A (B)}:= S_{A}+S_{(B)}+S_{A|B}   \qquad {\rm with} \quad  S_{A|B}:=  \sum_{a_1<a_2 \in A, b\in B }s_{a_1 a_2 b}\,,
 \ee 
 which can also be written as $S_{(B)A}$. Now the remaining $4$ facets are
 \be\label{ajfshdk}
 S_{12(34)}, S_{23(45)}; \quad S_{(12)34}, S_{(23)45}\,.
 \ee
We emphasize that the $16$ facets/poles of \eqref{fasiduhf} and \eqref{fjadsfi} are result from Minkowski sum, which express them as linear combinations of $X$'s and $c$'s, and here are a few examples with rewriting in our new notation
\ba \label{jafsd}
  X_1&=0 , \qquad S_{3456} 
\nl
  c_{235}+c_{236}+c_{245}+c_{246}+c_{256}-X_1&=0 ,\qquad S_{234} 
  \nl
c_{346}+c_{356}+X_1-X_3&=0 , \qquad S_{345}
\nl 
c_{356}+X_1-X_3+X_4&=0. \qquad S_{12(34)} \,
  \ea  
For example for the last one we indeed find 
\be
c_{356}+X_1-X_3+X_4= (s_{125}+s_{256}+s_{561}+s_{612})+(s_{123}+s_{124})=S_{12(34)}\,.
\ee

Remarkably, these facets, and the polytope cut out by them, can be obtained using SE map \eqref{sesss}. For example, the equation for $X_1$ is 
\begin{align}
X_1=  x_1 & \left(\frac{x_3 c_{235}}{x_1 x_3+x_3+1}+\frac{\left(x_2  x_3x_4+ x_3x_4+x_3\right)
   c_{236}}{x_1 x_2  x_3 x_4 +x_1  x_3x_4+x_1 x_3+ x_3x_4+x_3+1}+\frac{c_{245}}{x_1+1}
  \right. \nl& \left.
   +\frac{\left(x_2
   x_4+x_4+1\right) c_{246}}{x_1x_2 x_4 +x_1x_4 +x_1+x_4+1}+\frac{\left(x_2+1\right) c_{256}}{x_1x_2
   +x_1+1}\right).
   \label{se36}
\end{align}
The first two facets in \eqref{jafsd} can be obtained by taking the limit of $x_1\to0$ or $\infty$ respectively. More generally, a facet is mapped from taking limits for multiple  web variables. For example, the limit $x_1= {\hat x}_1 \,\epsilon$,  $x_4= {\hat x}_4 \,\epsilon$, $x_3= {\hat x}_3 \,\epsilon^{-1}$ with $\epsilon\to  0$ gives the last facet in \eqref{jafsd}. 

\begin{figure}[!htb]
\centering

\begin{tikzpicture} [xscale=1.55,yscale=1.55]

\begin{scope}[xshift=0]
\draw[thick] (1.59,-5.27) --(1.73,-4.15) --(2.91,-2.56)--(3.35,-2.44)--(3.87,-2.67)--(4.67,-4.48)--(4.36,-5.50)--(3.15,-6.39)-- (1.59,-5.27)  
 (3.15,-6.39)--(3.28,-5.87)--(2.80,-4.81) --(1.73,-4.15)  
 (3.28,-5.87)--(3.89,-4.96)--(3.42,-3.93) --(2.80,-4.81)
(3.89,-4.96)--(4.67,-4.48)
 (3.42,-3.93)--(3.44,-2.82)--(3.87,-2.67)
 (3.44,-2.82)--(2.91,-2.56);
 \draw[thick,blue,dashed] (1.59,-5.27) --(2.88,-4.58)--(4.36,-5.50)
 (2.88,-4.58)--(3.35,-2.44) ;

 \node at (2.55,-5.30) {$S_{1234}$};
  \node[blue] at (2.34,-3.93) {$S_{4561}$};
  \node at (2.67,-3.41) {$S_{561}$};
   \node at (3.37,-2.28) {$S_{(23)45}$};
    \node at (3.76,-3.52) {$S_{345}$};
      \node[blue] at (3.98,-4.14) {$S_{6123}$};
          \node at (3.86,-5.52) {$S_{3456}$};
 \node[blue] at (2.79,-5.65) {$S_{456}$};
 \node at (3.28,-5.09) {$S_{12(34)}$};
   \node at (3.39,-6.70) {(a) Facet  $S_{123}$};
 \end{scope}
 \begin{scope}[xshift=-.6cm]
 \draw[thick] (5.65,-5.46) --(5.46,-3.69)--(5.75,-2.93)
 node{\tikz{\filldraw (0,0)  circle (2pt);}} node[above=1pt]{non-simple vertex}
 --(6.65,-2.81)--(8.83,-3.38)--(8.68,-4.85)--(8.21,-5.54)--(7.13,-6.02)--(5.65,-5.46)
 (5.46,-3.69)--(8.28,-4.55)--(8.21,-5.54)
 (8.28,-4.55)--(8.58,-3.70)--(8.83,-3.38)
 (8.58,-3.70)--(5.75,-2.93);
  \draw[thick,blue,dashed] (5.65,-5.46)--(5.94,-5.16)--(6.70,-5.01)--(7.64,-5.35)--(7.13,-6.02)
 (7.64,-5.35)--(8.68,-4.85)
(5.94,-5.16)-- (5.75,-2.93) 
(6.70,-5.01)--(6.65,-2.81);

 \node[blue] at (7.90,-5.46) {$S_{4561}$};
 \node[blue] at (6.58,-5.49) {$S_{123}$};
  \node[blue] at (5.74,-4.40) {$S_{3456}$};
  \node at (8.49,-4.62) {$S_{234}$};
    \node at (7.00,-4.80) {$S_{456}$};
   \node at (7.40,-3.19) {$S_{5612}$};
    \node at (7.24,-3.83) {$S_{(12)34}$};
      \node[blue] at (6.27,-4.18) {$S_{12(34)}$};
         \node at (7.17,-6.70) {(b) Facet  $S_{1234}$};
    \end{scope}       

 \begin{scope}[xshift=-.9cm]
 \draw[thick] (9.55,-5.72)-- (9.19,-4.20)--(10.31,-3.76)--(12.15,-3.79)--(12.16,-5.32)--(11.24,-5.82)--(9.55,-5.72)
(9.19,-4.20)--(11.10,-4.25)-- (12.15,-3.79)
(11.10,-4.25)--(11.24,-5.82);
  \draw[thick,blue,dashed] (9.55,-5.72)--(10.53,-5.24)--(12.16,-5.32)
  (10.53,-5.24)--(10.31,-3.76)
  ;

  \node[blue] at (9.76,-4.64) {$S_{3456}$};
 \node[blue] at (11.49,-4.52) {$S_{345}$};
  \node[blue] at (10.81,-5.55) {$S_{123}$};
  \node at (9.95,-5.08) {$S_{1234}$};
   \node at (10.70,-3.97) {$S_{5612}$};
    \node at (11.58,-4.99) {$S_{561}$};
       \node at (10.63,-6.70) {(c) Facet $S_{12(34)}$};
\end{scope}

\end{tikzpicture}

\caption{\label{fmfafqoiejf} Three kinds of facets in ${\cal P}(3,6)$. Here  a 2-plane is an intersection of two facets. For example, the bottom square in (a) is the intersection of facets $S_{456}$ and $S_{123}$. }
\end{figure}
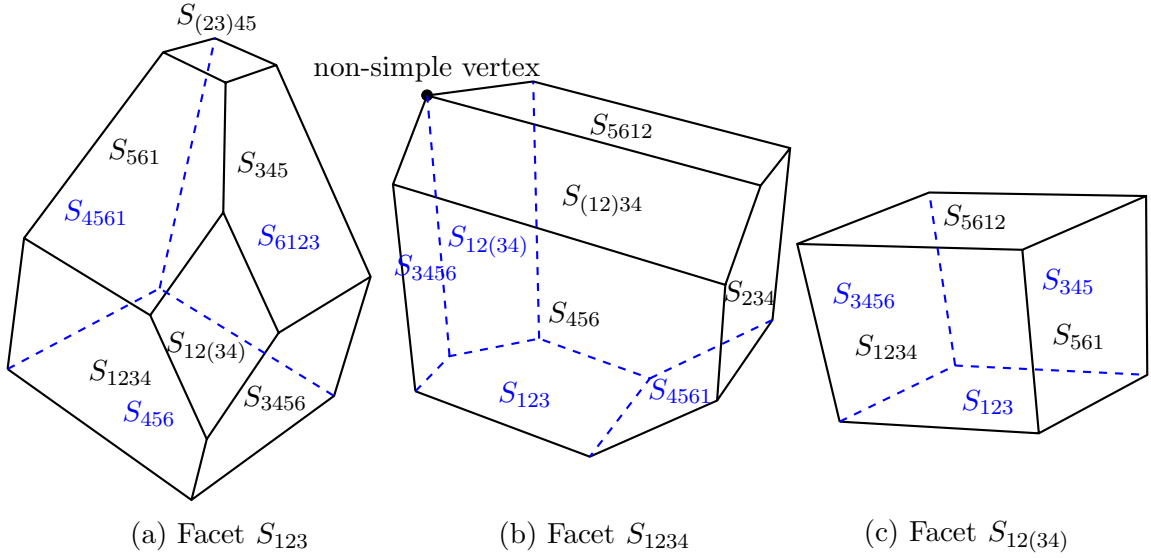

Before proceeding, let us comment on identities satisfied by the last type of facets. On the support of ``momentum conservation" \eqref{fqfeq}, 
 we have an identity, 
\ba\label{eqidentity3}
  & S_{A(B)}=   S_{B(C)}=   S_{C(A)} \quad \forall A\sqcup B \sqcup C= \{1,\cdots,n\}\,.
\ea
Thus {\it e.g.} we have $S_{12(34)}=S_{34(56)}=S_{56(12)}$, $S_{23(45)}=S_{45(61)}=S_{61(23)}$, so only $2$ of these $6$ cyclic images are distinct; the other orientation also gives $2$ distinct facets, which are the $4$ in \eqref{ajfshdk}. 
The list of these $16$ facets, which can be expanded in $X$'s and $c$'s, is a rather compact and invariant way to describe the polytope. Note that there are three kinds of facets for ${\cal P}(3,6)$, which are all three-dimensional polytopes and we draw them in figure  \ref{fmfafqoiejf}. 
 
 Last but not least, let us also describe the $48$ vertices which can be computed from the facets. Out of them, $46$ of them are simple vertices, and the remaining two are non-simple ones. The former correspond to facets of the dual polytope ${\cal P}^\circ(3,6)$ that are simplices, and the latter correspond to non-simplicial ones. 
Examples for these two kinds are given in figure \ref{fmfafqoiejf} (b).   The $f$-vector  for the polytope ${\cal P}(3,6)$ is  $(1,\,48,\,98,\,66,\,16,\,1)$. 

\begin{figure}[!htb]
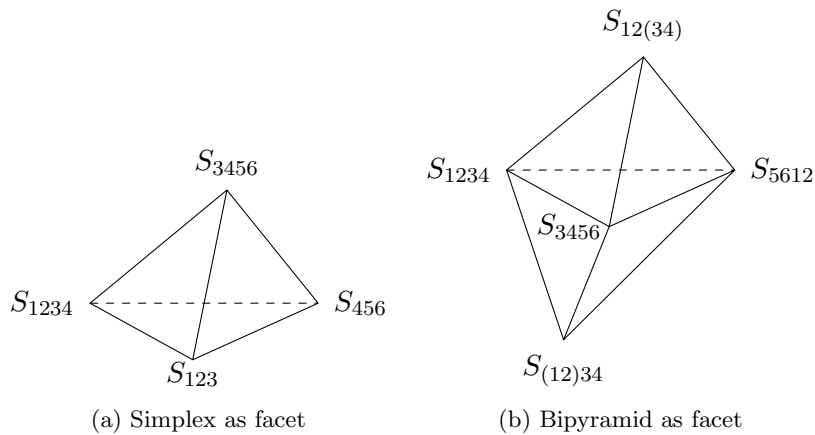

  \centering
  \subfloat[Simplex as facet]{
  \tikz[scale=1.5]{
  \draw[dashed] (0,0)--(2,0);
   \draw (0,0) node [left=2pt] {$S_{1234}$} (2,0)--(1.2,1) node [above=2pt] {$S_{3456}$} --(0,0)-- (.9,-.5) node [below=-3pt] {$S_{123}$} --(2,0) node [right=2pt] {$S_{456}$}   (.9,-.5)--(1.2,1);
  }
  }  \subfloat[Bipyramid as facet]{
  \tikz[scale=1.5]{
  \draw[dashed] (0,0)--(2,0);
   \draw (0,0) node [left=2pt] {$S_{1234}$}--(.5,-1.5) node [below=2pt] {$ S_{(12)34}$} --(2,0)--(1.2,1) node [above=2pt] {$S_{12(34)}$} --(0,0)-- (.9,-.5)  --(2,0) node [right=2pt] {$S_{5612}$}  (.5,-1.5)-- (.9,-.5)--(1.2,1);
   \node at (.55,-.5) {$S_{3456}$};
  }
  }
   \caption{\label{fhausfkj} Two examples of facets of the dual polytope ${\cal P}^\circ(3,6)$ }
    \end{figure}

As we have mentioned, the canonical function, ${\cal A}_{3,6}$ is given by the volume of the dual polytope $\dcalP(3,6)$, which can be triangulated using the reference point $(1,0,0,0,0)$ in ${\mathbb{CP}}^4$ and written as a sum over its facets. For a simplex such as the one in figure \ref{fhausfkj} (a), the contribution to the volume is simply the product of inverse facets $\frac{1}{S_{123}S_{456}S_{1234}S_{3456}}$.  For a non-simplicial facet of the dual polytope such as the one in shown in figure \ref{fhausfkj} (b), one can triangulate it further. Note that it is a bipyramid with $5$ vertices, which in the original polytope ${\cal P}(3,6)$ correspond to the $5$ facets touching the vertex, namely $S_{1234},S_{3456},S_{5612},S_{12(34)},S_{(12)34}$. One way to triangulate it is to cut it into two simplices by the 2-plane passing the vertices   $S_{1234},S_{3456},S_{5612}$ in the dual polytope. Therefore, the contribution to the  volume is given by \be 
\frac{1}{S_{1234}S_{3456}S_{5612}S_{12(34)}}+\frac{1}{S_{1234}S_{3456}S_{5612}S_{(12)34}}\,.
\ee 
  
By adding up the contribution from $48$ facets of ${\cal P}^\circ(3,6)$, we obtain its  volume, or equivalently the canonical function of ${\cal P}(3,6)$, which we also refer to as the generalized amplitudes ${\cal A}_{3,6}$. We emphasize that there are numerous ways to compute ${\cal A}_{3,6}$ by triangulating the dual (or the polytope itself). The invariant object is the polytope ${\cal P}(3,6)$ cut out by the $16$ facets as we discussed. 

\section{Polytopes, amplitudes and boundary configurations for $k=3$ }

Using the methods of Minkowski sum and SE map, we have computed the polytope ${\cal P}(3,n)$ up to $n=10$; since the subspace is already given in the above section (setting all $c_I$ in \eqref{fqe3fq} as positive constants), the knowledge of all facets (or equivalently all the poles of the canonical form) suffices to determine ${\cal P}(3,n)$ completely.  As we discussed, a natural basis for the subspace ${\bm X}$ is given for all multiplicity,
\be\label{xbase}
{\bm X}= \bigcup_{j=3}^{n-2} \big\{S_{1,2,\cdots,j}, \,  S_{(1,2,\cdots,j-1)} \big\}\,.
\ee 
We emphasize that by expanding the facets/poles into the basis ${\bm X}$ and the constants in \eqref{fqe3fq} cuts out the polytope ${\cal P}(3,n)$.  
 
As we have seen, any facet of ${\cal P}(k, n)$ is the image of certain boundary configurations of $G_+(k,n)/T$ under SE map. Given a parametrization, such a boundary configuration can be described by taking certain limit of the web variables, which is controlled by one parameter $\epsilon \to 0$. Recall that the SE map is given by ${\rm d}\log R_{k, n}=0$, and under the limit that approaches a co-dimension 1 boundary, ${\rm d}\log R_{k,n}$ becomes proportional to ${\rm d}\log \epsilon$, and the residue at $\epsilon \to 0$ is nothing but the linear function for the corresponding facet \footnote{In particular, for $k=4$, which will be discussed in the next section, we also denote the map as ${\bm S}[\gamma]$.}
\be \label{map}
S[\gamma]:=\frac{1}{\a'}{\rm Res}_{\epsilon=0} \,{\rm d}\log R_{k,n}(\gamma,\epsilon) \,,
\ee
where we have denoted the boundary configuration as $\gamma$, and the corresponding facet $S[\gamma]$. This is a direct consequence of our SE map, and has also been shown in the context of higher-$k$ CHY formula in \cite{Cachazo:2019ngv}. Therefore, once we have obtained a boundary configuration we can directly read off the facet by computing the residue of ${\rm d}\log R_{k,n}$. 

On the other hand, already in the simplest examples of $n=5,6$ with $k=3$, we have seen that in order to describe these facets in an invariant fashion, {\it i.e.} independent of parametrization, it is natural to introduce the notation $S_A$, $S_{(B)}$  and $S_{A(B)}$. As we will see shortly, such a notation exactly reflects the underlying boundary configurations for general $n$, and the latter all have very natural geometric interpretations. Restricting to $k=3$ case for now, note that ${\rm d}\log R_{3,n} =
 \sum_{a<b<c} s_{a\,b\,c} \, {\rm d} \log (a, b, c)$, and one way to describe a boundary configuration $\gamma$ is to list at which order every minor $(a,b,c)$ vanishes in the limit $\epsilon\to 0$. We will show that each facet for ${\cal P}(3,n)$ corresponds to a boundary configuration of $G_+(3, n)/T$, which can be encoded geometrically by how the $n$ points on a plane (columns of the $3\times n$ matrix with first components fixed to be $1$) become collinear or colliding with each other. 
 
The simplest example is the configuration for the facet $S_A$ which turns out to correspond to a set of points becoming collinear, {\it i.e.}
 \be 
 (a_1,a_2,a_3)={\cal O}(\epsilon), \quad \forall\, a_1,a_2,a_3 \in A\,,
 \ee 
where $a_1,a_2,a_3$ are any $3$ different labels in $A$, and we will use a line to denote this configuration. For example, for $A=\{1,2,3,4\}$, we have
 \raisebox{-.35cm}{
\begin{tikzpicture}[scale=.3]
\filldraw (-1.5,0) node{\tikz{\filldraw (0,0)  circle (1pt);}} node[below]{$1$} circle (0pt);
\filldraw (0,0) node{\tikz{\filldraw (0,0)  circle (1pt);}} node[below]{$2$} circle (0pt);
\filldraw (1.5,0) node{\tikz{\filldraw (0,0)  circle (1pt);}} node[below]{$3$}  circle (0pt);
\filldraw (3,0)  node{\tikz{\filldraw (0,0)  circle (1pt);}}  node[below]{$4$}  circle (0pt);
\draw (-1.5,0) -- (3,0) ;
\end{tikzpicture}}\,, and under the SE map \eqref{map} it gives
\be 
S\left[\!\!\!
\raisebox{-.45cm}{
 \begin{tikzpicture}[scale=.3]
\filldraw (-1.5,0) node{\tikz{\filldraw (0,0)  circle (1pt);}} node[below]{$1$} circle (0pt);
\filldraw (0,0) node{\tikz{\filldraw (0,0)  circle (1pt);}} node[below]{$2$} circle (0pt);
\filldraw (1.5,0) node{\tikz{\filldraw (0,0)  circle (1pt);}} node[below]{$3$}  circle (0pt);
\filldraw (3,0)  node{\tikz{\filldraw (0,0)  circle (1pt);}}  node[below]{$4$}  circle (0pt);
\draw (-1.5,0) -- (3,0) ;
\end{tikzpicture}}
\right]= {\rm Res}_{\epsilon=0}\left( 
\frac{{\rm d} \epsilon}{\epsilon}   \sum_{1\leq a_1<a_2<a_3\leq 4} s_{a_1\,a_2\,a_3} + {\cal O}(\epsilon^0)\right)
= S_{1234}\,. 
\ee 
 
Similarly, we can have a set of points in $B$ collide or coincide with each other, which means that the minors involving $3$ or $2$ of these labels vanish as ${\cal O}(\epsilon^2)$ or ${\cal O}(\epsilon)$, respectively
\be 
 (b_1,b_2,b_3)={\cal O}(\epsilon^2), \quad \forall\, b_1,b_2,b_3 \in B\,,
 \qquad 
  (b_1,b_2,i)={\cal O}(\epsilon), \quad \forall\, b_1,b_2 \in B, i\notin B\,.
 \ee 
We denote the configuration by a black dot \raisebox{-.35cm}{
\begin{tikzpicture}[scale=.3]
\filldraw (0,0) node{\tikz{\filldraw (0,0)  circle (2pt);}} node[below]{$B$} circle (0pt);
\end{tikzpicture}}, and the corresponding facet reads
\be 
S\left[\!\!\!
\raisebox{-.25cm}{
\begin{tikzpicture}[scale=.3]
\filldraw (0,0) node{\tikz{\filldraw (0,0)  circle (2pt);}} node[below]{$B$} circle (0pt);
\end{tikzpicture}} \right]= {\rm Res}_{\epsilon=0}\left( 
\frac{{\rm d} \epsilon}{\epsilon}   \sum_{b_1<b_2<b_3\in B} 2 s_{b_1\,b_2\,b_3} +\frac{{\rm d} \epsilon}{\epsilon}   \sum_{b_1<b_2\in B,\,i\notin B}  s_{b_1\,b_2\,i}+ {\cal O}(\epsilon^0)\right)\equiv S_{(B)}
\,. 
\ee 
By momentum conservation it is easy to show that in fact $S_{(B)}$ is equal to $S_{\bar B}$ where ${\bar B}$ denotes the complementary set of $B$:
\be 
S_{(B)}=2 \sum_{b_1<b_2<b_3\in B}  s_{b_1\,b_2\,b_3} +  \sum_{b_1<b_2\in B,\,i\notin B}  s_{b_1\,b_2\,i}
  = \sum_{a_1<a_2<a_3\in {\bar B}} s_{a_1\, a_2\, a_3}=S_{\bar B} 
\,. 
\ee 
Thus we say that the configuration where points in $B$ collide is equivalent to the one where points in ${\bar B}$ become collinear $
\raisebox{-.25cm}{
\begin{tikzpicture}[scale=.3]
\filldraw (0,0) node{\tikz{\filldraw (0,0)  circle (2pt);}} node[below]{$B$} circle (0pt);
\end{tikzpicture}} \simeq \raisebox{-.3cm}{
 \begin{tikzpicture}[scale=.6 
 ]
\draw (-1,0) node{\tikz{\filldraw (0,0)  circle (1pt);}}--(-.5,0) node{\tikz{\filldraw (0,0)  circle (1pt);}}--(0,0) node[below] {${\bar B}$}-- (1,0)node{\tikz{\filldraw (0,0)  circle (1pt);}}  ;
\end{tikzpicture}}\,$. 
We emphasize that such an equivalence between boundary configurations of $G_+(k,n)/T$ can also be verified by direct computation, but here we see how it directly follows from momentum conservation. For example \be
S\left[\!\!\!
\raisebox{-.45cm}{
 \begin{tikzpicture}[scale=.3]
\filldraw (0,0) node{\tikz{\filldraw (0,0)  circle (2pt);}} node[below]{$1,2,3,4$} circle (0pt);
\end{tikzpicture}}
\right]= S_{(1234)}=S_{56\cdots n}\,. 
\ee 

The last class of configurations which have appeared for $G_+(3,6)/T$ are those for the facets $S_{A(B)}$. As the notation suggests, such a configuration corresponds to the case where   points in $B$ collide and at the same time they are also collinear with points in $A$. For example,  for $S_{12(34)}$, we use \raisebox{-.5cm}{
 \begin{tikzpicture}[scale=.6 
 ]
\filldraw (-1.5,0) node{\tikz{\filldraw (0,0)  circle (1pt);}}  node[below]{$1$} circle (1pt);
\filldraw (0,0) node{\tikz{\filldraw (0,0)  circle (1pt);}} node[below]{$2$} circle (1pt);
\filldraw (1.5,0) node{\tikz{\filldraw (0,0)  circle (2pt);}} node[below]{$3,4$}  circle (2pt);
\draw (-1.5,0) -- (1.5,0) ;
\end{tikzpicture}} to  denote its configuration. Here $\{1,2,3\}$ and $\{1,2,4\}$ are collinear respectively, which means $(1,2,3)={\cal O}(\epsilon)$, $(1,2,4)={\cal O}(\epsilon)$. In addition, $3$ and $4$ collide, thus $(3,4, i)={\cal O}(\epsilon)$ for any $i\neq 3,4$. Using the map \eqref{map}, we have   
\be 
S\left[
\raisebox{-.5cm}{
 \begin{tikzpicture}[scale=.6 
 ]
\filldraw (-1.5,0) node{\tikz{\filldraw (0,0)  circle (1pt);}}  node[below]{$1$} circle (1pt);
\filldraw (0,0) node{\tikz{\filldraw (0,0)  circle (1pt);}} node[below]{$2$} circle (1pt);
\filldraw (1.5,0) node{\tikz{\filldraw (0,0)  circle (2pt);}} node[below]{$3,4$}  circle (2pt);
\draw (-1.5,0) -- (1.5,0) ;
\end{tikzpicture}}
\right]= s_{123}+s_{124}+ \sum_{i\neq 3,4}  s_{3\,4\,i}\,,
\ee 
which is exactly the definition of $S_{A(B)}$ given in \eqref{fjadsfi}.

Note that we also have equivalence relation for this class of configurations, which can be  diagrammatically shown by
\be \label{jfasd}
\raisebox{-.7cm}{
 \begin{tikzpicture}[scale=.5]
\filldraw (-1.5,0) node[below]{} circle (0pt);
\filldraw (0,0) node[below]{$A$} circle (0pt);
\filldraw (1.5,0) node{\tikz{\filldraw (0,0)  circle (2pt);}}  node[below]{$B$}  circle (2pt);
\draw (-1.5,0) -- (1.5,0) ;
\node at (-1,1.5) {$C$};
\end{tikzpicture}}
\simeq
\raisebox{-.7cm}{
 \begin{tikzpicture}[scale=.5]
\filldraw (-1.5,0) node[below]{} circle (0pt);
\filldraw (0,0) node[below]{$B$} circle (0pt);
\filldraw (1.5,0)node{\tikz{\filldraw (0,0)  circle (2pt);}}  node[below]{$C$}  circle (2pt);
\draw (-1.5,0) -- (1.5,0) ;
\node at (-1,1.5) {$A$};
\end{tikzpicture}}
\simeq
\raisebox{-.7cm}{
 \begin{tikzpicture}[scale=.5]
\filldraw (-1.5,0) node[below]{} circle (0pt);
\filldraw (0,0) node[below]{$C$} circle (0pt);
\filldraw (1.5,0) node{\tikz{\filldraw (0,0)  circle (2pt);}}  node[below]{$A$}  circle (2pt);
\draw (-1.5,0) -- (1.5,0) ;
\node at (-1,1.5) {$B$};
\end{tikzpicture}}
\,,
\ee
where $A \sqcup B \sqcup C=\{ 1, 2, \cdots, n\}$, which is consistent to the identity 
\eqref{eqidentity3} for $S_{A(B)}$ due to momentum conservation.
For example, for $n=6$, we have 
 \be 
\raisebox{-.7cm}{
 \begin{tikzpicture}[scale=.5]
\filldraw (-1.5,0)node{\tikz{\filldraw (0,0)  circle (1pt);}}  node[below]{1} circle (1pt);
\filldraw (0,0) node{\tikz{\filldraw (0,0)  circle (1pt);}}  node[below]{$2$} circle (1pt);
\filldraw (1.5,0)
node{\tikz{\filldraw (0,0)  circle (2pt);}} 
node[below]{$3,4$}  circle (2pt);
\draw (-1.5,0) -- (1.5,0) ;
\filldraw(-.4,1.5) node[below]{$5$}  circle (2pt);
\filldraw  (-1,1.2) node[below]{$6$}  circle (2pt);
\end{tikzpicture}}
\simeq
\raisebox{-.7cm}{
 \begin{tikzpicture}[scale=.5]
\filldraw (-1.5,0)node{\tikz{\filldraw (0,0)  circle (1pt);}}  node[below]{5} circle (1pt);
\filldraw (0,0) node{\tikz{\filldraw (0,0)  circle (1pt);}}  node[below]{$6$} circle (1pt);
\filldraw (1.5,0)
node{\tikz{\filldraw (0,0)  circle (2pt);}} 
node[below]{$1,2$}  circle (2pt);
\draw (-1.5,0) -- (1.5,0) ;
\filldraw(-.4,1.5) node[below]{$3$}  circle (2pt);
\filldraw  (-1,1.2) node[below]{$4$}  circle (2pt);
\end{tikzpicture}}
\simeq
\raisebox{-.7cm}{
 \begin{tikzpicture}[scale=.5]
\filldraw (-1.5,0)node{\tikz{\filldraw (0,0)  circle (1pt);}}  node[below]{3} circle (1pt);
\filldraw (0,0) node{\tikz{\filldraw (0,0)  circle (1pt);}}  node[below]{$4$} circle (1pt);
\filldraw (1.5,0)
node{\tikz{\filldraw (0,0)  circle (2pt);}} 
node[below]{$5,6$}  circle (2pt);
\draw (-1.5,0) -- (1.5,0) ;
\filldraw(-.4,1.5) node[below]{$1$}  circle (2pt);
\filldraw  (-1,1.2) node[below]{$2$}  circle (2pt);
\end{tikzpicture}}
\,.
\ee
Here we comment that we have assumed there are at least two points in each of the set $A,B,C$ in \eqref{fjadsfi} and \eqref{jfasd}. However when any of them are empty, such as $B=\emptyset$, the identity \eqref{eqidentity3} as well as equivalence relation \eqref{jfasd} still holds and will reduce to the case where $S_A=S_{(\bar A)}$. Besides, when there is only one point in a set, such as $B=\{b\}$, then $S_{A(B)}$ is nothing but $S_{\tilde{A}}$ with $\tilde{A}=A\cup \{b\}$ as implied by the geometry of the boundary configuration \eqref{jfasd}.  

In the rest of the section, we will present all the facets of ${\cal P}(3,n)$ and corresponding boundary configurations of $G_+(3,n)/T$ up to $n=9$, where more types of configurations appear as $n$ increases. It turns out that all boundary configurations up to $n=8$ can be nicely described using lines and black dots, which we will refer to as \textit{normal configurations}, and via \eqref{map} they give all facets of the polytopes. Starting at $n=9$ we will see even more complicated configurations, which requires ``blowing up" the black dot {\it etc.}. We will refer to those configurations which are not normal ones as \textit{exceptional configurations}. We also got 
all of the $314\times 10$ facets of ${\cal P}(3,10)$  by SE map, which are  put   in the auxiliary file \textsf{G310facets.txt}.   As we have emphasized repeatedly, once the facets are known, the polytope can be cut out directly. We will also comment on other aspects of the result, such as various types of vertices, the computation of the dual volume and connections to cluster polytopes up to $n=8$. 

\subsection{${\cal P}(3,7)$}
Using Minkowski sum or SE map, we obtain all $6\times 7=42$ facets for the $(3,7)$ case. All of them can be obtained by cyclic permutations of point labels of the following $6$ seeds
\be 
S_{123},\ S_{1234},\ S_{12345},\
 S_{12(34)},\ S_{(12)34},\
 S_{(12)3(45)}\,.
\ee 
The first $5$ types of facets have been explained before, so we will focus on the last type. As the notation suggests, the configuration can be drawn as
\raisebox{-.45cm}{
 \begin{tikzpicture}[scale=.5]
\filldraw (-1.5,0) node{\tikz{\filldraw (0,0)  circle (2pt);}}  node[below]{$1,2$} circle (2pt);
\filldraw (0,0) node{\tikz{\filldraw (0,0)  circle (1pt);}}  node[below]{$3$} circle (1pt);
\filldraw (1.5,0) node{\tikz{\filldraw (0,0)  circle (2pt);}}  node[below]{$4,5$}  circle (2pt);
\draw (-1.5,0) -- (1.5,0) ;
\end{tikzpicture}}. This means $\{1,3,4\},\{1,3,5\},\{2,3,4\},\{2,3,5\}$ are collinear, and both $\{1,2\}$ and $\{4,5\}$ collide.  Using the map \eqref{map}, we obtain
\be\label{jfaisj}
S_{(12)3(45)}=S\left[\!\!\!
\raisebox{-.45cm}{
 \begin{tikzpicture}[scale=.5]
\filldraw (-1.5,0) node{\tikz{\filldraw (0,0)  circle (2pt);}}  node[below]{$1,2$} circle (2pt);
\filldraw (0,0) node{\tikz{\filldraw (0,0)  circle (1pt);}}  node[below]{$3$} circle (1pt);
\filldraw (1.5,0) node{\tikz{\filldraw (0,0)  circle (2pt);}}  node[below]{$4,5$}  circle (2pt);
\draw (-1.5,0) -- (1.5,0) ;
\end{tikzpicture}}
\right]=s_{134}+s_{135}+s_{234}+s_{235} +\sum_{i\neq 1,2} s_{12i}+\sum_{i\neq 4,5} s_{45i}\,.
\ee 
In fact, on the support of momentum conservation, one can check $S_{(12)3(45)}=S_{45(67)12}$, where the latter comes from the configuration
\be
  \raisebox{-.7 cm}{
 \begin{tikzpicture}[scale=0.5]
\filldraw (-2,-1) node{\tikz{\filldraw (0,0)  circle (1pt);}}   circle (1pt);
\filldraw (-1,0) node{\tikz{\filldraw (0,0)  circle (1pt);}}  circle (1pt) node[left]{$5$};
\filldraw (1,0) node{\tikz{\filldraw (0,0)  circle (1pt);}}  circle (1pt) node[right]{$1$};
\filldraw (2,-1)  node{\tikz{\filldraw (0,0)  circle (1pt);}}  circle (1pt);
\filldraw (0,1) node{\tikz{\filldraw (0,0)  circle (2pt);}}  circle  (2pt);
\draw (-2,-1) node[left] {$4$} -- (0,1) node[above] {$6,7$} -- (2,-1) node[right] {$2$};
\node at (-3.5,0) {~};
\node at (3.5,0) {~};
\end{tikzpicture}}\,.
 \ee 
Here the configuration means $(456),(457),(626),(217)= {\cal O}(\epsilon)$ and $(67i)= {\cal O}(\epsilon)$ for any $i\neq 6,7$. Using the map \eqref{map},  we have
  \ba \label{jfaosdf}
  S\left[
  \raisebox{-.7 cm}{
 \begin{tikzpicture}[scale=0.4]
\filldraw (-2,-1)  node{\tikz{\filldraw (0,0)  circle (1pt);}} circle (1pt);
\filldraw (-1,0)  node{\tikz{\filldraw (0,0)  circle (1pt);}} circle (1pt) node[left]{$5$};
\filldraw (1,0)  node{\tikz{\filldraw (0,0)  circle (1pt);}} circle (1pt) node[right]{$1$};
\filldraw (2,-1)  node{\tikz{\filldraw (0,0)  circle (1pt);}} circle (1pt);
\filldraw (0,1)  node{\tikz{\filldraw (0,0)  circle (2pt);}}  circle (2pt);
\draw (-2,-1) node[left] {$4$} -- (0,1) node[above] {$6,7$} -- (2,-1) node[right] {$2$};
\end{tikzpicture}
}
\right]
=& s_{456}+s_{457}+ s_{216}+s_{217}+ \sum_{i\neq 3,4} s_{67i}
\\ 
=&
 S\left[
  \raisebox{-.7 cm}{
 \begin{tikzpicture}[scale=0.4]
\filldraw (-2,-1)  node{\tikz{\filldraw (0,0)  circle (1pt);}} circle (1pt);
\filldraw (-1,0)   node{\tikz{\filldraw (0,0)  circle (1pt);}} circle (1pt) node[left]{$5$};
\filldraw (0,1)  node{\tikz{\filldraw (0,0)  circle (2pt);}} circle (2pt);
\draw (-2,-1) node[left] {$4$} -- (0,1) node[above] {$6,7$} ;
\end{tikzpicture}
}
\right]+
 S\left[
  \raisebox{-.7 cm}{
 \begin{tikzpicture}[scale=0.4]
\filldraw (1,0)  node{\tikz{\filldraw (0,0)  circle (1pt);}}  circle (1pt) node[right]{$1$};
\filldraw (2,-1) node{\tikz{\filldraw (0,0)  circle (1pt);}}  circle (1pt);
\filldraw (0,1)  node{\tikz{\filldraw (0,0)  circle (2pt);}} circle (2pt);
\draw  (0,1) node[above] {$6,7$} -- (2,-1) node[right] {$2$};
\end{tikzpicture}
}
\right]- S\left[
  \raisebox{-.7 cm}{
 \begin{tikzpicture}[scale=0.4]
\filldraw (0,1)  node{\tikz{\filldraw (0,0)  circle (2pt);}} circle (2pt);
\node at (0,-1.3) {};
\draw (0,1) -- (0,1) node[above] {$6,7$} ;
\end{tikzpicture}
}
\right]\,,
\nonumber
 \ea
where the last line explicitly reads $S_{45(67)}+S_{(67)12}-S_{(67)}$. By momentum conservation, it is easy to show that \eqref{jfaisj} equals to \eqref{jfaosdf}. Again, one can check that the two configurations are equivalent: 
\be \label{fahjkafld}
\raisebox{-.8cm}{
 \begin{tikzpicture}[scale=.5]
\filldraw (-1.5,0)  node{\tikz{\filldraw (0,0)  circle (2pt);}} node[below]{$1,2$} circle (2pt);
\filldraw (0,0)  node{\tikz{\filldraw (0,0)  circle (1pt);}}  node[below]{$3$} circle (1pt);
\filldraw (1.5,0)  node{\tikz{\filldraw (0,0)  circle (2pt);}} node[below]{$4,5$}  circle (2pt);
\filldraw (1,1)  node{\tikz{\filldraw (0,0)  circle (1pt);}} node[above]{$6$}  circle (1pt);
\filldraw (.4,1.3)  node{\tikz{\filldraw (0,0)  circle (1pt);}} node[above]{$7$}  circle (1pt);
\draw (-1.5,0) -- (1.5,0) ;
\end{tikzpicture}}
\simeq
  \raisebox{-.65 cm}{
 \begin{tikzpicture}[scale=0.4]
\filldraw (-2,-1)  node{\tikz{\filldraw (0,0)  circle (1pt);}} circle (1pt);
\filldraw (-1,0)  node{\tikz{\filldraw (0,0)  circle (1pt);}} circle (1pt) node[left]{$5$};
\filldraw (1,0)  node{\tikz{\filldraw (0,0)  circle (1pt);}} circle (1pt) node[right]{$1$};
\filldraw (2,-1)  node{\tikz{\filldraw (0,0)  circle (1pt);}} circle (1pt);
\filldraw (0,1)  node{\tikz{\filldraw (0,0)  circle (2pt);}} circle (2pt);
\draw (-2,-1) node[left] {$4$} -- (0,1) node[above] {$6,7$} -- (2,-1) node[right] {$2$};
\filldraw (-.4,-1.7)  node{\tikz{\filldraw (0,0)  circle (1pt);}}  node[above]{$3$}  circle (1pt);
\end{tikzpicture}
}\,.
\ee 

Let us spell out this type of facets and corresponding configurations  more generally. We start with the configuration where points in $A$ and $C$ collide respectively, and additionally points in $A$ and $B$ are collinear, as well as those in $B$ and $C$. Using the map we obtain the facet
\be \label{eqdefsABC}
S_{(A)B(C)}=S_{(A)}+S_{B}+S_{(C)}+S_{B|C}+S_{C|A}+\sum\limits_{a\in A,b\in B,c\in C}s_{a\,b\,c}\,,
\ee
where we have used the notation $S_{B|C}$ defined in \eqref{fjadsfi}. It is an interesting exercise to show that on the support of momentum conservation, it  equals to the facet of a different configuration: 
\be
\label{eqequa3}
S_{(A)B(C)}=S_{C(D)A},\quad  \forall\,A\sqcup B\sqcup C\sqcup D=\{1,2,\cdots,n\}\,,
\ee where we have defined the new facet as
\be
S_{C(D)A}=S_{C(D)}+S_{(D)A}-S_{(D)}\,.  
\ee
Diagrammatically, the equivalence of the two configurations reads
\be \label{fahudsjf}
\raisebox{-.45cm}{
 \begin{tikzpicture}[scale=.5]
\filldraw (-1.5,0) node{\tikz{\filldraw (0,0)  circle (2pt);}} node[below]{$A$} circle (2pt);
\filldraw (0,0) node[below]{$B$} circle (0pt);
\filldraw (1.5,0) node{\tikz{\filldraw (0,0)  circle (2pt);}} node[below]{$C$}  circle (2pt);
\draw (-1.5,0) -- (1.5,0) ;
\node at (0.5,1.5) {$D$};
\end{tikzpicture}}
\simeq
  \raisebox{-.9 cm}{
 \begin{tikzpicture}[scale=0.4]
\filldraw (-2,-1) circle (0pt);
\filldraw (-1,0) circle (0pt) node[left]{$C$};
\filldraw (1,0) circle (0pt) node[right]{$A$};
\filldraw (2,-1) circle (0pt);
\filldraw (0,1) node{\tikz{\filldraw (0,0)  circle (2pt);}} circle (2pt);
\draw (-2,-1) node[left] {} -- (0,1) node[above] {$D$} -- (2,-1) node[right] {};
\node at (-1,-2) {$B$};
\end{tikzpicture}
}\,.
\ee 
Here we comment that we have assumed there are at least two points in each of the sets $A,C,D$ in \eqref{eqequa3} and \eqref{fahudsjf}.  However when $A$ or $C$ are empty,  the identity \eqref{eqequa3} as well as the equivalence relation \eqref{fahudsjf} still holds and will reduce to the case \eqref{jfasd} or even simpler one.  When there is only one point in the set $A$ or $C$, such as $A=\{a\}$,  $S_{(A)B(C)}$ is nothing but  $S_{{\tilde B}(C)}$ with $\tilde{B}=\{a\}\cup B$ as implied by the geometry of the boundary configuration  \eqref{fahudsjf}.  The case where $D$ only contains a single point won't appear in  the boundary configuration of $G_+(3,n)/T$.

Let's turn back to the polytope ${\cal P}(3,7)$.
Its $f-$vector is $(1,\ 693,\ 2163,\ 2583,\ 1463,\ 392,$ $\ 42,\ 1)$.  Out of 693 vertices, 595 of them are simple ones. The remaining non-simple vertices are intersected by 7, 8 and 9 facets respectively.  The numbers of each kind of vertices are put in  
 table~\ref{tabg37vert}.
 \begin{table}[!htbp]
	\centering  

	\begin{tabular}{|c|c|c|c|}  
		\hline  
		 6-vertex & 7-vertex &8-vertex &9-vertex \\  
		\hline
		595 & 63 & 28 & 7 \\
		\hline
	\end{tabular}
		\caption{The numbers for various types of vertices in ${\cal P}(3,7)$, classified by the number of facets adjacent to each vertex.}  
	\label{tabg37vert}  
\end{table}

 Each simple vertex would correspond to a simplex as a facet on dual polytope $\dcalP(3,7)$. Likewise in $\dcalP(3,6)$, their contribution to the whole amplitudes is the volume of the convex hull of this facet and the reference point $(1,0,0,0,0,0,0)$, and just equals to the inverse of the production of the facets involved.

As for those remaining non-simple vertices, their corresponding facets on the dual polytope $\dcalP(3,7)$ are non-simplicial. Similarly to those two non-simplicial facets in $\dcalP(3,6)$, we can triangulate these non-simplicial facets in $\dcalP(3,7)$ into simplices to calculate their contribution to the whole amplitudes. Here we give one example where a vertex of ${\cal P}(3,7)$ is intersected by such 7 facets:
\begin{equation}
\{S_{123},\ S_{4567},\ S_{12345},\ S_{34567},\ S_{67123},\ S_{45(67)},\ S_{(45)67} \},	
\end{equation}

There are  two ways to triangulate the non-simplicial facet on $\dcalP(3,7)$ corresponding to such a vertex,
\begin{equation}\label{exp37nons1}
\begin{aligned}
&\frac{1 }{  S_{123}\,S_{ 4567}\,S_{ 12345}\,S_{ 34567}\,S_{  67123} }\left(\,\frac{1}{S_{45(67)}}+ \,\frac{1}{S_{ (45)67}} \right)
\nl
=&
\frac{1}{S_{123}S_{34567}S_{45(67)}S_{(45)67}}\Big(\frac{1}{S_{12345}S_{34567}}+\frac{1}{S_{4567}S_{12345}}&+\frac{1}{S_{4567}S_{67123}}\Big)\,,
\end{aligned}
\end{equation}
which are the same due to the identity:
\begin{equation}\label{g37identity1}
S_{45(67)}+S_{(45)67}=S_{4567}+S_{67123}+S_{12345},
\end{equation}
which means that once we set $S_{4567}=S_{67123}=S_{12345}=S_{45(67)}=0$, we automatically get $S_{(45)67}$=0. Therefore on the polytope $\calP(3,7)$, the vertex intersected by $\{S_{123}$, $S_{4567}$, $S_{12345}$, $S_{34567}$, $S_{67123}$, and $S_{45(67)}\}$ also touches the facet $S_{(45)67}$, forming a 7-vertex.

Summing over all 693 facets of the dual polytope ${\cal P}^\circ(3,7)$ and we obtain the generalized amplitude $\calA_{3,7}$. The result is given in the auxiliary file \textsf{G37.nb}. In~\cite{Arkani-Hamed:2019mrd}, it has been shown that $\calP(3,7)$ is a degeneration of the generalized associahedron of type $\calE_6$, which we denote as ${\cal P}(\calE_6)$. The latter is a simple polytope, with in total $833$ simple vertices. A set of extra identities among the boundaries of $\calP(3,7)$ such as (\ref{g37identity1}) lead to the degeneration from ${\cal P}(\calE_6)$ to $\calP(3,7)$ by making various simple vertices merge into non-simple ones.

\subsection{${\cal P}(3,8)$ }\label{sect38}

The same construction is also valid to build the polytope $\calP(3,8)$. From Minkowski sum or SE map, we found that there are in all $15\times 8=120$ facets. All of them can be obtained by cyclic permutation of point labels of the following 12 seeds,
\ba \label{fajkslfd} 
&S_{123},\,S_{1234},\,S_{12345},\, S_{123456},\, S_{(12)34},\,S_{12(34)},\,S_{(12)345},\,S_{123(45)},\
\nl 
&
S_{(12)3(45)},\,S_{(123)4(56)}, \,S_{(12)3(456)}, \,S_{(12)34(56)},\ 
\nl 
&
S\left[\!\!\!
\raisebox{-1cm}{
\begin{tikzpicture}[scale=.8]
	\fill[black] (0,1.5) node{\tikz{\filldraw (0,0)  circle (1pt);}} circle (1pt) node[above]{1};
	\fill[black] (-0.5,0.75) node{\tikz{\filldraw (0,0)  circle (1pt);}} circle (1pt) node[left]{2};		\fill[black] (-1,0) node{\tikz{\filldraw (0,0)  circle (2pt);}} circle (2pt) node[below]{3,4};
	\fill[black] (0,0) node{\tikz{\filldraw (0,0)  circle (1pt);}} circle (1pt) node[below]{5};
	\fill[black] (1,0) node{\tikz{\filldraw (0,0)  circle (2pt);}} circle (2pt) node[below]{6,7};
	\draw (0,1.5) -- (-1,0) -- (1,0) ;
\end{tikzpicture}
}
\right]
,
\quad
S\left[\!\!\!
\raisebox{-1cm}{
\begin{tikzpicture}[scale=.8]
	\fill[black] (0,1.5) node{\tikz{\filldraw (0,0)  circle (1pt);}} circle (1pt) node[above]{8};
	\fill[black] (-0.5,0.75) node{\tikz{\filldraw (0,0)  circle (1pt);}} circle (1pt) node[left]{7};		\fill[black] (-1,0) circle (2pt) node{\tikz{\filldraw (0,0)  circle (2pt);}} node[below]{6,5};
	\fill[black] (0,0) node{\tikz{\filldraw (0,0)  circle (1pt);}}  circle (1pt) node[below]{4};
	\fill[black] (1,0) node{\tikz{\filldraw (0,0)  circle (2pt);}} circle (2pt) node[below]{3,2};
	\draw (0,1.5) -- (-1,0) -- (1,0) ;
\end{tikzpicture}
}
\right]
,
\quad
S\left[\!\!\!
\raisebox{-1cm}{
\begin{tikzpicture}[scale=.8]
	\fill[black] (0,1.5) node{\tikz{\filldraw (0,0)  circle (1pt);}}  circle (1pt) node[above]{1};
	\fill[black] (-0.5,0.75) node{\tikz{\filldraw (0,0)  circle (1pt);}}  circle (1pt) node[left]{2};
	\fill[black] (-1,0) node{\tikz{\filldraw (0,0)  circle (2pt);}}  circle (2pt) node[below]{3,4};
	\fill[black] (0,0) 
	node{\tikz{\filldraw (0,0)  circle (1pt);}} circle (1pt) node[below]{5};
	\fill[black] (1,0) 
	node{\tikz{\filldraw (0,0)  circle (2pt);}} circle (2pt) node[below]{6,7};
	\fill[black] (0.5,0.75) 
	node{\tikz{\filldraw (0,0)  circle (1pt);}} circle (1pt) node[right]{8};
	\draw (0,1.5) -- (-1,0) -- (1,0) -- (0,1.5);
\end{tikzpicture}
}
\right]\,,
\ea 
where the facets in the first two lines has been explained before, so we focus on the three new types in the third line.  For convenience we denote them as 
$S_{12(34)5(67)}$, $S_{(23)4(56)78}$ and $S_{12(34)5(67)81}$ respectively.  The first case, $S_{12(34)5(67)}$, means that $\{ 1,2,3 \},\{ 1,2,4 \},\{ 3,5,6 \}$, $\{ 3,5,7 \},\{ 4,5,6 \},\{ 4,5,7 \}$ are collinear respectively, and both $\{1,2\}$ and $\{4,5\}$ collide. Using the map \eqref{map}, we have
\be 
S_{12(34)5(67)}=S_{(34)}+S_{(67)}+(s_{123}+s_{124})+(s_{356}+s_{357}+s_{456}+s_{457}).
\ee

The second case, $S_{(23)4(56)78}$, is just the reflection of point labels, $i\leftrightarrow 9-i$ for any $i$, of $S_{12(34)5(67)}$. The third case, $S_{12(34)5(67)81}$, means that $\{1,2,3\},\{1,2,4\},\{3,5,6\},\{3,5,7\},$ $\{4,5,6\}$, $\{4,5,7\},\{6,7,8\},\{6,7,1\}$ are collinear respectively, and both $\{3,4\}$ and $\{6,7\}$ collide. Using the map \eqref{map}, we have


\begin{equation}\label{defS1234567}
S_{12(34)5(67)81}=S_{(34)}+S_{(67)}+(s_{123}+s_{124})+(s_{356}+s_{357}+s_{456}+s_{457})+(s_{168}+s_{178}).
\end{equation}

Let's spell out these types of facets and corresponding configurations more generally. We start with the configurations where points in $B$ and $D$ collide respectively, and additionally points in $A$ and $B$ are collinear, as well as those in $B$ and $C$, and those in $C$ and $D$. Using the map \eqref{map} we obtain the facet,
\begin{equation}
S_{A(B)C(D)}=S_{A(B)}+S_{(B)C(D)}-S_{(B)}. 
\end{equation}
Likewise, with the reflection of labels, we have,
\begin{equation}
S_{(A)B(C)D}=S_{(A)B(C)}+S_{(C)D}-S_{(C)}.
\end{equation}

We can also have the configurations where points in $B$ and $D$ collide respectively, and additionally points in $A$ and $B$ are collinear, as well as those in $B$ and $C$, those in $C$ and $D$, and those in $D$ and $E$. Using the map \eqref{map} we obtain the facet,
\begin{equation}
S_{A(B)C(D)E}=S_{A(B)}+S_{(B)C(D)}+S_{(D)E}-S_{(B)}-S_{(D)}.
\end{equation}
Here $A$ and $E$ could share one point. See \eqref{defS1234567}
as an example.

Let's go back to the polytope $\calP(3,8)$. Its $f-$vector is $(1, \ 13612,\ 57768,\ 100852,$ $\ 93104,\ 48544,\ 14088,$ $\ 2072,\ 120,\ 1)$. 
Out of all the 13612 vertices, only 9672 vertices are simple ones, which are intersected by 8 facets. The other non-simple vertices range from 9-vertex to 17-vertex. The numbers of each kind of vertices are put in table~\ref{tabg38vert}.
\begin{table}[!htb]
\centering
	\begin{tabular}{|c|c|c|c|c|}  
		\hline  
		8-vertex & 9-vertex & 10-vertex &11-vertex &12-vertex \\  
		\hline
		9672 & 1696& 1092& 480& 416\\
		\hline
		13-vertex &14-vertex &15-vertex & 16-vertex &17-vertex\\
		\hline
		104& 88& 32& 24& 8 \\
		\hline
	\end{tabular}
		\caption{The numbers for various types of vertices in ${\cal P}(3,8)$, classified by the number of facets adjacent to each vertex.}  
	\label{tabg38vert}
\end{table}

As in the case of $(3,6)$ and $(3,7)$, each simple vertex corresponds to a simplex as a facet on dual polytope $\dcalP(3,8)$, and their contribution to the generalized amplitude is the volume of the convex hull of this facet and the reference point $(1,0,0,0,0,0,0,0,0)$, which equals to the inverse of the production of the facets involved. 

For those non-simple vertices, their corresponding facets on $\dcalP(3,8)$ are non-simplicial facets. Similarly to the case of $\dcalP(3,6)$ and $\dcalP(3,7)$, we can triangulate those non-simplicial facets into simplices to calculate their contribution to the generalized amplitude. Here we give one example where a vertex of ${\cal P}(3,8)$ is intersected by 9 facets as follows:
\begin{equation}
\{ S_{123},\ S_{12(34)},\ S_{123456},\ S_{345},\ S_{34(56)},\ S_{345678},\ S_{5678},\ S_{(78)12(34)},\ S_{56(78)} \}.
\end{equation} There are two natural ways to triangulate its corresponding facets of $\dcalP(3,8)$,
\begin{equation}
\begin{aligned}
&\frac{1}{S_{123} S_{12(34)} S_{123456} S_{345} S_{34(56)} S_{345678} S_{5678}}
 \left(\frac{1}{S_{(78)12(34)}}+\frac{1}{S_{56(78)}}\right),\\
&\frac{1}{S_{123} S_{12(34)}  S_{345} S_{345678} S_{(78)12(34)} S_{56(78)}}
 \left(\frac{1}{S_{5678}S_{34(56)}}+\frac{1}{S_{5678}S_{123456}}+\frac{1}{S_{34(56)}S_{123456}}\right),\\
\end{aligned}
\end{equation}
which are the same due to the identity:
\begin{equation}
S_{(78)12(34)}+S_{56(78)}=S_{5678}+S_{34(56)}+S_{123456}.
\end{equation}

Summing over all the 13612 facets of the dual polytope $\dcalP(3,8)$ as above, we get its entire volume, which gives the generalized amplitude $\calA_{3,8}$. The result is given in the auxiliary file \textsf{G38.nb}. As shown in~\cite{Arkani-Hamed:2019mrd}, $\calP(3,8)$ is a degeneration of the generalized associahedron of type $\calE_8$, which we denote as $\calP(\calE_8)$. The latter is a simple polytope with 128 facets and 25080 vertices. Note that there are only 120 facets for $\calP(3,8)$, so the degeneration from $\calP(\calE_8)$ to $\calP(3,8)$ has made 8 facets disappear. We find that the intersection of two facets $S_{(12)345}, S_{456(78)}$ of $\calP(3,8)$,  which is a codim-2 boundary,  can actually be obtained by shrinking a facet of $\calP(\calE_8)$.  The remaining seven disappearing facets are completely analogous. 
See Appendix~\ref{appd1}  for more details.

\subsection{${\cal P}(3,9)$}

Now we move to the polytope $\calP(3,9)$, and by the construction of Minkowski sum or SE map, we have obtained 471 facets in total, which are put in the auxiliary file \textsf{G39facets.txt}.  Out of these facets, there are $38\times 9+ 3\times 3= 351$ facets which correspond to normal configurations. All of them can be obtained by cyclic permutation of 41 seeds. Let's first list the 38 cyclic classes of length 9, where the seeds are,
\begin{equation}\label{fdjasf}
\begin{aligned}
&S_{123},\ S_{1234},\ S_{12345},\ S_{123456},\ S_{1234567},\,\\
&S_{12(34)},\ S_{123(45)},\ S_{1234(56)},\ S_{(12)34},\ S_{(12)345},\ S_{(12)3456},\ ,\\
&S_{(12)3(45)},\ S_{(123)4(56)},\ S_{(12)34(56)},\ S_{(12)3(456)},\ S_{(1234)5(67)},\\
&S_{(123)45(67)},\ S_{(123)4(567)},\ S_{(12)345(67)},\ S_{(12)34(567)},\ S_{(12)3(4567)},\ \\
&S_{12(34)5(67)},\ S_{123(45)6(78)},\ S_{12(345)6(78)},\ S_{12(34)56(78)},\ S_{12(34)5(678)},\ \\
&S_{(12)3(45)67},\ S_{(123)4(56)78},\ S_{(12)34(56)78},\ S_{(12)3(456)78},\ S_{(12)3(45)678},\ \\
&S_{12(34)5(67)89},\ S_{123(45)6(78)91},\ S_{12(345)6(78)91},\ S_{12(34)56(78)91},\
 \\
& S_{12(34)5(678)91}, S_{12(34)5(67)891},S_{(12)3(45)6(78)}. 
\end{aligned}
\end{equation}
There are 3 more cyclic classes of length 3, where the seeds are
\begin{equation}\label{fdjasf2}
\begin{aligned}
S_{123(456)},\ S_{(123)456},\ 
S_{(12)3(45)6(78)9(12)}\,. 
\end{aligned}
\end{equation}
All the facets above have be explained before except for $S_{(12)3(45)6(78)}$ and $S_{(12)3(45)6(78)9(12)}$. As the notation suggest, their configurations can be drawn as
\ba
&
\raisebox{-1cm}{
\begin{tikzpicture}[scale=.8]
	\fill[black] (0,1.5) node{\tikz{\filldraw (0,0)  circle (2pt);}} circle (2pt) node[above]{1,2};
	\fill[black] (-0.5,0.75) node{\tikz{\filldraw (0,0)  circle (1pt);}}  circle (1pt) node[left]{3};		\fill[black] (-1,0) 
	node{\tikz{\filldraw (0,0)  circle (2pt);}} circle (2pt) node[below]{4,5};
	\fill[black] (0,0) node{\tikz{\filldraw (0,0)  circle (1pt);}}  circle (1pt) node[below]{6};
	\fill[black] (1,0) node{\tikz{\filldraw (0,0)  circle (2pt);}}  circle (2pt) node[below]{7,8};
	\draw (0,1.5) -- (-1,0) -- (1,0) ;
\end{tikzpicture}
}
,
\raisebox{-1cm}{
\begin{tikzpicture}[scale=.8]
	\fill[black] (0,1.5) node{\tikz{\filldraw (0,0)  circle (2pt);}} circle (2pt) node[above]{1,2};
	\fill[black] (-0.5,0.75) node{\tikz{\filldraw (0,0)  circle (1pt);}} circle (1pt) node[left]{3};
	\fill[black] (-1,0) node{\tikz{\filldraw (0,0)  circle (2pt);}} circle (2pt) node[below]{4,5};
	\fill[black] (0,0) node{\tikz{\filldraw (0,0)  circle (1pt);}} circle (1pt) node[below]{6};
	\fill[black] (1,0) node{\tikz{\filldraw (0,0)  circle (2pt);}} circle (2pt) node[below]{7,8};
	\fill[black] (0.5,0.75) 
	node{\tikz{\filldraw (0,0)  circle (1pt);}} circle (1pt) node[right]{9};
	\draw (0,1.5) -- (-1,0) -- (1,0) -- (0,1.5);
\end{tikzpicture}
}
.
\ea  
The first one means that $\{1,3,4\},\{1,3,5\},\{2,3,4\},\{2,3,5\},\{4,6,7\},\{4,6,8\},\{5,6,7\}$ and $\{5,6,8\}$ being collinear, and that $\{1,2\}$, $\{4,5\}$ and $\{7,8\}$ collide. Using the map, we have
\be
S_{(12)3(45)6(78)}=S_{(12)3(45)}+S_{(45)6(78)}-S_{(45)}\,.
\ee
The second one means the combination of the first one and the relation that $\{7,9,1\},\{7,9,2\},$ $\{8,9,1\}$ and $\{8,9,2\}$ are collinear. Using the map \eqref{map}, we obtain the facet
\ba  
&S_{(12)3(45)6(78)9(12)}=S_{(12)3(45)}+S_{(45)6(78)}+S_{(78)9(12)}-S_{(12)}-S_{(45)}-S_{(78)}\,.
\ea 

Apart from these 351 facets, there are 120 remaining facets which come from configurations that require ``blowup". Before, when we have 
two colliding points $a,b$ and another point $i$ that doesn't collide to them, their minor $(a,b,i)$ vanishes at order ${\cal O}(\epsilon)$.  However their minor can also vanish at order ${\cal O}(\epsilon^2)$  when $b$ collides to $a$ in the direction determined by $a$ and $i$.   New method is needed to study this feature. 

Let's start by introducing a dashed circle  to blow up a black spot. For example
\be \label{hafjld}
S\left[
\raisebox{-.5cm}{
 \begin{tikzpicture}[scale=.9 
 ]
\filldraw (-1,0) node{\tikz{\filldraw (0,0)  circle (2pt);}}  node[below]{$1,2,3$} circle (2pt);
\filldraw (0,0)  node{\tikz{\filldraw (0,0)  circle (1pt);}} node[below]{$4$} circle (1pt);
\filldraw (1,0) node{\tikz{\filldraw (0,0)  circle (1pt);}}  node[below]{$5$}  circle (1pt);
\draw (-1,0) -- (1,0) ;
\end{tikzpicture}}
\right]= S\left[
\raisebox{-.5cm}{
 \begin{tikzpicture}[scale=.9 
 ]
\filldraw (0,0) node{\tikz{\filldraw (0,0)  circle (1pt);}}  node[below]{$4$} circle (1pt);
\filldraw (1,0) node{\tikz{\filldraw (0,0)  circle (1pt);}}  node[below]{$5$}  circle (1pt);
\draw (-1,0) -- (1,0) ;
\draw[dashed] (-1-.75,0) circle (0.75);
\filldraw (-1-.9,0.3) node{\tikz{\filldraw (0,0)  circle (1pt);}} node[left]{$1$} circle (1pt);
\filldraw (-1-.5,.15) node{\tikz{\filldraw (0,0)  circle (1pt);}} node[below]{$3$} circle (1pt);
\filldraw (-1-.9,-.3) node{\tikz{\filldraw (0,0)  circle (1pt);}} node[left]{$2$} circle (1pt);
\end{tikzpicture}
}
\right]= S_{(123)45}\,.
\ee
Here the radius of the circle is at order ${\cal O}(\epsilon)$. Let's use a parallel line to denote the configuration where 2 collide with 3 in the direction determined by the line $\overline{4 5}$\,,
\be\label{newnewe}
 \raisebox{-.5cm}{
 \begin{tikzpicture}[scale=.9 
 ]
\filldraw (0,0) node{\tikz{\filldraw (0,0)  circle (1pt);}} node[below]{$4$} circle (1pt);
\filldraw (1,0) node{\tikz{\filldraw (0,0)  circle (1pt);}} node[below]{$5$}  circle (1pt);
\draw (-1,0) -- (1,0) (-1-.5,-.3)--(-1-.9,-.3);
\draw[dashed] (-1-.75,0) circle (0.75);
\filldraw (-1-.9,0.3)node{\tikz{\filldraw (0,0)  circle (1pt);}}
node[left]{$1$} circle (1pt);
\filldraw (-1-.5,-.3)  node{\tikz{\filldraw (0,0)  circle (1pt);}} 
node[above]{$3$} circle (1pt);
\filldraw (-1-.9,-.3) node{\tikz{\filldraw (0,0)  circle (1pt);}} 
node[left]{$2$} circle (1pt);
\end{tikzpicture}
}\,\,.
\ee
The distance of 2 and 3 is at order ${\cal O}(\epsilon)$ because they are inside the dashed circle. What's more, the distance of 4 and the line $\overline{23}$ is also at order ${\cal O}(\epsilon)$, thus the area of the triangle formed by $2,3,4$ is at order ${\cal O}(\epsilon^2)$, {\it i.e.} $(2,3,4)= {\cal O}(\epsilon^2)$. Similarly $(2,3,5)= {\cal O}(\epsilon^2)$.

Considering the residue of ${\rm d}
\log R_{3,n}(\epsilon)$ for the above new configuration \eqref{newnewe},  one can easily find that it is directly related to that of  normal configuration \eqref{hafjld} by adding several generalized Mandelstam variables,
\ba \label{hafasuhfjld}
S\left[
\raisebox{-.5cm}{
 \begin{tikzpicture}[scale=.9 
 ]
\filldraw (0,0) node{\tikz{\filldraw (0,0)  circle (1pt);}} node[below]{$4$} circle (1pt);
\filldraw (1,0)node{\tikz{\filldraw (0,0)  circle (1pt);}}  node[below]{$5$}  circle (1pt);
\draw (-1,0) -- (1,0) (-1-.5,-.3)--(-1-.9,-.3);
\draw[dashed] (-1-.75,0) circle (0.75);
\filldraw (-1-.9,0.3) node{\tikz{\filldraw (0,0)  circle (1pt);}}   node[left]{$1$} circle (1pt);
\filldraw (-1-.5,-.3) node{\tikz{\filldraw (0,0)  circle (1pt);}} node[above]{$3$} circle (1pt);
\filldraw (-1-.9,-.3) node{\tikz{\filldraw (0,0)  circle (1pt);}}  node[left]{$2$} circle (1pt);
\end{tikzpicture}
}
\right]= S\left[
\raisebox{-.5cm}{
 \begin{tikzpicture}[scale=.9 
 ]
\filldraw (-1,0) node{\tikz{\filldraw (0,0)  circle (2pt);}}  node[below]{$1,2,3$} circle (2pt);
\filldraw (0,0)  node{\tikz{\filldraw (0,0)  circle (1pt);}} node[below]{$4$} circle (1pt);
\filldraw (1,0) node{\tikz{\filldraw (0,0)  circle (1pt);}}  node[below]{$5$}  circle (1pt);
\draw (-1,0) -- (1,0) ;
\end{tikzpicture}}
\right]+ s_{2\,3\,4}+ s_{2\,3\,5}
\,.
\ea 
Note that the coefficient of $s_{234}$  (similar for $s_{235}$) is 1 on the RHS via $(2,3,4)={\cal O}(\epsilon^2)$, since another piece of $s_{234}$ is already included in $S_{(123)45}$. 

In this particular case,  under the momentum conservation of nine points, one can find 
\be \label{fadjf}
S\left[
\raisebox{-.5cm}{
 \begin{tikzpicture}[scale=.9 
 ]
\filldraw (0,0) node{\tikz{\filldraw (0,0)  circle (1pt);}} node[below]{$4$} circle (1pt);
\filldraw (1,0)node{\tikz{\filldraw (0,0)  circle (1pt);}}  node[below]{$5$}  circle (1pt);
\draw (-1,0) -- (1,0) (-1-.5,-.3)--(-1-.9,-.3);
\draw[dashed] (-1-.75,0) circle (0.75);
\filldraw (-1-.9,0.3) node{\tikz{\filldraw (0,0)  circle (1pt);}}   node[left]{$1$} circle (1pt);
\filldraw (-1-.5,-.3) node{\tikz{\filldraw (0,0)  circle (1pt);}} node[above]{$3$} circle (1pt);
\filldraw (-1-.9,-.3) node{\tikz{\filldraw (0,0)  circle (1pt);}}  node[left]{$2$} circle (1pt);
\end{tikzpicture}
}
\right]= S_{23(45)6789}= S_{(6789)1(23)}.
\ee 
Hence no new facet is produced.  
Correspondingly, their boundary configurations  are also equivalent,
\be \label{fahjkafld3}
\raisebox{-.5cm}{
 \begin{tikzpicture}[scale=.9 
 ]
\filldraw (0,0) node{\tikz{\filldraw (0,0)  circle (1pt);}} node[below]{$4$} circle (1pt);
\filldraw (1,0)node{\tikz{\filldraw (0,0)  circle (1pt);}}  node[below]{$5$}  circle (1pt);
\draw (-1,0) -- (1,0) (-1-.5,-.3)--(-1-.9,-.3);
\draw[dashed] (-1-.75,0) circle (0.75);
\filldraw (-1-.9,0.3) node{\tikz{\filldraw (0,0)  circle (1pt);}}   node[left]{$1$} circle (1pt);
\filldraw (-1-.5,-.3) node{\tikz{\filldraw (0,0)  circle (1pt);}} node[above]{$3$} circle (1pt);
\filldraw (-1-.9,-.3) node{\tikz{\filldraw (0,0)  circle (1pt);}}  node[left]{$2$} circle (1pt);
\filldraw (0.7,0.6) node{\tikz{\filldraw (0,0)  circle (1pt);}} node[above]{$6$} circle (1pt);
\filldraw (0.27,0.4) node{\tikz{\filldraw (0,0)  circle (1pt);}} node[above]{$7$} circle (1pt);
\filldraw (-0.43,0.3) node{\tikz{\filldraw (0,0)  circle (1pt);}} node[above]{$8$} circle (1pt);
\filldraw (-0.9,0.9) node{\tikz{\filldraw (0,0)  circle (1pt);}} node[below]{$9$} circle (1pt);
\end{tikzpicture}
}
\simeq
\raisebox{-.8cm}{
 \begin{tikzpicture}[scale=.5]
\filldraw (-2.5,0)  node{\tikz{\filldraw (0,0)  circle (2pt);}} node[below]{$6,7,8,9$} circle (2pt);
\filldraw (0,0)  node{\tikz{\filldraw (0,0)  circle (1pt);}}  node[below]{$1$} circle (1pt);
\filldraw (1.5,0)  node{\tikz{\filldraw (0,0)  circle (2pt);}} node[below]{$2,3$}  circle (2pt);
\filldraw (1,1)  node{\tikz{\filldraw (0,0)  circle (1pt);}} node[above]{$4$}  circle (1pt);
\filldraw (.4,1.3)  node{\tikz{\filldraw (0,0)  circle (1pt);}} node[above]{$5$}  circle (1pt);
\draw (-2.5,0) -- (1.5,0) ;
\end{tikzpicture}}
\simeq
  \raisebox{-.65 cm}{
 \begin{tikzpicture}[scale=0.4]
\filldraw (-2,-1)  node{\tikz{\filldraw (0,0)  circle (1pt);}} circle (1pt);
\filldraw (-1,0)  node{\tikz{\filldraw (0,0)  circle (1pt);}} circle (1pt) node[left]{$3$};
\filldraw (1,0)  node{\tikz{\filldraw (0,0)  circle (1pt);}} circle (1pt) node[above right=-2.5pt]{$7$};
\filldraw (.5,.5)  node{\tikz{\filldraw (0,0)  circle (1pt);}} circle (1pt) node[below left=-2.5pt]{$6$};
\filldraw (1.5,-.5)  node{\tikz{\filldraw (0,0)  circle (1pt);}} circle (1pt) node[below left=-2.5pt]{$8$};
\filldraw (2,-1)  node{\tikz{\filldraw (0,0)  circle (1pt);}} circle (1pt);
\filldraw (0,1)  node{\tikz{\filldraw (0,0)  circle (2pt);}} circle (2pt);
\draw (-2,-1) node[left] {$2$} -- (0,1) node[above] {$4,5$} -- (2,-1) node[above right=-2.5pt] {$9$};
\filldraw (-.4,-1.7)  node{\tikz{\filldraw (0,0)  circle (1pt);}}  node[above]{$1$}  circle (1pt);
\end{tikzpicture}
}\,.
\ee

However  some ``blowup'' of the normal configurations for  $S_{12(345)6(78)}, S_{12(345)6(78)91}$ and $ S_{(12)3(45)6(78)9(12)}$  indeed produce  the seeds of new boundary  configurations 
for the remaining 120 facets of (3,9).  First we have six kinds of configurations containing the new structure \eqref{newnewe},
\ba \label{fajkfaodijfslfd}
&
\raisebox{-1.1cm}{
\begin{tikzpicture}[scale=.8]
\node[circle,draw=black, fill=black, inner sep=0pt,minimum size=1.5pt] (v4) at (-2.15,2.15) {};
\node[circle,draw=black, fill=black, inner sep=0pt,minimum size=1.5pt] (v5) at (-1.5,2.15) {};
\node[circle,draw=black, fill=black, inner sep=0pt,minimum size=1.5pt] (v3) at (-1.9,1.76) {};
\node[circle,draw=black, fill=black, inner sep=0pt,minimum size=1.5pt] (v2) at (-1.25,0.875) {};
\node[circle,draw=black, fill=black, inner sep=0pt,minimum size=1.5pt] (v1) at (-0.75,0.125) {};
\node[circle,draw=black, fill=black, inner sep=0pt,minimum size=1.5pt] (v6) at (-0.5,2) {};
\node[circle,draw=black, fill=black, inner sep=0pt,minimum size=2.5pt] (v7) at (0.5,2) {};
\draw (v1) -- (-1.85,1.775);
\draw (-2,2) -- (v7);
\filldraw[fill=white!,draw=black,dashed] (-1.75,2.1) circle (0.75);
\filldraw (v4) node{\tikz{\filldraw (0,0)  circle (1pt);}} circle (1pt) node[above] {$4$};
\filldraw (v5) node{\tikz{\filldraw (0,0)  circle (1pt);}}  circle (1pt) node[above] {$5$};
\filldraw (v1) node{\tikz{\filldraw (0,0)  circle (1pt);}}  circle (1pt) node[left] {$1$};
\filldraw (v3) node{\tikz{\filldraw (0,0)  circle (1pt);}}  circle (1pt) node[right] {$3$};
\filldraw (v2)node{\tikz{\filldraw (0,0)  circle (1pt);}}  circle (1pt) node[left] {$2$};
\filldraw (v6) node{\tikz{\filldraw (0,0)  circle (1pt);}} circle (1pt) node[above] {$6$};
\filldraw (v7) node{\tikz{\filldraw (0,0)  circle (2pt);}}  circle (1.5pt) node[above] {$7,8$};
\draw  (v4) -- (v5);
\end{tikzpicture}
}, \ 
\raisebox{-1.1cm}{
\begin{tikzpicture}[scale=.8]
\node[circle,draw=black, fill=black, inner sep=0pt,minimum size=1.5pt] (v4) at (-2.15,2.15) {};
\node[circle,draw=black, fill=black, inner sep=0pt,minimum size=1.5pt] (v5) at (-1.5,2.15) {};
\node[circle,draw=black, fill=black, inner sep=0pt,minimum size=1.5pt] (v3) at (-1.9,1.76) {};
\node[circle,draw=black, fill=black, inner sep=0pt,minimum size=1.5pt] (v2) at (-1.25,0.875) {};
\node[circle,draw=black, fill=black, inner sep=0pt,minimum size=1.5pt] (v1) at (-0.75,0.125) {};
\node[circle,draw=black, fill=black, inner sep=0pt,minimum size=1.5pt] (v6) at (-0.5,2) {};
\node[circle,draw=black, fill=black, inner sep=0pt,minimum size=2.5pt] (v7) at (0.5,2) {};
\draw (v1) -- (-1.85,1.775);
\draw (-2,2) -- (v7);
\filldraw[fill=white!,draw=black,dashed] (-1.75,2.1) circle (0.75);
\filldraw (v4) node{\tikz{\filldraw (0,0)  circle (1pt);}}  circle (1pt) node[above] {$4$};
\filldraw (v5) node{\tikz{\filldraw (0,0)  circle (1pt);}}  circle (1pt) node[above] {$5$};
\filldraw (v1) node{\tikz{\filldraw (0,0)  circle (1pt);}} circle (1pt) node[left] {$1$};
\filldraw (v3) node{\tikz{\filldraw (0,0)  circle (1pt);}} circle (1pt) node[right] {$3$};
\filldraw (v2)node{\tikz{\filldraw (0,0)  circle (1pt);}} circle (1pt) node[left] {$2$};
\filldraw (v6)node{\tikz{\filldraw (0,0)  circle (1pt);}} circle (1pt) node[above] {$6$};
\filldraw (v7)node{\tikz{\filldraw (0,0)  circle (2pt);}}  circle (1.5pt) node[above] {$7,8$};
\draw  (v4) -- (v3);
\end{tikzpicture}
}
,\ 
\raisebox{-1.1cm}{
\begin{tikzpicture}[scale=.8]
\node[circle,draw=black, fill=black, inner sep=0pt,minimum size=1.5pt] (v4) at (-2.15,2.15) {};
\node[circle,draw=black, fill=black, inner sep=0pt,minimum size=1.5pt] (v5) at (-1.5,2.15) {};
\node[circle,draw=black, fill=black, inner sep=0pt,minimum size=1.5pt] (v3) at (-1.9,1.76) {};
\node[circle,draw=black, fill=black, inner sep=0pt,minimum size=1.5pt] (v2) at (-1.25,0.875) {};
\node[circle,draw=black, fill=black, inner sep=0pt,minimum size=1.5pt] (v1) at (-0.75,0.125) {};
\node[circle,draw=black, fill=black, inner sep=0pt,minimum size=1.5pt] (v6) at (-0.5,2) {};
\node[circle,draw=black, fill=black, inner sep=0pt,minimum size=2.5pt] (v7) at (0.5,2) {};
\draw (v1) -- (-1.85,1.775);
\draw (-2,2) -- (v7);
\filldraw[fill=white!,draw=black,dashed] (-1.75,2.1) circle (0.75);
\filldraw (v4) node{\tikz{\filldraw (0,0)  circle (1pt);}} circle (1pt) node[above] {$4$};
\filldraw (v5) node{\tikz{\filldraw (0,0)  circle (1pt);}}  circle (1pt) node[above] {$5$};
\filldraw (v1) node{\tikz{\filldraw (0,0)  circle (1pt);}} circle (1pt) node[left] {$1$};
\filldraw (v3) node{\tikz{\filldraw (0,0)  circle (1pt);}}  circle (1pt) node[right] {$3$};
\filldraw (v2) node{\tikz{\filldraw (0,0)  circle (1pt);}}  circle (1pt) node[left] {$2$};
\filldraw (v6) node{\tikz{\filldraw (0,0)  circle (1pt);}} circle (1pt) node[above] {$6$};
\filldraw (v7) node{\tikz{\filldraw (0,0)  circle (2pt);}}  circle (1.5pt) node[above] {$7,8$};
\draw (v3) -- (v4) -- (v5);
\node[circle,draw=black, fill=black, inner sep=0pt,minimum size=1.5pt] (v9) at (-0.125,1.0625) {};
\draw (v9) node{\tikz{\filldraw (0,0)  circle (1pt);}}  node[right]{$9$};
\draw (v1) -- (v7);
\end{tikzpicture}
},
\nl
& 
\raisebox{-1.1cm}{
\begin{tikzpicture}[scale=.8]
\node[circle,draw=black, fill=black, inner sep=0pt,minimum size=1.5pt] (v4) at (-2.15,2.15) {};
\node[circle,draw=black, fill=black, inner sep=0pt,minimum size=1.5pt] (v5) at (-1.5,2.15) {};
\node[circle,draw=black, fill=black, inner sep=0pt,minimum size=1.5pt] (v3) at (-1.9,1.76) {};
\node[circle,draw=black, fill=black, inner sep=0pt,minimum size=1.5pt] (v2) at (-1.25,0.875) {};
\node[circle,draw=black, fill=black, inner sep=0pt,minimum size=1.5pt] (v1) at (-0.75,0.125) {};
\node[circle,draw=black, fill=black, inner sep=0pt,minimum size=1.5pt] (v6) at (-0.5,2) {};
\node[circle,draw=black, fill=black, inner sep=0pt,minimum size=2.5pt] (v7) at (0.5,2) {};
\draw (v1) -- (-1.85,1.775);
\draw (-2,2) -- (v7);
\filldraw[fill=white!,draw=black,dashed] (-1.75,2.1) circle (0.75);
\filldraw (v4) node{\tikz{\filldraw (0,0)  circle (1pt);}} circle (1pt) node[above] {$4$};
\filldraw (v5) node{\tikz{\filldraw (0,0)  circle (1pt);}} circle (1pt) node[above] {$5$};
\filldraw (v1)node{\tikz{\filldraw (0,0)  circle (1pt);}}  circle (1pt) node[left] {$1$};
\filldraw (v3)node{\tikz{\filldraw (0,0)  circle (1pt);}}  circle (1pt) node[right] {$3$};
\filldraw (v2) node{\tikz{\filldraw (0,0)  circle (1pt);}}  circle (1pt) node[left] {$2$};
\filldraw (v6) node{\tikz{\filldraw (0,0)  circle (1pt);}}  circle (1pt) node[above] {$6$};
\filldraw (v7) node{\tikz{\filldraw (0,0)  circle (2pt);}} circle (1.5pt) node[above] {$7,8$};
\draw  (v4) -- (v5);
\node[circle,draw=black, fill=black, inner sep=0pt,minimum size=1.5pt] (v9) at (-0.125,1.0625) {};
\draw (v9) node{\tikz{\filldraw (0,0)  circle (1pt);}} node[right]{$9$};
\draw (v1) -- (v7);
\end{tikzpicture}
},
\ 
\raisebox{-1.1cm}{
\begin{tikzpicture}[scale=.8]
\node[circle,draw=black, fill=black, inner sep=0pt,minimum size=1.5pt] (v4) at (-2.15,2.15) {};
\node[circle,draw=black, fill=black, inner sep=0pt,minimum size=1.5pt] (v5) at (-1.5,2.15) {};
\node[circle,draw=black, fill=black, inner sep=0pt,minimum size=1.5pt] (v3) at (-1.9,1.76) {};
\node[circle,draw=black, fill=black, inner sep=0pt,minimum size=1.5pt] (v2) at (-1.25,0.875) {};
\node[circle,draw=black, fill=black, inner sep=0pt,minimum size=1.5pt] (v1) at (-0.75,0.125) {};
\node[circle,draw=black, fill=black, inner sep=0pt,minimum size=1.5pt] (v6) at (-0.5,2) {};
\node[circle,draw=black, fill=black, inner sep=0pt,minimum size=2.5pt] (v7) at (0.5,2) {};
\draw (v1) -- (-1.85,1.775);
\draw (-2,2) -- (v7);
\filldraw[fill=white!,draw=black,dashed] (-1.75,2.1) circle (0.75);
\filldraw (v4) node{\tikz{\filldraw (0,0)  circle (1pt);}}  circle (1pt) node[above] {$4$};
\filldraw (v5)node{\tikz{\filldraw (0,0)  circle (1pt);}}  circle (1pt) node[above] {$5$};
\filldraw (v1)node{\tikz{\filldraw (0,0)  circle (1pt);}}  circle (1pt) node[left] {$1$};
\filldraw (v3)node{\tikz{\filldraw (0,0)  circle (1pt);}}  circle (1pt) node[right] {$3$};
\filldraw (v2)node{\tikz{\filldraw (0,0)  circle (1pt);}}  circle (1pt) node[left] {$2$};
\filldraw (v6)node{\tikz{\filldraw (0,0)  circle (1pt);}}  circle (1pt) node[above] {$6$};
\filldraw (v7)node{\tikz{\filldraw (0,0)  circle (2pt);}}  circle (1.5pt) node[above] {$7,8$};
\draw  (v4) -- (v3);
\node[circle,draw=black, fill=black, inner sep=0pt,minimum size=1.5pt] (v9) at (-0.125,1.0625) {};
\draw (v9) node{\tikz{\filldraw (0,0)  circle (1pt);}} node[right]{$9$};
\draw (v1) -- (v7);
\end{tikzpicture}
},\ 
\raisebox{-1.1cm}{
\begin{tikzpicture}[scale=.8]
\node[circle,draw=black, fill=black, inner sep=0pt,minimum size=1.5pt] (v4) at (-2,2.2) {};
\node[circle,draw=black, fill=black, inner sep=0pt,minimum size=1.5pt] (v5) at (-1.5,2.2) {};
\node[circle,draw=black, fill=black, inner sep=0pt,minimum size=1.5pt] (v3) at (-1.85,1.775) {};
\node[circle,draw=black, fill=black, inner sep=0pt,minimum size=1.5pt] (v2) at (-1.375,1.0625) {};
\node[circle,draw=black, fill=black, inner sep=0pt,minimum size=2.5pt] (v1) at (-0.75,0.125) {};
\node[circle,draw=black, fill=black, inner sep=0pt,minimum size=1.5pt] (v6) at (-0.5,2) {};
\node[circle,draw=black, fill=black, inner sep=0pt,minimum size=2.5pt] (v7) at (0.5,2) {};
\draw (v1) -- (v3);
\draw (-2,2) -- (v7);
\filldraw[fill=white!,draw=black,dashed] (-1.75,2.1) circle (0.75);
\filldraw (v4)node{\tikz{\filldraw (0,0)  circle (1pt);}}  circle (1pt) node[above] {$4$};
\filldraw (v5)node{\tikz{\filldraw (0,0)  circle (1pt);}}  circle (1pt) node[above] {$5$};
\filldraw (v1)node{\tikz{\filldraw (0,0)  circle (1pt);}}  circle (1pt) node[left] {$1,2$};
\filldraw (v2)node{\tikz{\filldraw (0,0)  circle (1pt);}}  circle (1pt) node[left] {$3$};
\filldraw (v6)node{\tikz{\filldraw (0,0)  circle (1pt);}}  circle (1pt) node[above] {$6$};
\filldraw (v7)node{\tikz{\filldraw (0,0)  circle (2pt);}}  circle (1.5pt) node[above] {$7,8$};
\draw (v4) -- (v5);
\node[circle,draw=black, fill=black, inner sep=0pt,minimum size=1.5pt] (v9) at (-0.125,1.0625) {};
\draw (v9) node{\tikz{\filldraw (0,0)  circle (1pt);}} node[right]{$9$};
\draw (v1) -- (v7);
\end{tikzpicture}
},
\ea  
which via the map give the following six seeds of facets, 
\ba 
&S_{12(345)6(78)}+s_{456}+s_{457}+s_{458}\,,
\nl
&
S_{12(345)6(78)}+s_{431}+s_{432}\,,
\nl 
&S_{12(345)6(78)91}+(s_{456}+s_{457}+s_{458})+(s_{134}+s_{234})\,,
\nl
&S_{12(345)6(78)91}+s_{456}+s_{457}+s_{458}\,
\nl 
&S_{12(345)6(78)91}+s_{431}+s_{432}\,,
\nl
&S_{(12)3(45)6(78)9(12)}+s_{456}+s_{457}+s_{458},
\ea  
respectively. Under cyclic permutations, the first seed gives a cyclic class of length 3, and the rest give 5 cyclic classes of length 9. By reflection $i\to 10-i$ we have 6 more classes, thus  we get 96 facets in total through \eqref{fajkfaodijfslfd}. The last two kinds of seeds are given by
\ba \label{ffadsojfjasf}
&
S_{12(345)6(78)91}+(s_{456}+s_{457}+s_{458})+(s_{134}+s_{234})+(s_{467}+s_{567})\,,
\nl 
&S_{(12)3(45)6(78)9(12)}+(s_{456}+s_{457}+s_{458})+(s_{467}+s_{567})\,,
\ea 
whose  boundary  configurations 
can be obtained by further ``blowing up'' the remaining black dots of the two boundary configurations in the last column of \eqref{fajkfaodijfslfd}. See details  in appendix \ref{fahfo}. 
 Under the cyclic permutations, they give 2  classes of length 9 and 3 respectively. By reflection, we obtain another 2  classes, thus we get 24 facets in total through \eqref{ffadsojfjasf}. 

Given all the facets $F_a$ of $\calP(3,9)$ through \eqref{fdjasf}, \eqref{fdjasf2}, \eqref{fajkfaodijfslfd} and \eqref{ffadsojfjasf}, the polytope can be cut out by the inequalities $F_a\geq 0$ for $a = 1,2,...,471$.  We have found out the number of vertices of ${\cal P}(3,9)$ is 346 710.

\section{Polytopes, amplitudes and boundary configurations for $k=4$}

In this section, we move to $k=4$ and present our results for ${\cal P}(4,n)$ up to $n=9$ using Minkowski sum or SE map. Again since the subspace is given, it suffices to provide linear functions for all the facets (or poles of the amplitudes), which cut out the polytope ${\cal P}(4,n)$. Similar to the case in $k=3$, each facet of $\calP(4,n)$ corresponds to a boundary configuration of $G_+(4,n)/T$ where we take limit of one or more web variables controlled by a parameter $\epsilon\to 0$. Each facet is given by the residue at $\epsilon \to 0$ via the SE map, \eqref{map}. 

We will again provide a gauge-invariant way to describe these facets and the corresponding boundary configurations, which are now configurations of $n$ points in $3$-dimensional space.  As a warm-up, we will first take a look at these configurations in $G_+(4,7)/T$. The ``parity" duality between $G_+(4,7)/T$ and $G_+(3,7)/T$ allows us to directly obtain all the facets of $\calP(4,7)$ from those of $\calP(3,7)$ by the isomorphism $s_{i\,j\,k}\mapsto s_{a\,b\,c\,d}$ where $\{i,\,j,\,k,\,a,\,b,\,c,\,d\}=\{1,2,...,7\}$.  

This isomorphism, when applied to the facets of $\calP(3,7)$, provides the dictionary between their facets in table~\ref{tab3747}. We will explain the notation for facets of $\calP(4,7)$ which is encoding the corresponding boundary configurations of $G_+(4,7)/T$. This will serve as our starting point for presenting facets and configurations for higher points with $k=4$. 

\begin{table}[!h]
\centering
\begin{tabular}{c|c}
$\calP(3,7)$ & $\calP(4,7)$\\
\hline
$S_{123}$ & ${\bm S}_{4567}$ \\
$S_{1234}$ & ${\bm S}_{[567]}$ \\ 
$S_{12345}$ & ${\bm S}_{(67)}$ \\
$S_{(12)34}=S_{(34)567}=S_{(567)12}$ & ${\bm S}_{567(12)}={\bm S}_{[12(34)]}={\bm S}_{34[567]}$\\
$S_{(12)3(45)}=S_{45(67)12}$ & ${\bm S}_{(45)67(12)}={\bm S}_{[12(3)45]}$
\end{tabular}
\caption{Dictionary between facets in $\calP(3,7)$ and those in $\calP(4,7)$.}
\label{tab3747}
\end{table}

The simplest facets of $\calP(3,7)$ are like $S_{123}=s_{123}$, which map to $s_{4567}$ for $\calP(4,7)$. We denote the facet as ${\bm S}_{4567}$, and in general ${\bm S}_A$ for a set $A$ is defined as:
\begin{equation}
\bms_A:=\sum\limits_{a_1<a_2<a_3<a_4\in A}s_{a_1\,a_2\,a_3\,a_4}.
\end{equation}
The boundary configuration of $G_+(4, n)/T$ corresponding to ${\bm S}_A$ is given by a configuration where all points in $A$ are coplanar. This means
\begin{equation}
(a_1,\,a_2,\,a_3,\,a_4)=\mathcal{O}(\epsilon),\ \forall a_1,\,a_2,\,a_3,\,a_4\in A,
\end{equation}
where $a_1,\,a_2,\,a_3,\,a_4$ are $4$ labels in $A$. We draw this configuration by putting all points in $A$ on a plane. For example, for $A=\{1,2,3,4,5\}$, we have
\ba 
{\bm S}\left[
\raisebox{-.45cm}{
\begin{tikzpicture}
\draw[blue] (0,0)--++(2.2,0)--+(70:1)
(0,0)--++(70:1)--+(2.2,0);
\filldraw (.5,.5) circle (1pt) node [below] {1}  ;
\filldraw (.9,.8) circle (1pt) node [below] {2}  ;
\filldraw (1.5,.1) circle (1pt) node [above] {3}  ;
\filldraw (1.9,.7) circle (1pt) node [left] {4}  ;
\filldraw (2.0,.2) circle (1pt) node [above] {5}  ;
\end{tikzpicture} 
}
\right]
= \sum_{1\leq a_1<a_2<a_3<a_4\leq 5} s_{a_1\,a_2\,a_3\,a_4} ={\bm S}_{12345}\,,
\ea

The next case concerns facets like $S_{1234}$ in $\calP(3,7)$. Under the isomorphism, it becomes a facet of $\calP(4,7)$, which reads
\begin{equation}
s_{123}+s_{124}+s_{134}+s_{234}\mapsto s_{4567}+s_{3567}+s_{2567}+s_{1567}\,.
\end{equation}
We denote the facet as $\bms_{[567]}$, where in general $\bms_{[B]}$ for a set $B$ is defined as:
\begin{equation}
\bms_{[B]}:=\sum\limits_{b_1<b_2<b_3<b_4\in A}2s_{b_1\,b_2\,b_3\,b_4}+\sum\limits_{\substack{b_1<b_2<b_3\in B\\i\notin B}}s_{b_1\,b_2\,b_3\,i}.
\end{equation}
The boundary configuration corresponding to the facet $\bms_{[B]}$ is one where the points in $B$ are collinear, {\it i.e.} for any $b_1, \cdots, b_4 \in B$ and $i \in \bar{B}$,
\begin{equation}
\begin{aligned}
(b_1,\,b_2,\,b_3,\,b_4)=\mathcal{O}(\epsilon^2),\quad 
(b_1,\,b_2,\,b_3,\,i)=\mathcal{O}(\epsilon)\,.
\end{aligned}
\end{equation}
We draw points in $B$ on a line (which is of course on a plane) to denote the configuration. For example, for $B=\{1,2,3,4,5\}$, we have 
\ba 
{\bm S}\left[
\raisebox{-.6cm}{
\begin{tikzpicture}
\draw[blue] (0,0)--++(2.2,0)--+(70:1.5)
(0,0)--++(70:1.5)--+(2.2,0);
\draw[thick]
(.4,.5)
node {\tikz{\filldraw (0,0) circle (1pt) ;}}
node [below] {1} 
--++($.25*(70:1)+.25*(2.2,0)-.5*(.4,.2)$)
node {\tikz{\filldraw (0,0) circle (1pt) ;}}
node [below] {2} 
--++($.25*(70:1)+.25*(2.2,0)-.5*(.4,.2)$)
node {\tikz{\filldraw (0,0) circle (1pt) ;}}
node [below] {3} 
--++($.25*(70:1)+.25*(2.2,0)-.5*(.4,.2)$)
node {\tikz{\filldraw (0,0) circle (1pt) ;}}
node [below] {4} 
--+($.25*(70:1)+.25*(2.2,0)-.5*(.4,.2)$)
node {\tikz{\filldraw (0,0) circle (1pt) ;}}
node [below] {5} 
;
\end{tikzpicture} 
}
\right]
= 2 \sum_{1\leq b_1<b_2<b_3<b_4\leq 5} s_{b_1\,b_2\,b_3\,b_4} +\sum_{1\leq b_1<b_2<b_3\leq 5<i\leq n} s_{b_1\,b_2\,b_3\,i} ={\bm S}_{[12345]}\,.
\ea

Next we move to facets like $S_{(12)}$, which maps to a facet of $\calP(4,7)$ which reads
\begin{equation}
\begin{aligned}
 &s_{123}+s_{124}+s_{125}+s_{234}+s_{235}+s_{245}\\
\mapsto& s_{4567}+s_{3567}+s_{3467}+s_{1567}+s_{1467}+s_{1367},
\end{aligned}
\end{equation}
We denote such a facet as $\bms_{(12)}$, and in general $\bms_{(C)}$ for a set $C$ is defined as:
\begin{equation}
\bms_{(C)}:=\sum\limits_{\substack{c_1,c_2,c_3,c_4\in A}}3s_{c_1\,c_2\,c_3\,c_4}+\sum\limits_{\substack{c_1,c_2,c_3\in A\\i\notin A}}2s_{c_1\,c_2\,c_3\,i}+\sum\limits_{\substack{c_1,c_2\in A\\ i,j\notin A}}s_{c_1\,c_2\,i\,j}.
\end{equation}
The configuration corresponding to $\bms_C$ is that all the points in $C$ collide, {\it i.e.} for any $c_1, \cdots, c_4 \in C$ and $i,j \in \bar{C}$,
\begin{equation}
(c_1,\,c_2,\,c_3,\,c_4)=\mathcal{O}(\epsilon^3),\quad 
(c_1,\,c_2,\,c_3,\,i)=\mathcal{O}(\epsilon^2),\ \quad
(c_1,\,c_2,\,i,\,j)=\mathcal{O}(\epsilon). \ 
\end{equation}
We draw all points in $C$ on a black dot to denote this configuration. For example. for $C=\{1,2,3,4\}$, we have 
 \ba 
{\bm S}\left[
\raisebox{-.45cm}{
\begin{tikzpicture}
\draw[blue] (0,0)--++(1.5,0)--+(70:1)
(0,0)--++(70:1)--+(1.5,0);
\filldraw ($.5*(70:1)+.5*(1.5,0)$) circle (2pt) node [below] {1,2,3,4}  ;
\end{tikzpicture} 
}
\right]
=
3\, s_{1234} +2\sum_{1\leq c_1<c_2<c_3\leq 4<i\leq n} s_{c_1\,c_2\,c_3\,i} + \sum_{1\leq c_1<c_2\leq 4<i<j\leq n} s_{c_1\,c_2\,i\,j}={\bm S}_{(1234)}\,.
\ea  

Similar to the identity $S_{(B)}=S_{\bar B}$  for $k=3$, here for $k=4$ we have the following identities on the support of momentum conservation $\sum\limits_{a,b,c\neq i}s_{a\,b\,c\,i}=0$ (for $i=1,2,\cdots, n$): 
\begin{equation}
\bms_{A}=\bms_{(\bar{A})},\ \bms_{[A]}=\bms_{[\bar{A}]}.
\end{equation}
The corresponding configurations are equivalent for $k=4$: points in $A$ are coplanar is equivalent to the points in $\bar{A}$ colliding, and points in $A$ collinear is equivalent to points in $\bar{A}$ collinear. Before proceeding, note that a natural basis where the polytope lives can be written using these facets already. We have ${\bm X}=\{{\bm S}_{1234},{\bm S}_{12345},{\bm S}_{[123]},{\bm S}_{[1234]},{\bm S}_{(12)},{\bm S}_{(123)}\}$ for $(4,7)$, and for general $n$ the basis ${\bm X}=\{X_1,\cdots, X_{3(n-5)}\}$ can be written as
  \be \label{jfask}
{\bm X}=  \bigcup_{j=4}^{n-2} \big\{{\bm S}_{1,2,\cdots,j},{\bm S}_{[1,2,\cdots,j-1]},{\bm S}_{(1,2,\cdots,j-2)}\big\} \,.
  \ee 

Now let's present the remaining facets which correspond to the last two types in $\calP(3,7)$. Consider the facet $S_{(12)34}$ of $\calP(3,7)$, under the isomorphism it becomes one for $\calP(4,7)$:
\begin{equation}
S_{(12)34}=S_{(12)}+s_{134}+s_{234}\mapsto \bms_{(12)}+s_{2567}+s_{1567},
\end{equation}
which corresponds to a configuration where points $1, 2$ collide and they become coplanar with points $ 5,6,7$ at the same rate, and we denote the facet as $\bms_{567(12)}$. Recall that for $\calP(3,7)$, we have the identity $S_{(12)34}=S_{(34)567}=S_{(567)12}$, and the latter two facets become 
\begin{equation}
\begin{aligned}
&S_{(34)567}=S_{(34)}+s_{356}+s_{357}+s_{367}+s_{456}+s_{457}+s_{467}+s_{567}\\
\mapsto &S_{(34)}+s_{1247}+s_{1246}+s_{1245}+s_{1237}+s_{1236}+s_{1235}+s_{1234}:=\bms_{[12(34)]};\\
 &S_{(567)12}=S_{(567)}+s_{125}+s_{126}+s_{127}\\
\mapsto &S_{[567]}+s_{3467}+s_{3457}+s_{3456}:=\bms_{34[567]}\,.\\
\end{aligned}
\end{equation}
As the notation suggests, the first configuration denotes that points $3,4$ collide and they become collinear with $1$ and $2$;  the second configuration denotes that points $5,6,7$ are collinear, and become coplanar with  $3$ and $4$. Thus we have an identity for $(4,7)$ as well:
\begin{equation}\label{eqidentity471}
\bms_{567(12)}=\bms_{[12(34)]}=\bms_{34[567]}.
\end{equation}

More generally, we can have a  boundary configuration for $G_+(4,n)/T$ like this
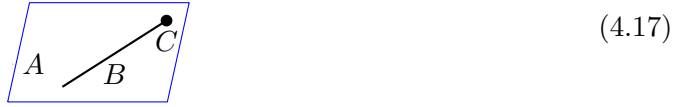
\be 
\raisebox{-.95cm}{
\begin{tikzpicture}[xscale=.6]
\draw[blue] (0.3,0.3)--++(3.5,0)--+(70:1.4)
(0.3,0.3)--++(70:1.4)--+(3.5,0);
\draw[thick]
 ($(1.5,.5)$)
node {}
node [below] {} 
--++($.5*(70:2)+.5*(4,0)-(1.2,.5)$)
node {\tikz{\filldraw (0,0) circle (0pt) ;}}
node [below] {$B$} 
--+($.5*(70:2)+.5*(4,0)-(1.2,.5)$)
node {\tikz{\filldraw (0,0) circle (2pt) ;}}
node [below] {$C$} 
;
\filldraw (.4,.8) circle (0pt) node [right] {$A$}  ;
\end{tikzpicture}
}
\ee 
where 
 points in $C$ collide, and they are collinear with points in $B$, and finally points in $B$, $C$ altogether become coplanar with points in another set $A$.   Via the map \eqref{map},
the facet corresponding for such a configuration becomes
\ba \label{jaklsf;}
{\bm S}_{A[B(C)]}= & {\bm S}_{A}+{\bm S}_{[B(C)]}+\!\!\sum_{\substack{a_1<a_2<a_3\in A\\i\in B\cup C}}\!\!s_{a_1\,a_2\,a_3\,i}
 +\!\!\sum_{\substack{a_1<a_2\in A\\b_1<b_2\in B}} \!\!s_{a_1\,a_2\,b_1\,b_2}
 +\!\!\sum_{\substack{a_1,a_2\in A\\b\in B,c\in C}}\!\!s_{a_1\,a_2\,b\,c}\,,
\ea
where  ${\bm S}_{[B(C)]}$ is also a facet given by
\be \label{faiuefj}
{\bm S}_{[B(C)]}=  {\bm S}_{[B]}+ {\bm S}_{(C)}+  \sum_{\substack{b_1<b_2\in B\\c\in C, i\notin B\cup C}} s_{b_1\,b_2\,c\,i}+  \sum_{\substack{b_1<b_2<b_3\in B\\c\in C}} s_{b_1\,b_2\,b_3\,c}+  \sum_{\substack{b_1<b_2\in B\\c_1<c_2\in C}} s_{b_1\,b_2\,c_1\,c_2}\,.
\ee
${\bm S}_{A[B(C)]}$  can also be written as ${\bm S}_{[(C)B]A}$. 
Here each of the set $A,B,C$ could be empty and the facet ${\bm S}_{A[B(C)]}$ would reduce to ${\bm S}_{[B(C)]}, {\bm S}_{A(C)}$ and ${\bm S}_{A[B]}$ or even ${\bm S}_{A},{\bm S}_{[B]}$ and ${\bm S}_{(C)}$. When there is only one point in $A,B$ or $C$, the facet can also be simplified as implied by the geometry of their configurations. For example, if $B=\{b\}$, we have
${\bm S}_{A[B(C)]}={\bm S}_{{\tilde A}(C)}$ with $\tilde{A}=A\cup\{b\}$.  Without ambiguity, we can just denote ${\bm S}_{{\tilde A}(C)}$ as ${\bm S}_{Ab(C)}$.

We denote the set of remaining points as $D$ and 
on the support of momentum conservation one can find 
\ba\label{eqidentity3fasioj}
{\bm S}_{A[B(C)]}={\bm S}_{B[C(D)]}={\bm S}_{C[D(A)]}={\bm S}_{D[A(B)]},\ \quad  \forall A\sqcup B\sqcup C\sqcup D=\{1,2,...,n\}\,.
\ea
 This means that the four corresponding configurations of $G_+(4,n)/T$ are equivalent. Diagrammatically we write
\ba 
\raisebox{-.75cm}{
\begin{tikzpicture}[xscale=.6]
\draw[blue] (0.3,0.3)--++(3.5,0)--+(70:1.4)
(0.3,0.3)--++(70:1.4)--+(3.5,0);
\draw[thick]
 ($(1.5,.5)$)
node {}
node [below] {} 
--++($.5*(70:2)+.5*(4,0)-(1.2,.5)$)
node {\tikz{\filldraw (0,0) circle (0pt) ;}}
node [below] {$B$} 
--+($.5*(70:2)+.5*(4,0)-(1.2,.5)$)
node {\tikz{\filldraw (0,0) circle (2pt) ;}}
node [below] {$C$} 
;
\filldraw (.4,.8) circle (0pt) node [right] {$A$}  ;
\filldraw (3,2) circle (0pt) node [right] {$D$}  ;
\end{tikzpicture} 
}
\simeq
\raisebox{-.75cm}{
\begin{tikzpicture}[xscale=.6]
\draw[blue] (0.3,0.3)--++(3.5,0)--+(70:1.4)
(0.3,0.3)--++(70:1.4)--+(3.5,0);
\draw[thick]
 ($(1.5,.5)$)
node {}
node [below] {} 
--++($.5*(70:2)+.5*(4,0)-(1.2,.5)$)
node {\tikz{\filldraw (0,0) circle (0pt) ;}}
node [below] {$C$} 
--+($.5*(70:2)+.5*(4,0)-(1.2,.5)$)
node {\tikz{\filldraw (0,0) circle (2pt) ;}}
node [below] {$D$} 
;
\filldraw (.4,.8) circle (0pt) node [right] {$B$}  ;
\filldraw (3,2) circle (0pt) node [right] {$A$}  ;
\end{tikzpicture} 
}
\simeq
\raisebox{-.75cm}{
\begin{tikzpicture}[xscale=.6]
\draw[blue] (0.3,0.3)--++(3.5,0)--+(70:1.4)
(0.3,0.3)--++(70:1.4)--+(3.5,0);
\draw[thick]
 ($(1.5,.5)$)
node {}
node [below] {} 
--++($.5*(70:2)+.5*(4,0)-(1.2,.5)$)
node {\tikz{\filldraw (0,0) circle (0pt) ;}}
node [below] {$D$} 
--+($.5*(70:2)+.5*(4,0)-(1.2,.5)$)
node {\tikz{\filldraw (0,0) circle (2pt) ;}}
node [below] {$A$} 
;
\filldraw (.4,.8) circle (0pt) node [right] {$C$}  ;
\filldraw (3,2) circle (0pt) node [right] {$B$}  ;
\end{tikzpicture} 
}
\simeq
\raisebox{-.75cm}{
\begin{tikzpicture}[xscale=.6]
\draw[blue] (0.3,0.3)--++(3.5,0)--+(70:1.4)
(0.3,0.3)--++(70:1.4)--+(3.5,0);
\draw[thick]
 ($(1.5,.5)$)
node {}
node [below] {} 
--++($.5*(70:2)+.5*(4,0)-(1.2,.5)$)
node {\tikz{\filldraw (0,0) circle (0pt) ;}}
node [below] {$A$} 
--+($.5*(70:2)+.5*(4,0)-(1.2,.5)$)
node {\tikz{\filldraw (0,0) circle (2pt) ;}}
node [below] {$B$} 
;
\filldraw (.4,.8) circle (0pt) node [right] {$D$}  ;
\filldraw (3,2) circle (0pt) node [right] {$C$}  ;
\end{tikzpicture} 
}\,.
\ea  
Eq. \eqref{eqidentity471} is an application of this identity where $A=\{5,6,7\},B=\emptyset,C=\{1,2\},D=\{3,4\}$.

Finally, for the last type of facets of $\calP(3,7)$, $S_{(12)3(45)}$, we have its parity dual
\begin{equation}
\begin{aligned}
 &S_{(12)3(45)}=S_{(12)}+S_{(45)}+s_{134}+s_{135}+s_{234}+s_{235}\\
\mapsto& S_{(12)}+S_{(45)}+s_{2567}+s_{2467}+s_{1567}+s_{1467}.
\end{aligned}
\end{equation}
We denote such a facet as $\bms_{(45)67(12)}$, and the corresponding configuration is that points $1, 2$ and $4,5$ collide respectively at the same rate, and both of them become coplanar with points $6$ and $7$. Diagrammatically the configuration is denoted as $\raisebox{-.6cm}{
\begin{tikzpicture}[xscale=.3]
\draw[blue] ($(0.5,0.3)-(3.5,0)$)--++(2*3.5,0)--+(70:1.4)
($(0.5,0.3)-(3.5,0)$)--++(70:1.4)--+(2*3.5,0);
\draw
($(.8,.5)+(70:2)+(4,0)-2*(1.2,.5)$)
--($(.8,.5)+(70:2)+(4,0)-2*(1.2,.5)$)
node {\tikz{\filldraw (0,0) circle (2pt) ;}}
node [below ] {$1,2$} 
;
\draw
($(0.5,.5)+(180-70:2)+(-4,0)-2*(-1.2,.5)$)
--($(0.5,.5)+(180-70:2)+(-4,0)-2*(-1.2,.5)$)
node {\tikz{\filldraw (0,0) circle (2pt) ;}}
node [below ] {$4,5$}
(.7,1.2)
--(.7,1.2)
node {\tikz{\filldraw (0,0) circle (1pt) ;}}
node [left ] {$7$}
(.4,.8)
--(.4,.8)
node {\tikz{\filldraw (0,0) circle (1pt) ;}}
node [below ] {$6$}
  ($(.2,1.1)+(70:2)+(4,0)-2*(1.2,.5)$)
;
\end{tikzpicture} 
}$. 

Recall that for $(3,7)$ we have $S_{(12)3(45)}=S_{45(67)12}$, and the facet of $(4,7)$ corresponds to the latter reads
\begin{equation}
\begin{aligned}
&S_{45(67)12}=(s_{456}+s_{457}+s_{467}+s_{567})+(s_{126}+s_{127}+s_{167}+s_{267})+s_{367}\\
\mapsto &(s_{1237}+s_{1236}+s_{1235}+s_{1234})+(s_{3457}+s_{3456}+s_{2345}+s_{1345})+s_{1245}\\
 =& \bms_{[123]}+\bms_{[345]}+s_{1245}\,.\\
\end{aligned}
\end{equation}
We denote the facet as $\bms_{[12(3)45]}$, and the corresponding configuration is given by two lines, one with points $1,2,3$, and the other with points $3,4,5$. Diagrammatically we have $\raisebox{-.9cm}
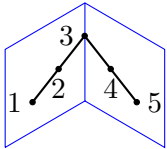
{
\begin{tikzpicture}[xscale=.3]
\draw[blue] (0.3,-.1)--++(90:1.3)--++(10:4-.45)--++(-10:4-.45)--++(-90:1.3)--+ (170:4-.45)
(0.3,-.1)--++(10:4-.45)--+(90:1.3);
\draw[thick]
 ($(1.5,.5)$)
node {\tikz{\filldraw (0,0) circle (1pt) ;}}
node [left] {$1$} 
--++($.5*(70:2)+.5*(4,0)-(1.2,.5)$)
node {\tikz{\filldraw (0,0) circle (1pt) ;}}
node [below] {$2$} 
--++($.5*(70:2)+.5*(4,0)-(1.2,.5)$)
node {\tikz{\filldraw (0,0) circle (1pt) ;}}
node [left] {$3$} 
--++($.5*(-70:2)+.5*(4,0)-(1.2,-.5)$)
node {\tikz{\filldraw (0,0) circle (1pt) ;}}
node [below] {$4$} 
--++($.5*(-70:2)+.5*(4,0)-(1.2,-.5)$)
node {\tikz{\filldraw (0,0) circle (1pt) ;}}
node [right] {$5$} 
 ($(5.5,1.9)$)
;
\end{tikzpicture} 
}$. 
The identity \eqref{fahjkafld} in ${\cal P}(3,7)$ now becomes an identity for ${\cal P}(4,7)$:
\begin{equation}\label{fasfijd}
\bms_{(45)67(12)}={\bm S}\left[
\raisebox{-.75cm}{
\begin{tikzpicture}[xscale=.3]
\draw[blue] ($(0.5,0.3)-(3.5,0)$)--++(2*3.5,0)--+(70:1.4)
($(0.5,0.3)-(3.5,0)$)--++(70:1.4)--+(2*3.5,0);
\draw
($(.8,.5)+(70:2)+(4,0)-2*(1.2,.5)$)
--($(.8,.5)+(70:2)+(4,0)-2*(1.2,.5)$)
node {\tikz{\filldraw (0,0) circle (2pt) ;}}
node [below ] {$1,2$} 
;
\draw
($(0.5,.5)+(180-70:2)+(-4,0)-2*(-1.2,.5)$)
--($(0.5,.5)+(180-70:2)+(-4,0)-2*(-1.2,.5)$)
node {\tikz{\filldraw (0,0) circle (2pt) ;}}
node [below ] {$4,5$}
(.7,1.2)
--(.7,1.2)
node {\tikz{\filldraw (0,0) circle (1pt) ;}}
node [left ] {$7$}
(.4,.8)
--(.4,.8)
node {\tikz{\filldraw (0,0) circle (1pt) ;}}
node [below ] {$6$}
  ($(.2,1.1)+(70:2)+(4,0)-2*(1.2,.5)$)-- ($(.2,1.1)+(70:2)+(4,0)-2*(1.2,.5)$) node {\tikz{\filldraw (0,0) circle (1pt) ;}}
node [left] {$3$}
;
\end{tikzpicture} 
}
\right]=
{\bm S}\left[
\raisebox{-.95cm}{
\begin{tikzpicture}[xscale=.3]
\draw[blue] (0.3,-.1)--++(90:1.3)--++(10:4-.45)--++(-10:4-.45)--++(-90:1.3)--+ (170:4-.45)
(0.3,-.1)--++(10:4-.45)--+(90:1.3);
\draw[thick]
 ($(1.5,.5)$)
node {\tikz{\filldraw (0,0) circle (1pt) ;}}
node [left] {$1$} 
--++($.5*(70:2)+.5*(4,0)-(1.2,.5)$)
node {\tikz{\filldraw (0,0) circle (1pt) ;}}
node [below] {$2$} 
--++($.5*(70:2)+.5*(4,0)-(1.2,.5)$)
node {\tikz{\filldraw (0,0) circle (1pt) ;}}
node [left] {$3$} 
--++($.5*(-70:2)+.5*(4,0)-(1.2,-.5)$)
node {\tikz{\filldraw (0,0) circle (1pt) ;}}
node [below] {$4$} 
--++($.5*(-70:2)+.5*(4,0)-(1.2,-.5)$)
node {\tikz{\filldraw (0,0) circle (1pt) ;}}
node [right] {$5$} 
 ($(5.5,1.9)$)-- ($(5.5,1.9)$)
node {\tikz{\filldraw (0,0) circle (1pt) ;}}
node [left] {$7$} 
 ($(6.2,1.7)$)-- ($(6.2,1.7)$)
node {\tikz{\filldraw (0,0) circle (1pt) ;}}
node [right] {$6$} 
;
\end{tikzpicture} 
}
\right]
=\bms_{[12(3)45]}.
\end{equation}

More generally we could have two facets in the form ${\bm S}_{[(A)B]C[D(E)]}, {\bm S}_{D[E(F)A]B}$ 
whose configurations can be diagrammatically shown by 
\be \label{fasshfjdap}
\raisebox{-.7cm}{
\begin{tikzpicture}[xscale=.6]
\draw[blue] ($(0.3,0.3)-(3.5,0)$)--++(2*3.5,0)--+(70:1.4)
($(0.3,0.3)-(3.5,0)$)--++(70:1.4)--+(2*3.5,0);
\draw[thick]
 ($(1.5,.5)$)
node {\tikz{\filldraw (0,0) circle (0pt) ;}}
node [left] {} 
--++($.5*(70:2)+.5*(4,0)-(1.2,.5)$)
node {\tikz{\filldraw (0,0) circle (0pt) ;}}
node [below] {$D$} 
--+($.5*(70:2)+.5*(4,0)-(1.2,.5)$)
node {\tikz{\filldraw (0,0) circle (2pt) ;}}
node [left=2pt] {$E$} 
;
\draw[thick]
 ($(0.5,.5)$)
node {\tikz{\filldraw (0,0) circle (0pt) ;}}
node [left=2pt] {} 
--++($.5*(180-70:2)+.5*(-4,0)-(-1.2,.5)$)
node {\tikz{\filldraw (0,0) circle (0pt) ;}}
node [below] {$B$} 
--+($.5*(180-70:2)+.5*(-4,0)-(-1.2,.5)$)
node {\tikz{\filldraw (0,0) circle (2pt) ;}}
node [below left=-2pt] {$A$} 
;
\filldraw (.7,1.2) circle (0pt) node [right] {$C$}  ;
\end{tikzpicture} 
}
\,,\quad 
\raisebox{-1.05cm}{
\begin{tikzpicture}[xscale=.6]
\draw[blue] (0.3,-.1)--++(90:1.6)--++(10:4-.45)--++(-10:4-.45)--++(-90:1.6)--+ (170:4-.45)
(0.3,-.1)--++(10:4-.45)--+(90:1.6);
\draw[thick]
 ($(1.5,.5)$)
node {\tikz{\filldraw (0,0) circle (0pt) ;}}
node [left] {} 
--++($.5*(70:2)+.5*(4,0)-(1.2,.5)$)
node {\tikz{\filldraw (0,0) circle (0pt) ;}}
node [below] {$E$} 
--++($.5*(70:2)+.5*(4,0)-(1.2,.5)$)
node {\tikz{\filldraw (0,0) circle (2pt) ;}}
node [left] {$F$} 
--++($.5*(-70:2)+.5*(4,0)-(1.2,-.5)$)
node {\tikz{\filldraw (0,0) circle (0pt) ;}}
node [below] {$A$} 
--++($.5*(-70:2)+.5*(4,0)-(1.2,-.5)$)
node {\tikz{\filldraw (0,0) circle (0pt) ;}}
node [right] {} 
;
\filldraw (1,1.1) circle (0pt) node [right] {$D$}  ;
\filldraw (6,1.3) circle (0pt) node [right] {$B$}  ;
\end{tikzpicture} 
}\,,
\ee  
respectively. Via the map \eqref{map}, we have 
\ba 
{\bm S}_{[(A)B]C[D(E)]}&={\bm S}_{[(A)B]CDE}+{\bm S}_{ABC[D(E)]}-{\bm S}_{ABCDE}+\!\!\sum_{\substack{e_1,e_2\in E,a_1,a_2\in A}}\!\!s_{e_1\,e_2\,a_1\,a_2}\,.
\nl 
{\bm S}_{D[E(F)A]B}&= {\bm S}_{D[E(F)]}+ {\bm S}_{[(F)A]B} -{\bm S}_{(F)} +\!\!\sum_{\substack{e_1,e_2\in E,a_1,a_2\in A}}\!\!s_{e_1\,e_2\,a_1\,a_2}\,.
\ea 
Besides, there is one identity as an analog of \eqref{eqequa3} on the support of the momentum conservation,
 \ba \label{fahjsf}
{\bm S}_{[(A)B]C[D(E)]}=
{\bm S}_{D[E(F)A]B}\,, \quad \forall\, A\sqcup B\sqcup C\sqcup D\sqcup E\sqcup F =\{1,2,\cdots,n\}\,. 
\ea 
Correspondingly, one can check their  configurations are equivalent,
\be  \label{fahjsf2}
\raisebox{-.7cm}{
\begin{tikzpicture}[xscale=.6]
\draw[blue] ($(0.3,0.3)-(3.5,0)$)--++(2*3.5,0)--+(70:1.4)
($(0.3,0.3)-(3.5,0)$)--++(70:1.4)--+(2*3.5,0);
\draw[thick]
 ($(1.5,.5)$)
node {\tikz{\filldraw (0,0) circle (0pt) ;}}
node [left] {} 
--++($.5*(70:2)+.5*(4,0)-(1.2,.5)$)
node {\tikz{\filldraw (0,0) circle (0pt) ;}}
node [below] {$D$} 
--+($.5*(70:2)+.5*(4,0)-(1.2,.5)$)
node {\tikz{\filldraw (0,0) circle (2pt) ;}}
node [left=2pt] {$E$} 
;
\draw[thick]
 ($(0.5,.5)$)
node {\tikz{\filldraw (0,0) circle (0pt) ;}}
node [left=2pt] {} 
--++($.5*(180-70:2)+.5*(-4,0)-(-1.2,.5)$)
node {\tikz{\filldraw (0,0) circle (0pt) ;}}
node [below] {$B$} 
--+($.5*(180-70:2)+.5*(-4,0)-(-1.2,.5)$)
node {\tikz{\filldraw (0,0) circle (2pt) ;}}
node [below left=-2pt] {$A$} 
;
\filldraw (.7,1.2) circle (0pt) node [right] {$C$}  ;
\filldraw (1.2,2) circle (0pt) node [right] {$F$}  ;
\end{tikzpicture} 
}
\simeq
\raisebox{-1.05cm}{
\begin{tikzpicture}[xscale=.6]
\draw[blue] (0.3,-.1)--++(90:1.6)--++(10:4-.45)--++(-10:4-.45)--++(-90:1.6)--+ (170:4-.45)
(0.3,-.1)--++(10:4-.45)--+(90:1.6);
\draw[thick]
 ($(1.5,.5)$)
node {\tikz{\filldraw (0,0) circle (0pt) ;}}
node [left] {} 
--++($.5*(70:2)+.5*(4,0)-(1.2,.5)$)
node {\tikz{\filldraw (0,0) circle (0pt) ;}}
node [below] {$E$} 
--++($.5*(70:2)+.5*(4,0)-(1.2,.5)$)
node {\tikz{\filldraw (0,0) circle (2pt) ;}}
node [left] {$F$} 
--++($.5*(-70:2)+.5*(4,0)-(1.2,-.5)$)
node {\tikz{\filldraw (0,0) circle (0pt) ;}}
node [below] {$A$} 
--++($.5*(-70:2)+.5*(4,0)-(1.2,-.5)$)
node {\tikz{\filldraw (0,0) circle (0pt) ;}}
node [right] {} 
;
\filldraw (1,1.1) circle (0pt) node [right] {$D$}  ;
\filldraw (6,1.3) circle (0pt) node [right] {$B$}  ;
\filldraw (5,2.2) circle (0pt) node [right] {$C$}  ;
\end{tikzpicture} 
}\,.
\ee 
Here we comment that  some of the sets $A,B,C,D,E,F$ could  be empty and the above identity \eqref{fahjsf} and equivalence relation \eqref{fahjsf2} still hold.   When $A,B,C$ or $D$ only contain one point, the facets can also be simplified as implied by the geometry of their configurations.  For example, if $D=\{d\}$, ${\bm S}_{[(A)B]C[D(E)]}$ and  ${\bm S}_{D[E(F)A]B}$ are nothing but ${\bm S}_{[A(B)]Cd(E)}$ and  ${\bm S}_{[E(F)A]B }$ respectively.  If $A=\{a\}$ and $D=\{d\}$, we have 
\be 
{\bm S}_{[(A)B]C[D(E)]}={\bm S}_{[aB]Cd(E)}, \quad {\bm S}_{D[E(F)A]B}={\bm S}_{[E(F)]aB}\,,
\ee
which are the same 
on the support of momentum conservation. Here we point out that points in $E,F,\tilde{B}$ with $\tilde{B}=\{a\}\cup B$  are on two planes instead of one and hence we cannot write ${\bm S}_{[E(F)]\tilde{B}}$ as ${\bm S}_{[(F)E]\tilde{B}}$ whose expression is given in \eqref{jaklsf;}.

When $F$  only contain a point $f$, the two facets cannot be simplified and we still denote them as ${\bm S}_{[(A)B]C[D(E)]}$ and  ${\bm S}_{D[E(f)A]B}$.  The identity \eqref{fasfijd} for $(4,7)$ is an application of \eqref{fahjsf2} where $A=\{4,5\},B=\emptyset,C=\{6,7\},D=\emptyset,E=\{1,2\},F=\{3\}$.

Up to now we have assumed $B$ and $D$  in \eqref{fasshfjdap} don't intersect. When $C$ is empty,  and $B,D$ contain a common point $c$, we get a minor 
 variation of  \eqref{fahjsf2},
\begin{equation}
    \raisebox{-0.7cm}{
    \begin{tikzpicture}[xscale=.6]
    \draw[blue] ($(0.3,0.3)-(3.5,0)$)--++(2*3.5,0)--+(70:1.4)
    ($(0.3,0.3)-(3.5,0)$)--++(70:1.4)--+(2*3.5,0);
    \draw[thick]
     ($(1,.5)$)
    node {\tikz{\filldraw (0,0) circle (0pt) ;}}
    node [left] {} 
    --++($.5*(70:2)+.5*(4,0)-(1.2,.5)$)
    node {\tikz{\filldraw (0,0) circle (0pt) ;}}
    node [below] {$\hat{D}$} 
    --+($.5*(70:2)+.5*(4,0)-(1.2,.5)$)
    node {\tikz{\filldraw (0,0) circle (2pt) ;}}
    node [left=2pt] {$E$} 
    ;
    \draw[thick]
     ($(1,.5)$)
    node {\tikz{\filldraw (0,0) circle (1pt) ;}}
    node [left=2pt] {$c$} 
    --++($.5*(180-70:2)+.5*(-4,0)-(-1.2,.5)$)
    node {\tikz{\filldraw (0,0) circle (0pt) ;}}
    node [below left=-2pt] {$\hat{B}$} 
    --+($.5*(180-70:2)+.5*(-4,0)-(-1.2,.5)$)
    node {\tikz{\filldraw (0,0) circle (2pt) ;}}
    node [below left=-2pt] {$A$} 
    ;
    \filldraw (1.2,2) circle (0pt) node [right] {$F$}  ;
    \end{tikzpicture}
    }
\simeq
    \raisebox{-1.05cm}{
    \begin{tikzpicture}[xscale=.6]
    \draw[blue] (0.3,-.1)--++(90:1.6)--++(10:4-.45)--++(-10:4-.45)--++(-90:1.6)--+ (170:4-.45)
    (0.3,-.1)--++(10:4-.45)--+(90:1.6*.75)
     node[black] {\tikz{\filldraw (0,0.3) circle (1pt) ;}}
    node [below right=-2pt,black] {$c$} 
    --+(90:1.6)
    ;
    \draw[thick]
     ($(1.5,.7-.35)$)
    node {\tikz{\filldraw (0,0) circle (0pt) ;}}
    node [left] {} 
    --++($.5*(70:2)+.5*(4,0)-(1.2,.5)$)
    node {\tikz{\filldraw (0,0) circle (0pt) ;}}
    node [below] {$E$} 
    --++($.5*(70:2)+.5*(4,0)-(1.2,.5)$)
    node {\tikz{\filldraw (0,0) circle (2pt) ;}}
    node [left] {$F$} 
    --++($.5*(-70:2)+.5*(4,0)-(1.2,-.5)$)
    node {\tikz{\filldraw (0,0) circle (0pt) ;}}
    node [below] {$A$} 
    --++($.5*(-70:2)+.5*(4,0)-(1.2,-.5)$)
    node {\tikz{\filldraw (0,0) circle (0pt) ;}}
    node [right] {} 
    ;
    \filldraw (1,1.1) circle (0pt) node [right] {$\hat{D}$}  ;
    \filldraw (6,1.3) circle (0pt) node [right] {$\hat{B}$}  ;
    \end{tikzpicture}
    }
    \,,
\end{equation}
 where we have defined 
 $\hat{B}=B\setminus\{c\}$, $\hat{D}=D\setminus\{c\}$ and we require $A\sqcup \hat{B} \sqcup \{c\} \sqcup \hat{D} \sqcup E \sqcup F=\{1,2,\cdots,n\}$.  
Correspondingly, via the map \eqref{map} their two facets read
\ba
{\bm S}_{[(A)\hat{B}(c)\hat{D}(E)]}&={\bm S}_{[(A)\Tilde{B}]c\hat{D}E}+{\bm S}_{A\hat{B}c[\hat{D}(E)]}-{\bm S}_{A\hat{B}c\hat{D}E}+\!\!\sum_{\substack{e_1,e_2\in E,a_1,a_2\in A}}\!\!s_{e_1\,e_2\,a_1\,a_2}
\nl 
{\bm S}_{c\hat{D}[E(F)A]\hat{B}c}&= {\bm S}_{c\hat{D}[E(F)]}+ {\bm S}_{[(F)A]\hat{B}c} -{\bm S}_{(F)} +\!\!\sum_{\substack{e_1,e_2\in E,a_1,a_2\in A}}\!\!s_{e_1\,e_2\,a_1\,a_2}\,.
\ea
which are equivalent on the support of momentum conservation.  For example,  when $n=8$ and $A=\{1\},{\hat B}=\{2\},C=\{3\}, {\hat D}=\{4\},E=\{5,6\},F=\{7,8\}$, we have
\begin{equation}\label{g48facet1flajkj}
{\bm S}_{[12(3)4(56)]}=\left[
    \raisebox{-1cm}{
    \begin{tikzpicture}[xscale=.6]
    \draw[blue] ($(0.3,0.3)-(3.5,0)$)--++(2*3.5,0)--+(70:1.4)
    ($(0.3,0.3)-(3.5,0)$)--++(70:1.4)--+(2*3.5,0);
    \draw[thick]
     ($(1,.5)$)
    node {\tikz{\filldraw (0,0) circle (0pt) ;}}
    node [left] {} 
    --++($.5*(70:2)+.5*(4,0)-(1.2,.5)$)
    node {\tikz{\filldraw (0,0) circle (1pt) ;}}
    node [below] {$4$} 
    --+($.5*(70:2)+.5*(4,0)-(1.2,.5)$)
    node {\tikz{\filldraw (0,0) circle (2pt) ;}}
    node [left=2pt] {$5,6$} 
    ;
    \draw[thick]
     ($(1,.5)$)
    node {\tikz{\filldraw (0,0) circle (1pt) ;}}
    node [left=2pt] {$3$} 
    --++($.5*(180-70:2)+.5*(-4,0)-(-1.2,.5)$)
    node {\tikz{\filldraw (0,0) circle (1pt) ;}}
    node [below left=-2pt] {$2$} 
    --+($.5*(180-70:2)+.5*(-4,0)-(-1.2,.5)$)
    node {\tikz{\filldraw (0,0) circle (2pt) ;}}
    node [below left=-2pt] {$1$} 
    ;
    \filldraw (1.6,2) circle (1pt) node [right] {$7$}  ;
        \filldraw (-.4,2.2) circle (1pt) node [right] {$8$}  ;
    \end{tikzpicture}
    }
\right]
=\left[
    \raisebox{-1.05cm}{
    \begin{tikzpicture}[xscale=.6]
    \draw[blue] (0.3,-.1)--++(90:1.6)--++(10:4-.45)--++(-10:4-.45)--++(-90:1.6)--+ (170:4-.45)
    (0.3,-.1)--++(10:4-.45)--+(90:1.6*.75)
     node[black] {\tikz{\filldraw (0,0.3) circle (1pt) ;}}
    node [right,black] {$3$} 
    --+(90:1.6)
    ;
    \draw[thick]
     ($(1.5,.7-.35)$)
    node {\tikz{\filldraw (0,0) circle (0pt) ;}}
        node {\tikz{\filldraw (0,0) circle (1pt) ;}}
    node [left] {$5$} 
    --++($.5*(70:2)+.5*(4,0)-(1.2,.5)$)
    node {\tikz{\filldraw (0,0) circle (1pt) ;}}
    node [below] {$6$} 
    --++($.5*(70:2)+.5*(4,0)-(1.2,.5)$)
    node {\tikz{\filldraw (0,0) circle (2pt) ;}}
    node [right] {$7,8$} 
    ;
    \filldraw (1,1.1) circle (1pt) node [right] {$4$}  ;
    \filldraw (6,1.3) circle (1pt) node [right] {$2$}  ;
        \filldraw (5,0.8) circle (1pt) node [right] {$1$}  ;
    \end{tikzpicture}
    }
    \right]
    ={\bm S}_{34[56(78)]123}
    \,.
\end{equation}

Now let's turn to all the facets and configurations for $\calP(4,8)$.

\subsection{$\calP(4,8)$}

Now let's present the results for the polytope $\calP(4,8)$ by listing all the 360 facets and the corresponding configurations of $G_+(4,8)/T$. Similar to the polytope $\calP(3,9)$, in addition to normal configurations which can be expressed in terms of coplanar, collinear and colliding points, there are also exceptional configurations which need ``blowup" for some black dots. Let's first present the $268$ facets corresponding to normal configurations. 

Out of these $268$ facets, there are $8\times 22+ 4 \times 1+2\times2= 184$ facets which have already been introduced. Let's first list the $22$ cyclic classes of length $8$, where the seeds are
\ba \label{g48facet1}
&{\bm S}_{(12)},\, {\bm S}_{(123)},\, {\bm S}_{ (1234)},\,{\bm S}_{[123]},\,\,{\bm S}_{[123]45},\, {\bm S}_{[123]456},\, {\bm S}_{[1234]56},\,{\bm S}_{12[3456]},\, {\bm S}_{123[456]},\, {\bm S}_{12[345]},\ \nl
&{\bm S}_{[1,2,3],4,(5,6)},\, {\bm S}_{(1,2),3,[4,5,6]},\, {\bm S}_{[(1,2),3,4],5,(6,7)},\, {\bm S}_{(1,2),3,[4,5,(6,7)]},\, {\bm S}_{(1,2),3,4,(5,6)},\, {\bm S}_{[1,2,3],[4,5,6]},\,\, \nl
&{\bm S}_{[1,2,(3),4,5]},\,{\bm S}_{[1,2,3,(4),5,6]},\, {\bm S}_{[1,2,(3),4,5,6]},\, {\bm S}_{[(1,2),3,(4),5,(6,7)]},\, {\bm S}_{[(1,2),3,(4),5,6]},\,{\bm S}_{[1,2,(3),4,(5,6)]}\,;\quad 
\ea 
there are $8$ more facets which read
\begin{equation}\label{g48facet12}
\begin{aligned} 
{\bm S}_{[1234]},\,{\bm S}_{[2345]},\,{\bm S}_{[3456]},\,{\bm S}_{[4567]};\,\quad 
{\bm S}_{[(12)34]56},\, {\bm S}_{[(23)45]67};\, \quad {\bm S}_{12[34(56)]},\, {\bm S}_{23[45(67)]}.
\end{aligned}
\end{equation}
In addition, there are $84$ facets which are new (but still come from normal configurations). First we introduce the following normal configurations,
\ba \label{fdhfasjkijfslfd} 
\raisebox{0cm}{
\begin{tikzpicture}[baseline={([yshift=-.5ex]current bounding box.center)},scale=.65]
	\draw[blue] (0,1.75) -- (-1.8,1) -- (-1.8,-2) -- (0,-1.25) -- (0,1.75) -- (1.8,1) -- (1.8,-2) -- (0,-1.25);
\draw[thick] (0,1) -- (0,-1);
	\fill (0,1) node {\tikz{\filldraw (0,0) circle (1pt) ;}}  circle (1.5pt) node[left]{$5$};
	\fill (0,-1) node {\tikz{\filldraw (0,0) circle (1pt) ;}}  circle (1.5pt) node[left]{$3$};
	\fill (0,0) node {\tikz{\filldraw (0,0) circle (1pt) ;}}  circle (1.5pt) node[left]{$4$};
	\fill (-1,-0.6) node {\tikz{\filldraw (0,0) circle (1pt) ;}} 
	circle (1.5pt) node[left]{$2$};
	\fill (-1,0.6) node {\tikz{\filldraw (0,0) circle (1pt) ;}}  circle (1.5pt) node[left]{$1$};
	\fill (1,0.6)  node {\tikz{\filldraw (0,0) circle (1pt) ;}} circle (1.5pt) node[right]{$6$};
	\fill (1,-0.6)  node {\tikz{\filldraw (0,0) circle (1pt) ;}} circle (1.5pt) node[right]{$7$};
\end{tikzpicture}
}
,\ 
\raisebox{0cm}{
\begin{tikzpicture}[baseline={([yshift=-.5ex]current bounding box.center)},scale=.65]
	\draw[blue] (0,1.75) -- (-1.8,1) -- (-1.8,-2) -- (0,-1.25) -- (0,1.75) -- (1.8,1) -- (1.8,-2) -- (0,-1.25);
	\draw[thick] (0,1) -- (0,-1);
	\fill (0,1) node {\tikz{\filldraw (0,0) circle (1pt) ;}}  circle (1.5pt) node[left]{$5$};
	\fill (0,-1)  node {\tikz{\filldraw (0,0) circle (1pt) ;}} circle (1.5pt) node[left]{$3$};
	\fill (0,0)  node {\tikz{\filldraw (0,0) circle (1pt) ;}} circle (1.5pt) node[right=-1pt]{$4$};
	\draw[thick] (-1,0.6) -- (0,-1);
	\draw[thick] (1,-0.6) -- (0,1);
	\fill (-0.5,-0.2)  node {\tikz{\filldraw (0,0) circle (1pt) ;}} circle (1.5pt) node[left]{$2$};
	\fill (-1,0.6) node {\tikz{\filldraw (0,0) circle (1pt) ;}}  circle (1.5pt) node[left]{$1$};
	\fill (0.5,0.2) node {\tikz{\filldraw (0,0) circle (1pt) ;}}  circle (1.5pt) node[right]{$6$};
	\fill (1,-0.6) node {\tikz{\filldraw (0,0) circle (1pt) ;}}  circle (1.5pt) node[right]{$7$};
\end{tikzpicture}
}
,\ 
\raisebox{0cm}{
\begin{tikzpicture}[baseline={([yshift=-.5ex]current bounding box.center)},scale=.65]
	\draw[blue] (0,1.75) -- (-1.8,1) -- (-1.8,-2) -- (0,-1.25) -- (0,1.75) -- (1.8,1) -- (1.8,-2) -- (0,-1.25);
	\fill (0,1)  node {\tikz{\filldraw (0,0) circle (1pt) ;}} circle (1.5pt) node[left]{$7$};
	\fill (0,-1) node {\tikz{\filldraw (0,0) circle (1pt) ;}}  circle (1.5pt) node[left]{$3$};
	\fill (-1.2,0)  node {\tikz{\filldraw (0,0) circle (1pt) ;}} circle (1.5pt) ;
	\node at (-1.35,0) {$1$};
	\fill (-0.6,-0.5) node {\tikz{\filldraw (0,0) circle (1pt) ;}}  circle (1.5pt) node[below left=-2pt]{$2$};
	\fill (-0.6,0.5) node {\tikz{\filldraw (0,0) circle (1pt) ;}}  circle (1.5pt) node[above left=-2pt]{$8$};
	\fill (0.6,-0.5) node {\tikz{\filldraw (0,0) circle (1pt) ;}}  circle (1.5pt) node[below]{$4$};
	\fill (1.2,0)  node {\tikz{\filldraw (0,0) circle (1pt) ;}} circle  (1.5pt) node[above]{$5$};
	\fill (0.6,0.5)  node {\tikz{\filldraw (0,0) circle (1pt) ;}}  circle  (1.5pt) node[above]{$6$};
	\draw[thick] (-1.2,0) -- (0,-1) -- (1.2,0) -- (0,1) -- (-1.2,0);
\end{tikzpicture}
}
,\ 
\raisebox{0cm}{
\begin{tikzpicture}[baseline={([yshift=-.5ex]current bounding box.center)},scale=.65]
	\draw[blue] (0,1.75) -- (-1.8,1) -- (-1.8,-2) -- (0,-1.25) -- (0,1.75) -- (1.8,1) -- (1.8,-2) -- (0,-1.25);
	\fill (0,-1) node {\tikz{\filldraw (0,0) circle (2pt) ;}}  circle (3pt) node[left]{$3,4$};
	\fill (0,1) node {\tikz{\filldraw (0,0) circle (2pt) ;}}  circle (3pt) node[left]{$7,8$};
	\fill (1,0.6)  node {\tikz{\filldraw (0,0) circle (1pt) ;}} circle (1.5pt) node[below]{$6$};
	\fill (1,-0.6)  node {\tikz{\filldraw (0,0) circle (1pt) ;}} circle (1.5pt) node[below]{$5$};
	\fill (-1,0.6) node {\tikz{\filldraw (0,0) circle (1pt) ;}}  circle (1.5pt) node[left]{$1$};
	\fill (-1,-0.6)  node {\tikz{\filldraw (0,0) circle (1pt) ;}} circle (1.5pt) node[left]{$2$};
\end{tikzpicture}
},
\ea 
which via the map gives the following $4$ seeds of facets:
\begin{equation}\label{g48facet2}
\begin{aligned}
{\bm S}_{12[345]67}&={\bm S}_{12[345]}+{\bm S}_{[345]67}-{\bm S}_{[345]},
\\
 {\bm S}_{[12(3)4(5)67]}&={\bm S}_{[12(3)45]}+{\bm S}_{[34(5)67]}-{\bm S}_{[345]},\\
{\bm S}_{[12(3)4(5)6(7)81]}&={\bm S}_{[12(3)45]}+{\bm S}_{56(7)81}+s_{1245}+s_{1568},\\ {\bm S}_{(78)12(34)56(78)}&={\bm S}_{(78)12(34)}+{\bm S}_{(34)56(78)}-{\bm S}_{(34)}-{\bm S}_{(78)}\,,
\end{aligned}
\end{equation}
respectively.  Performing cyclic permutations on these facets produce $4$ classes of  length $8,8, 2$ and $2$ respectively ($20$ in total). It is easy to see that these classes are invariant under reflection $i\leftrightarrow 9-i$. The last 
two classes of \eqref{fdhfasjkijfslfd} are also invariant under $i\to i+2$. It is intriguing that exactly these $4$ facets, ${\bm S}_{[12(3)4(5)6(7)81]}$,${\bm S}_{[23(4)5(6)7(8)12]}$, ${\bm S}_{(78)12(34)56(78)}$, ${\bm S}_{(81)23(45)67(81)}$, are directly related to algebraic functions containing square roots (as opposed to rational cluster variables) in the study of $G_+(4,8)/T$ in~\cite{Arkani-Hamed:2019rds,Henke:2019hve}.

Moreover, we have the following $4$ new types of normal configurations,
\ba \label{ufhia} 
\raisebox{0cm}{
\begin{tikzpicture}[baseline={([yshift=-.5ex]current bounding box.center)},scale=.65,rotate=180,yscale=-1]
	\draw[blue] (-0,1.75) -- (1.8,1) -- (1.8,-2) -- (0,-1.25) -- (0,1.75) -- (-1.8,1) -- (-1.8,-2) -- (0,-1.25);
	\fill (0,1)  node {\tikz{\filldraw (0,0) circle (1pt) ;}} circle (1.5pt) node[left]{$1$};
	\fill (0,-1) node {\tikz{\filldraw (0,0) circle (1pt) ;}}  circle (1.5pt) node[right]{$5$};
	\fill (0.6,-0.5) node {\tikz{\filldraw (0,0) circle (1pt) ;}}  circle (1.5pt) node[below]{$6$};
	\fill (1.2,0)  node {\tikz{\filldraw (0,0) circle (1pt) ;}} circle (1.5pt) node[below]{$7$};
	\fill (1.2,0.6) node {\tikz{\filldraw (0,0) circle (1pt) ;}}  circle (1.5pt) node[left]{$8$};
	\fill (-0.6,-0.5) node {\tikz{\filldraw (0,0) circle (1pt) ;}}  circle (1.5pt) node[right]{$4$};
	\fill (-1.2,0) node {\tikz{\filldraw (0,0) circle (1pt) ;}}  circle (1.5pt) node[above]{$3$};
	\fill (-0.6,0.5) node {\tikz{\filldraw (0,0) circle (1pt) ;}}  circle (1.5pt) node[above]{$2$};
	\draw[thick] (1.2,0) -- (0,-1) -- (-1.2,0) -- (0,1);
\end{tikzpicture}
},\ 
\raisebox{0cm}{
\begin{tikzpicture}[baseline={([yshift=-.5ex]current bounding box.center)},scale=.65,yscale=-1]
	\draw[blue] (-0,-1.75) -- (1.8,-1) -- (1.8,2) -- (0,1.25) -- (0,-1.75) -- (-1.8,-1) -- (-1.8,2) -- (0,1.25);
	\fill (0,1) node {\tikz{\filldraw (0,0) circle (2pt) ;}}  circle (3pt) node[left]{$5,6$};
	\fill (0,-1) node {\tikz{\filldraw (0,0) circle (1pt) ;}}  circle (1.5pt) node[left]{$1$};
	\fill (-1,-0.6) node {\tikz{\filldraw (0,0) circle (1pt) ;}}  circle (1.5pt) node[left]{$8$};
	\fill (-1,0.4) node {\tikz{\filldraw (0,0) circle (1pt) ;}}  circle (1.5pt) node[left]{$7$};
	\fill (0.6,-0.5) node {\tikz{\filldraw (0,0) circle (1pt) ;}}  circle (1.5pt) node[above]{$2$};
	\fill (1.2,0) node {\tikz{\filldraw (0,0) circle (1pt) ;}}  circle (1.5pt) node[below]{$3$};
	\fill (0.6,0.5) node {\tikz{\filldraw (0,0) circle (1pt) ;}}  circle (1.5pt) node[below]{$4$};
	\draw[thick] (0,-1) -- (1.2,0) -- (0,1);
\end{tikzpicture}
}
,\ 
\raisebox{0cm}{
\begin{tikzpicture}[baseline={([yshift=-.5ex]current bounding box.center)},scale=.65,rotate=180,yscale=-1]
	\draw[blue] (-0,1.75) -- (1.8,1) -- (1.8,-2) -- (0,-1.25) -- (0,1.75) -- (-1.8,1) -- (-1.8,-2) -- (0,-1.25);
	\fill (0,-1)  node {\tikz{\filldraw (0,0) circle (2pt) ;}}  
	circle (3pt) node[left]{$5,6$};
	\fill (0,1)  node {\tikz{\filldraw (0,0) circle (2pt) ;}}  circle (3pt) node[right]{$1,2$};
	\fill (1,0.6) node {\tikz{\filldraw (0,0) circle (1pt) ;}}  circle (1.5pt) node[left]{$8$};
	\fill (1,-0.4)  node {\tikz{\filldraw (0,0) circle (1pt) ;}} circle (1.5pt) node[left]{$7$};
	\fill (-1.2,0)  node {\tikz{\filldraw (0,0) circle (1pt) ;}} circle (1.5pt) node[below]{$3$};
	\fill (-0.6,-0.5) node {\tikz{\filldraw (0,0) circle (1pt) ;}}  circle (1.5pt) node[below]{$4$};
	\draw[thick] (-1.2,0) -- (0,-1);
\end{tikzpicture}
}
,\ 
\raisebox{0cm}{
\begin{tikzpicture}[baseline={([yshift=-.5ex]current bounding box.center)},scale=.65,rotate=180,yscale=-1]
	\draw[blue] (-0,1.75) -- (1.8,1) -- (1.8,-2) -- (0,-1.25) -- (0,1.75) -- (-1.8,1) -- (-1.8,-2) -- (0,-1.25);
	\fill (0,-1) node {\tikz{\filldraw (0,0) circle (1pt) ;}}  circle (1.5pt) node[left]{$6$};
	\fill (0,1)  node {\tikz{\filldraw (0,0) circle (2pt) ;}} circle (3pt) node[left]{$1,2$};
	\fill (1,0.4) node {\tikz{\filldraw (0,0) circle (1pt) ;}}  circle (1.5pt) node[left]{$8$};
	\fill (1,-0.6) node {\tikz{\filldraw (0,0) circle (1pt) ;}}  circle (1.5pt) node[left]{$7$};
	\fill (-0.6,-0.5)  node {\tikz{\filldraw (0,0) circle (1pt) ;}} circle (1.5pt) node[below]{$5$};
	\fill (-1.2,0)  node {\tikz{\filldraw (0,0) circle (1pt) ;}} circle (1.5pt) node[below]{$4$};
	\fill (-0.8,0.6) node {\tikz{\filldraw (0,0) circle (1pt) ;}} 
	circle (1.5pt) node[right]{$3$};
	\draw[thick] (0,-1) -- (-1.2,0);
\end{tikzpicture}
},
\ea  
which via the map give the following seeds for facets: 
\begin{equation}\label{g48facet3}
\begin{aligned}
{\bm S}_{[12(3)4(5)67]81}&={\bm S}_{[12(3)4(5)67]}+s_{1568}+s_{1578}+s_{1678},
\\
{\bm S}_{[12(3)4(56)]781}&={\bm S}_{[12(3)4(56)]}+{\bm S}_{(56)781}-{\bm S}_{(56)},\\
{\bm S}_{[34(56)]78(12)}&={\bm S}_{[34(56)]}+{\bm S}_{(56)78(12)}-{\bm S}_{(56)},\\
{\bm S}_{678(12)3[456]}&={\bm S}_{678(12)}+{\bm S}_{(12)3[456]}-{\bm S}_{(12)},
\end{aligned}
\end{equation}
respectively. Under cylclic permutations, they give $4$ classes of length $8$, and by reflection $4$ more classes, thus in total we have $64$ more facets. 

Apart from these $184+20+64$ facets, there are $92$ remaining facets which come from exceptional configurations that require ``blowup".

First, when  two points $a, b$ collide and at the same time they are collinear with two more points $i, j$ at different positions, we can require $a$ collide to $b$  in a {\it particular} direction paralleling to the plane $\overline{i \,j\,l}$ determined by $i, j$  and another point $l$ that is not collinear with them, which makes the minors $(a, b, i, l),(a, b, j, l)$ vanish at order $ {\cal O}(\epsilon^2)$
instead of $ {\cal O}(\epsilon)$.
Similarly to the case $(3,9)$, we introduce a ball to show the new structure,
\ba \label{fdff}
\raisebox{0cm}{
\begin{tikzpicture}[xscale=.7]
\draw[blue] (0,0)--++(5.5,0)--+(70:2)
(0,0)--++(70:2)--+(5.5,0);
\draw[thick] (1.9,1.2) --++ (2,0) node {\tikz{\filldraw (0,0) circle (1pt) ;}}  circle (1pt) node [above] {$i$} --++ (1,0) node {\tikz{\filldraw (0,0) circle (1pt) ;}}   circle (1pt) node [above] {$j$};
\draw
(1.9,.9)--(1.9,.9)
node {\tikz[scale=1.2]{\filldraw [dashed, fill=white](0,0) circle (.7) ;
\draw[blue] ($(-.5,-.43)+0*(180+70:.6)$)--++(.7,0)--+(70:.9)
($(-.5,-.43)+0*(180+70:.6)$)--++(70:.9)--+(.7,0);
\filldraw (.1,0.3)  node {\tikz{\filldraw (0,0) circle (1pt) ;}}  circle (0pt) node [below] {$b$}  ;
\filldraw (-.25,-0.2)   node {\tikz{\filldraw (0,0) circle (1pt) ;}}  circle (0pt) node [right] {$a$}  ;
}};
\filldraw ($(70:2)+(4,0)-(1,1.3)$)  node {\tikz{\filldraw (0,0) circle (1pt) ;}}   circle (1pt) node [right] {$l$}  ;
\end{tikzpicture} 
}
\ea  
where the radius of the ball is understood to be at order ${\cal O}(\epsilon)$.
We use a parallel plane to indicate that  $a$ is colliding to $b$ along a particular plane paralleling to $\overline{i\,j\,l}$.  
By computing the residue of ${\rm d}
\log R_{4,n}(\epsilon)$ for the above new configuration graph \eqref{fdff}, one can easily find that it is directly related to that of normal  configuration by adding several generalized Mandelstam variables,
\ba \label{fajsf}
{\bm S}\left[
\raisebox{-1cm}{
\begin{tikzpicture}[xscale=.7]
\draw[blue] (0,0)--++(5.5,0)--+(70:2)
(0,0)--++(70:2)--+(5.5,0);
\draw[thick] (1.9,1.2) --++ (2,0) node {\tikz{\filldraw (0,0) circle (1pt) ;}}  circle (1pt) node [above] {$i$} --++ (1,0) node {\tikz{\filldraw (0,0) circle (1pt) ;}}   circle (1pt) node [above] {$j$};
\draw
(1.9,.9)--(1.9,.9)
node {\tikz[scale=1.2]{\filldraw [dashed, fill=white](0,0) circle (.7) ;
\draw[blue] ($(-.5,-.43)+0*(180+70:.6)$)--++(.7,0)--+(70:.9)
($(-.5,-.43)+0*(180+70:.6)$)--++(70:.9)--+(.7,0);
\filldraw (.1,0.3)  node {\tikz{\filldraw (0,0) circle (1pt) ;}}  circle (0pt) node [below] {$b$}  ;
\filldraw (-.25,-0.2)   node {\tikz{\filldraw (0,0) circle (1pt) ;}}  circle (0pt) node [right] {$a$}  ;
}};
\filldraw ($(70:2)+(4,0)-(1,1.3)$)  node {\tikz{\filldraw (0,0) circle (1pt) ;}}   circle (1pt) node [right] {$l$}  ;
\end{tikzpicture} 
}
\right]
=
{\bm S}\left[
\raisebox{-1cm}{
\begin{tikzpicture}[xscale=.7]
\draw[blue] (0,0)--++(5.5,0)--+(70:2)
(0,0)--++(70:2)--+(5.5,0);
\draw[thick] (1.9,1.2) node{\tikz{\filldraw (0,0) circle (2pt);}} circle node[below] {$a,b$} --++ (2,0) node {\tikz{\filldraw (0,0) circle (1pt) ;}}  circle (1pt) node [above] {$i$} --++ (1,0) node {\tikz{\filldraw (0,0) circle (1pt) ;}}   circle (1pt) node [above] {$j$};
\filldraw ($(70:2)+(4,0)-(1,1.3)$)  node {\tikz{\filldraw (0,0) circle (1pt) ;}}   circle (1pt) node [right] {$l$}  ;
\end{tikzpicture} 
}
\right]+s_{a\,b\,i\,l}+s_{a\,b\,j\,l}\,.
\ea  

Similar to the case \eqref{fadjf},  the particular facet above can be reduced to an old one on the support of momentum conservation.  However we can indeed get 
3 new kinds of exceptional configurations by ``blowing up''  black dots in this way of the first normal configuration in \eqref{g48facet1flajkj}, and the second and third normal configurations in \eqref{ufhia}, respectively
\begin{equation}\label{iojpo} 
\begin{aligned}
\raisebox{0cm}{
\begin{tikzpicture}[baseline={([yshift=-.5ex]current bounding box.center)},scale=.8]
\draw[blue] (0,3) -- (-2.5,2) -- (-2.5,-1) -- (0,0) -- cycle;
\draw[blue] (0,3) -- (2,2) -- (2,-1) -- (0,0);
\draw[thick] (-1.8,1) -- (0,2.2) -- (1.5,1);
\filldraw[fill=black,draw=black] (-1.8,1)  circle (1.5pt) node[above]{$1$};
\filldraw[fill=black,draw=black] (-0.9,1.6) circle (1.5pt) node[above]{$2$};
\filldraw[fill=black,draw=black] (0,2.2) circle (1.5pt) node[above left=-2pt]{$3$};
\filldraw[fill=black,draw=black] (0.4,1.88) circle (1.5pt) node[above right=-2pt]{$4$};
\filldraw[fill=white,dashed] (1.2,1.16) circle (0.8);
\draw[blue] (1.2-1.2*0.3,1.16+1.2*0.6) -- (1.2-1.2*0.3,1.16-1.2*0.3) -- (1.2+1.2*0.3,1.16-1.2*0.6) -- (1.2+1.2*0.3,1.16+1.2*0.3) -- cycle;
\filldraw[fill=black,draw=black] (1.4,1.4) circle (1.5pt) node[left]{$5$};
\filldraw[fill=black,draw=black] (1,0.9) circle (1.5pt) node[right]{$6$};
\filldraw[fill=black,draw=black] (1,0.0) circle (1.5pt) node[right]{$7$};
\end{tikzpicture}
}
,\ 
\raisebox{0cm}{
\begin{tikzpicture}[baseline={([yshift=-.5ex]current bounding box.center)},scale=.8]
\draw[blue] (0,3) -- (-2.5,2) -- (-2.5,-1) -- (0,0) -- cycle;
\draw[blue] (0,3) -- (2,2) -- (2,-1) -- (0,0);
\draw[blue] (-2.5,-1) -- (-0.5,-2) -- (2,-1);

\filldraw[fill=black,draw=black] (1.2,-0.6) circle (1.5pt) node[below]{$4$};
\filldraw[fill=black,draw=black] (1.7,-0.85) circle (1.5pt) node[above]{$3$};
\draw[thick] (1.7,-0.85) -- (0,0);
\filldraw[fill=white,draw=black,dashed] (0,0) circle (0.9);
\draw[blue] (-0.9,0) -- (-0.1,-0.4) -- (0.9,0) -- (0.1,0.4) -- cycle;
\filldraw[fill=black,draw=black] (-0.2,0.1) circle (1.5pt) node[above]{$6$};
\filldraw[fill=black,draw=black] (0.2,-0.1) circle (1.5pt) node[below]{$5$};

\filldraw[fill=black,draw=black] (0,2.35) circle (1.5pt) node[right]{$1$};
\filldraw[fill=black,draw=black] (0.85,0.75) circle (1.5pt) node[right]{$2$};
\draw[thick] (0,2.35) -- (1.7,-0.85);

\filldraw[fill=black,draw=black] (-1.5,1) circle (1.5pt) node[right]{$8$};
\filldraw[fill=black,draw=black] (-1.875,-0.75) circle (1.5pt) node[below]{$7$};
\end{tikzpicture}
},\ 
\raisebox{0cm}{
\begin{tikzpicture}[baseline={([yshift=-.5ex]current bounding box.center)},scale=.8]
\draw[blue] (0,3) -- (-2.5,2) -- (-2.5,-1) -- (0,0) -- cycle;
\draw[blue] (0,3) -- (2,2) -- (2,-1) -- (0,0);
\draw[blue] (-2.5,-1) -- (-0.5,-2) -- (2,-1);

\filldraw[fill=black,draw=black] (1.2,-0.6) circle (1.5pt) node[above]{$4$};
\filldraw[fill=black,draw=black] (1.7,-0.85) circle (1.5pt) node[above]{$3$};
\draw[thick] (1.7,-0.85) -- (0,0);
\filldraw[fill=white,draw=black,dashed] (0,0) circle (0.9);
\draw[blue] (-0.9,0) -- (-0.1,-0.4) -- (0.9,0) -- (0.1,0.4) -- cycle;
\filldraw[fill=black,draw=black] (-0.2,0.1) circle (1.5pt) node[above]{$6$};
\filldraw[fill=black,draw=black] (0.2,-0.1) circle (1.5pt) node[below]{$5$};

\filldraw[fill=black,draw=black] (0,2.25) circle (2pt) node[left]{$1,2$};

\filldraw[fill=black,draw=black] (-1.5,1) circle (1.5pt) node[right]{$8$};
\filldraw[fill=black,draw=black] (-1.875,-0.75) circle (1.5pt) node[below]{$7$};
\end{tikzpicture}
}.
\end{aligned}
\end{equation}
Via the map these configurations give the following three facets
\be\label{fioafdj}   
{\bm S}_{[12(3)4(56)]}+s_{3567}+s_{4567}\,,\quad 
\ {\bm S}_{[12(3)4(56)]781}+s_{3567}+s_{4567}\,, \quad
{\bm S}_{[34(56)]78(12)}+s_{3567}+s_{4567}\,,
\ee
respectively. Under the cyclic permutations, they give 3 cyclic classes of length 4,8 and 8 respectively, and by reflection 3 more classes, thus in total we have 40 more facets.

Moreover, when three points $a,b,c$ are collinear, we can require $c$ to be collinear with $a$ and $b$ in a {\it particular} direction parallel to the plane $\overline{abi}$ determined by $a,b$ and another point $i$ that is not collinear with them. This makes 
the minor $(a,b,c,i)$ vanish at order ${\cal O}(\epsilon^2)$ instead of ${\cal O}(\epsilon)$.  
We still use a ball with the radius $\mathcal{O}(\epsilon)$ to show the new structure,
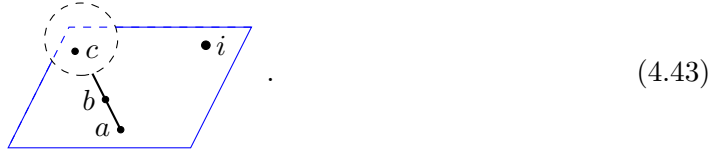
\ba\label{fjask}
\raisebox{0cm}{
\begin{tikzpicture}[baseline={([yshift=-.5ex]current bounding box.center)},scale=.8,rotate=-90]
\draw[blue] (0,3) -- (2,2) -- (2,-1) -- (0,0)--cycle;
\filldraw[fill=black,draw=black] (1.2,.6) circle (1.5pt) node[left]{$b$};
\filldraw[fill=black,draw=black] (1.7,.85) circle (1.5pt) node[left]{$a$};
\draw[thick] (1.7,.85) -- (0,0);
\filldraw[fill=white,draw=black,dashed] (0.2,0.2) circle (.6);
\draw[dashed,blue] (0,3) -- (0,0) -- (2,-1);
\filldraw[fill=black,draw=black] (0.4,0.1) circle (1.5pt) node[right]{$c$};
\filldraw[fill=black,draw=black] (0.3,2.25) circle (2pt) node[right]{$i$};
\end{tikzpicture}
}\,.
\ea 
where we ``blow up'' the point $c$ to a ball of radius $\epsilon$, and 
we see that the distance of $c$ to the plane $\overline{abi}$ is at the order $\mathcal{O}(\epsilon^2)$, thus as $\epsilon \to 0$, $c$ is still on the plane $\overline{abi}$. 
Similarly to \eqref{fajsf}, we have
\ba 
{\bm S}\left[
\raisebox{0cm}{
\begin{tikzpicture}[baseline={([yshift=-.5ex]current bounding box.center)},scale=.8,rotate=-90]
\draw[blue] (0,3) -- (2,2) -- (2,-1) -- (0,0)--cycle;
\filldraw[fill=black,draw=black] (1.2,.6) circle (1.5pt) node[left]{$b$};
\filldraw[fill=black,draw=black] (1.7,.85) circle (1.5pt) node[left]{$a$};
\draw[thick] (1.7,.85) -- (0,0);
\filldraw[fill=white,draw=black,dashed] (0.2,0.2) circle (.6);
\draw[dashed,blue] (0,3) -- (0,0) -- (2,-1);
\filldraw[fill=black,draw=black] (0.4,0.1) circle (1.5pt) node[right]{$c$};
\filldraw[fill=black,draw=black] (0.3,2.25) circle (2pt) node[right]{$i$};
\end{tikzpicture}
}
\right]
=
{\bm S}\left[
\raisebox{0cm}{
\begin{tikzpicture}[baseline={([yshift=-.5ex]current bounding box.center)},scale=.8,rotate=-90]
\draw[blue] (0,3) -- (2,2) -- (2,-1) -- (0,0)--cycle;
\filldraw[fill=black,draw=black] (1.2,.6) circle (1.5pt) node[left]{$b$};
\filldraw[fill=black,draw=black] (1.7,.85) circle (1.5pt) node[left]{$a$};
\draw[thick] (1.7,.85) -- (0,0);
\filldraw[fill=black,draw=black] (0.3,2.25) circle (2pt) node[right]{$i$};
\filldraw[fill=black,draw=black] (0,0) circle (2pt) node[above right]{$c$};
\end{tikzpicture}
}
\right]+s_{a\,b\,c\,i}\,.
\ea 
By ``blowing up'' black dots in this way of the third normal configuration in \eqref{fdhfasjkijfslfd} and the third one in \eqref{ufhia} respectively, we have two kinds of exceptional boundary configurations 
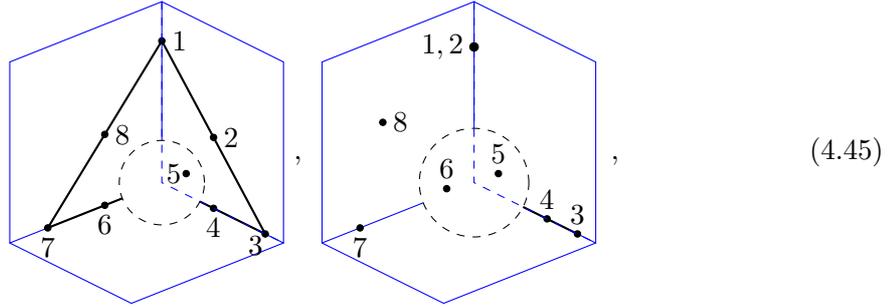
\begin{equation}\label{iojpo2} 
\raisebox{0cm}{
\begin{tikzpicture}[baseline={([yshift=-.5ex]current bounding box.center)},scale=.8]
\draw[blue] (0,3) -- (-2.5,2) -- (-2.5,-1) -- (0,0) -- cycle;
\draw[blue] (0,3) -- (2,2) -- (2,-1) -- (0,0);
\draw[blue] (-2.5,-1) -- (-0.5,-2) -- (2,-1);

\filldraw[fill=black,draw=black] (0.85,-0.425) circle (1.5pt) node[below]{$4$};
\filldraw[fill=black,draw=black] (1.7,-0.85) circle (1.5pt) node[below left =-3pt]{$3$};
\filldraw[fill=black,draw=black] (-0.9375,-0.375) circle (1.5pt) node[below]{$6$};
\filldraw[fill=black,draw=black] (-1.875,-0.75) circle (1.5pt) node[below]{$7$};
\filldraw[fill=black,draw=black] (-0.9375,-0.75*0.5+0.5*2.35) circle (1.5pt) node[right]{$8$};

\filldraw[fill=black,draw=black] (0,2.35) circle (1.5pt) node[right]{$1$};
\filldraw[fill=black,draw=black] (0.85,0.75) circle (1.5pt) node[right]{$2$};
\draw[thick] (0,2.35) -- (1.7,-0.85) -- (0,0) -- (-1.875,-0.75) -- (0,2.35);

\filldraw[fill=white,draw=black,dashed] (0,0) circle (0.7);
\draw[dashed,blue] (0,3) -- (0,0) -- (2,-1);
\filldraw[fill=black,draw=black] (0.4,0.15) circle (1.5pt) node[left=-2pt]{$5$};
\end{tikzpicture}
},\,
\raisebox{0cm}{
\begin{tikzpicture}[baseline={([yshift=-.5ex]current bounding box.center)},scale=.8]
\draw[blue] (0,3) -- (-2.5,2) -- (-2.5,-1) -- (0,0) -- cycle;
\draw[blue] (0,3) -- (2,2) -- (2,-1) -- (0,0);
\draw[blue] (-2.5,-1) -- (-0.5,-2) -- (2,-1);
\filldraw[fill=black,draw=black] (1.2,-0.6) circle (1.5pt) node[above]{$4$};
\filldraw[fill=black,draw=black] (1.7,-0.85) circle (1.5pt) node[above]{$3$};
\draw[thick] (1.7,-0.85) -- (0,0);
\filldraw[fill=white,draw=black,dashed] (0,0) circle (0.9);
\draw[dashed,blue] (0,3) -- (0,0) -- (2,-1);
\filldraw[fill=black,draw=black] (-0.45,-0.1) circle (1.5pt) node[above]{$6$};
\filldraw[fill=black,draw=black] (0.4,0.15) circle (1.5pt) node[above]{$5$};

\filldraw[fill=black,draw=black] (0,2.25) circle (2pt) node[left]{$1,2$};

\filldraw[fill=black,draw=black] (-1.5,1) circle (1.5pt) node[right]{$8$};
\filldraw[fill=black,draw=black] (-1.875,-0.75) circle (1.5pt) node[below]{$7$};
\end{tikzpicture}
}\,,
\end{equation}
where in the second configuration, 5 and 6 collide and at the same time they are collinear with 3 and 4, but 5 is collinear with them along a partuclar plane $\overline{134}$ determined by 1,3 and 4 (obviously 2 is also on the plane), while 6 is collinear with them along a generic direction.  
Compared to the last configuration in \eqref{iojpo}, we see the second configuration here involves a different way  to ``blow up''  the same black dot of the second configuration in \eqref{ufhia}. 
Via the map the two configurations in \eqref{iojpo2} give the following two seeds of facets: 
\be\label{fioafdj2}
{\bm S}_{[12(3)4(5)6(7)81]}+s_{1345}+s_{2345}\,,\quad {\bm S}_{[34(56)]78(12)}+s_{1345}+s_{2345}\,.
\ee
Under the cyclic permutations, they give 2 classes of length 8, and by reflection 2 more classes, thus in total we have 32 more facets.

Finally, we can have a combination of these two ways of ``blowup''. For example, for the last two configurations in \eqref{iojpo}, we can require additionally that point 5 is collinear to 3 and 4 along the particular plane $\overline{134}$ and that 6 still collides to 5 along the particular plane parallel to $\overline{347}$, which can be drawn as
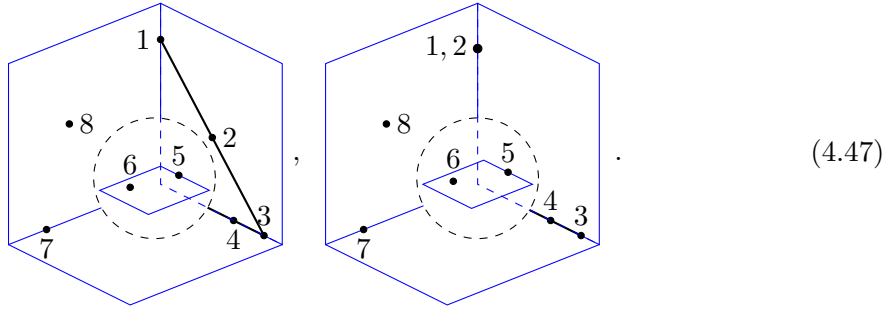
\begin{equation}\label{iojpfasfo} 
\begin{aligned}
\raisebox{0cm}{
\begin{tikzpicture}[baseline={([yshift=-.5ex]current bounding box.center)},scale=.8]
\draw[blue] (0,3) -- (-2.5,2) -- (-2.5,-1) -- (0,0) -- cycle;
\draw[blue] (0,3) -- (2,2) -- (2,-1) -- (0,0);
\draw[blue] (-2.5,-1) -- (-0.5,-2) -- (2,-1);

\filldraw[fill=black,draw=black] (1.2,-0.6) circle (1.5pt) node[below]{$4$};
\filldraw[fill=black,draw=black] (1.7,-0.85) circle (1.5pt) node[above]{$3$};
\draw[thick] (1.7,-0.85) -- (0,0);
\filldraw[fill=white,draw=black,dashed] (-.1,.1) circle (1);
\draw[blue] (-0.9-.1,0-.1) -- (-0.1-.1,-0.4-.1) -- (0.9-.1,0-.1) -- (0.1-.1,0.4-.1) -- cycle;
\draw[blue,dashed] (0,3) -- (0,0) -- (2,-1);
\filldraw[fill=black,draw=black] (-0.4-.1,0.05-.1) circle (1.5pt) node[above=1pt]{$6$};
\filldraw[fill=black,draw=black] (0.4-.1,.25-.1) circle (1.5pt) node[above]{$5$};

\filldraw[fill=black,draw=black] (0,2.4) circle (1.5pt) node[left]{$1$};
\filldraw[fill=black,draw=black] (0.85,1.2-0.85/2) circle (1.5pt) node[right]{$2$};
\draw[thick] (0,2.4) -- (1.7,-0.85);

\filldraw[fill=black,draw=black] (-1.5,1) circle (1.5pt) node[right]{$8$};
\filldraw[fill=black,draw=black] (-1.875,-0.75) circle (1.5pt) node[below]{$7$};
\end{tikzpicture}
},\ 
\raisebox{0cm}{
\begin{tikzpicture}[baseline={([yshift=-.5ex]current bounding box.center)},scale=.8]
\draw[blue] (0,3) -- (-2.5,2) -- (-2.5,-1) -- (0,0) -- cycle;
\draw[blue] (0,3) -- (2,2) -- (2,-1) -- (0,0);
\draw[blue] (-2.5,-1) -- (-0.5,-2) -- (2,-1);

\filldraw[fill=black,draw=black] (1.2,-0.6) circle (1.5pt) node[above]{$4$};
\filldraw[fill=black,draw=black] (1.7,-0.85) circle (1.5pt) node[above]{$3$};
\draw[thick] (1.7,-0.85) -- (0,0);
\filldraw[fill=white,draw=black,dashed] (0,.1) circle (1);
\draw[blue] (-0.9,0) -- (-0.1,-0.4) -- (0.9,0) -- (0.1,0.4) -- cycle;
\draw[blue,dashed] (0,3) -- (0,0) -- (2,-1);
\filldraw[fill=black,draw=black] (-0.4,0.05) circle (1.5pt) node[above=1pt]{$6$};
\filldraw[fill=black,draw=black] (0.5,0.2) circle (1.5pt) node[above]{$5$};

\filldraw[fill=black,draw=black] (0,2.25) circle (2pt) node[left]{$1,2$};

\filldraw[fill=black,draw=black] (-1.5,1) circle (1.5pt) node[right]{$8$};
\filldraw[fill=black,draw=black] (-1.875,-0.75) circle (1.5pt) node[below]{$7$};
\end{tikzpicture}
}\,.
\end{aligned}
\end{equation}
Via the map they give the following 2 seeds of facets,    
\ba  \label{fahdf;}
&{\bm S}_{[12(3)4(56)]781}+(s_{3567}+s_{4567})+(s_{1345}+s_{2345})\,, \nl
&{\bm S}_{[34(56)]78(12)}+(s_{3567}+s_{4567})+(s_{1345}+s_{2345})\,. 
\ea
Under the cyclic permutations, they give 2 classes of length 2 and 8 respecively, and by reflection 2 more classes, thus in total we have 20 such facets. 

Given all the facets $F_a$ of $\calP(4,8)$, \eqref{g48facet1},\eqref{g48facet12},\eqref{g48facet2}, \eqref{g48facet3},\eqref{fioafdj},\eqref{fioafdj2} and \eqref{fahdf;},  the polytope can be cut out by these inequalities $F_a \geq 0$ for $a=1,2, \cdots, 360$. The $f-$vector of $\calP(4,8)$ reads 
( 1, 90608, 444930, 922314, 1047200, 706042, 285948, 66740, 7984, 360, 1)
.
Out of 90608 vertices, 50356 are simple ones (adjacent to 9 facets), and the remaining non-simple vertices ranges from 10-vertex to 49-vertex. The numbers of these types of vertices are given in table~\ref{tabg48vert}. The volume of the dual polytope $\calP^\circ(4,8)$ gives the amplitudes ${\cal A}_{4,8}$, which are recorded in the auxiliary file \textsf{G48num.nb} for several sets of given numeric values of  facets.  The analytic expressions of all the facets are listed in another auxiliary file  \textsf{G48facets.txt}.

\begin{table}[!htbp]
	\centering  

	\begin{tabular}{|c|c|c|c|c|c|c|c|c|c|c|c|}  
		\hline  
		 9-vertex & 10-vertex &11-vertex &12-vertex & 13-vertex &14-vertex &15-vertex & 16-vertex \\
		\hline
		50356& 12320& 9116& 6064& 4448& 2332& 2176& 872\\
		\hline
		17-vertex &18-vertex &19-vertex &20-vertex &21-vertex & 22-vertex & 23-vertex & 24-vertex\\ 
		\hline
		976& 676& 384& 336& 200 & 48 & 8 & 80 \\
		\hline
		25-vertex & 26-vertex & 27-vertex & 29-vertex & 33-vertex & 34-vertex & 36-vertex & 49-vertex\\
		\hline
		72 & 24 & 48 & 16 & 20 & 16 & 16 & 4 \\
		\hline
	\end{tabular}
		\caption{The numbers of various types of vertices of ${\cal P}(4,8)$, classified by the number of facets adjacent to each vertex.}  
	\label{tabg48vert}  
\end{table}

\subsection{$\calP(4,9)$}

The same construction is also valid to build the polytope $\calP(4,9)$. From SE map, we found that there are in all $2155\times 9=19395$ facets, with 2155 cyclic classes of length 9. The seeds are listed in the auxiliary file \textsf{G49seeds.txt}. Given all these facets $F_a$ of $\calP(4,9)$,  the polytope can be cut out by these inequalities $F_a \geq 0$ for $a=1,2, \cdots, 19395$.

There are 138 cyclic classes of facets that come from normal configurations.  Here we present three examples, 
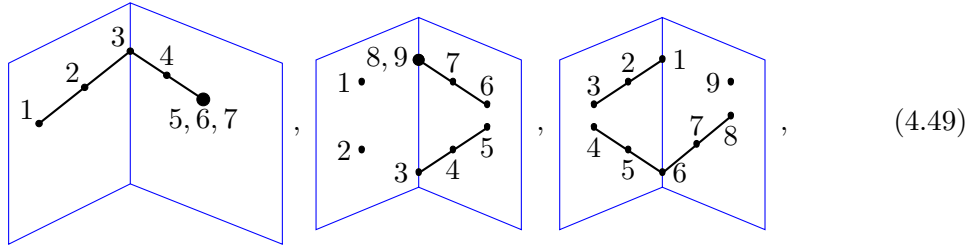
\be \label{fajifod;}
\raisebox{0cm}{
\begin{tikzpicture}[baseline={([yshift=-.5ex]current bounding box.center)},scale=0.8]
\draw[blue] (0,3) -- (2.5,2) -- (2.5,-1) -- (0,0) -- cycle;
\draw[blue] (0,3) -- (-2,2) -- (-2,-1) -- (0,0);
\draw[thick] (1.2,1.4) -- (0,2.2) -- (-1.5,1);
\filldraw[fill=black,draw=black] (0.6,1.8) circle (1.5pt) node[above]{$4$};
\filldraw[fill=black,draw=black] (0,2.2) circle (1.5pt) node[above left=-2pt]{$3$};
\filldraw[fill=black,draw=black] (1.2,1.4) circle (3pt) node[below]{$5,6,7$};
\filldraw[fill=black,draw=black] (-0.75,1.6) circle (1.5pt) node[above left=-2pt]{$2$};
\filldraw[fill=black,draw=black] (-1.5,1) circle (1.5pt) node[above left=-2pt]{$1$};
\end{tikzpicture}
},
\raisebox{0cm}{
\begin{tikzpicture}[baseline={([yshift=-.5ex]current bounding box.center)},scale=.75]
	\draw[blue] (0,1.75) -- (-1.8,1) -- (-1.8,-2) -- (0,-1.25) -- (0,1.75) -- (1.8,1) -- (1.8,-2) -- (0,-1.25);
	\fill (0,1) node {\tikz{\filldraw (0,0) circle (2pt) ;}}  circle (3pt) node[left]{$8,9$};
	\fill (0,-1) node {\tikz{\filldraw (0,0) circle (1pt) ;}}  circle (1.5pt) node[left]{$3$};
	\fill (-1,-0.6) node {\tikz{\filldraw (0,0) circle (1pt) ;}}  circle (1.5pt) node[left]{$2$};
	\fill (-1,0.6) node {\tikz{\filldraw (0,0) circle (1pt) ;}}  circle (1.5pt) node[left]{$1$};
	\fill (0.6,-0.6) node {\tikz{\filldraw (0,0) circle (1pt) ;}}  circle (1.5pt) node[below]{$4$};
	\fill (1.2,-0.2) node {\tikz{\filldraw (0,0) circle (1pt) ;}}  circle (1.5pt) node[below]{$5$};
	\fill (0.6,0.6) node {\tikz{\filldraw (0,0) circle (1pt) ;}}  circle (1.5pt) node[above]{$7$};
	\fill (1.2,0.2) node {\tikz{\filldraw (0,0) circle (1pt) ;}}  circle (1.5pt) node[above]{$6$};
	\draw[thick] (0,-1) -- (1.2,-.2) (1.2,.2) -- (0,1);
\end{tikzpicture}
}\,,
\raisebox{0cm}{
\begin{tikzpicture}[baseline={([yshift=-.5ex]current bounding box.center)},scale=.75]
	\draw[blue] (0,1.75) -- (-1.8,1) -- (-1.8,-2) -- (0,-1.25) -- (0,1.75) -- (1.8,1) -- (1.8,-2) -- (0,-1.25);
	\fill (0,1)  node {\tikz{\filldraw (0,0) circle (1pt) ;}} circle (1.5pt) node[right]{$1$};
	\fill (0,-1) node {\tikz{\filldraw (0,0) circle (1pt) ;}}  circle (1.5pt) node[right]{$6$};
	\fill (0.6,-0.5) node {\tikz{\filldraw (0,0) circle (1pt) ;}}  circle (1.5pt) node[above]{$7$};
	\fill (1.2,0)  node {\tikz{\filldraw (0,0) circle (1pt) ;}} circle (1.5pt) node[below]{$8$};
	\fill (1.2,0.6) node {\tikz{\filldraw (0,0) circle (1pt) ;}}  circle (1.5pt) node[left]{$9$};
	\fill (-0.6,-0.6) node {\tikz{\filldraw (0,0) circle (1pt) ;}}  circle (1.5pt) node[below]{$5$};
	\fill (-1.2,0.2) node {\tikz{\filldraw (0,0) circle (1pt) ;}}  circle (1.5pt) node[above]{$3$};
	\fill (-1.2,-0.2) node {\tikz{\filldraw (0,0) circle (1pt) ;}}  circle (1.5pt) node[below]{$4$};
	\fill (-0.6,0.6) node {\tikz{\filldraw (0,0) circle (1pt) ;}}  circle (1.5pt) node[above]{$2$};
	\draw[thick] (1.2,0) -- (0,-1) -- (-1.2,-0.2);
	\draw[thick] (-1.2,0.2)-- (0,1);
\end{tikzpicture}
}\,,
\ee
where the first one is an analog of the first configuration given in \eqref{g48facet1flajkj} while the remaining two do not have
analogs in $(4,8)$. Via the map the first configuration in above equation gives the facet  ${\bm S}_{[12(3)4(567)]}$ for $(4,9)$ and the remaining two give respectively
\ba
\bms_{[345][67(89)]123}&=\bms_{[345][67(89)]}+\bms_{(89)123}-\bms_{(89)}\,,
\nl
    {\bm S}_{[123][45(6)78]91}&={\bm S}_{[123]}+{\bm S}_{[45(6)78]91}+\sum\limits_{\substack{i<j\in\{1,2,3\}\\k<l\in\{4,5,6\}}}s_{i,j,k,l}.
\ea

The remaining $2017$ cyclic classes of facets come from exceptional configurations which can be obtained by ``blowing up'' some 
black dots of  normal configurations.  For example, for the first boundary configuration of \eqref{fajifod;}, we can require that point 6 and 7 collide to point 5 along a particular plane, such as the one parallel to the plane $\overline{348}$ determined by 3,4,8, which can be drawn as
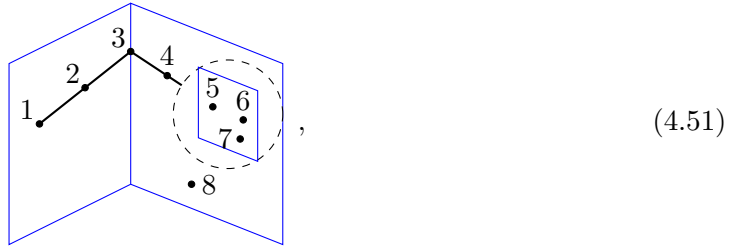
\begin{equation}\label{fiasjd;s}
\raisebox{0cm}{
\begin{tikzpicture}[baseline={([yshift=-.5ex]current bounding box.center)},scale=0.8]
\draw[blue] (0,3) -- (2.5,2) -- (2.5,-1) -- (0,0) -- cycle;
\draw[blue] (0,3) -- (-2,2) -- (-2,-1) -- (0,0);
\draw[thick] (1.8,1) -- (0,2.2) -- (-1.5,1);
\filldraw[fill=black,draw=black] (0.6,1.8) circle (1.5pt) node[above]{$4$};
\filldraw[fill=black,draw=black] (0,2.2) circle (1.5pt) node[above left=-2pt]{$3$};
\filldraw[fill=black,draw=black] (-0.75,1.6) circle (1.5pt) node[above left=-2pt]{$2$};
\filldraw[fill=black,draw=black] (-1.5,1) circle (1.5pt) node[above left=-2pt]{$1$};
\filldraw[fill=white,dashed] (0.4+1.2,1.16) circle (0.9);
\draw[blue] (0.4+1.2+1.3*0.375,1.16-1.3*0.6) -- (0.4+1.2+1.3*0.375,1.16+1.3*0.3) -- (0.4+1.2-1.3*0.375,1.16+1.3*0.6) -- (0.4+1.2-1.3*0.375,1.16-1.3*0.3) -- cycle;
\filldraw[fill=black,draw=black] (.3+1.2-0.4*0.375,1.5-1.3*0.5/3) circle (1.5pt) node[above]{$5$};
\filldraw[fill=black,draw=black] (0.5+1.2+0.4*0.375,1.5-1.3*0.5*2/3) circle (1.5pt) node[above]{$6$};
\filldraw[fill=black,draw=black] (1.8,0.75) circle (1.5pt) node[left=-1pt]{$7$};
\filldraw[fill=black,draw=black] (1,0.0) circle (1.5pt) node[right]{$8$};
\end{tikzpicture}
}\,,
\end{equation}
where the minors $(a,i,j,8)$ with $a=3,4$ and $5\leq i<j\leq 7$ vanish at ${\cal O}(\epsilon^{2})$ instead of ${\cal O}(\epsilon)$. 
The minors $(3,5,6,7),(4,5,6,7),(8,5,6,7)$  vanish at ${\cal O}(\epsilon^{3})$ instead of ${\cal O}(\epsilon^{2})$.  
Just as the first facet in \eqref{fioafdj}, via the map the above configuration gives a facet for $(4,9)$,
\be 
{\bm S}_{[12(3)4(567)]}+ \Bigg(s_{3567}+s_{4567}+s_{5678}+\sum\limits_{\substack{3\leq a\leq 4\\5\leq i<j\leq 7}}s_{a\,i\,j\,8}\Bigg). 
\ee 

Another way to ``blow up'' the same black dot is to require additionally that 6 collides to 5 in a particular direction parallel to the line $\overline{34}$, while 7 collides to 5 along a generic direction, which can be drawn as
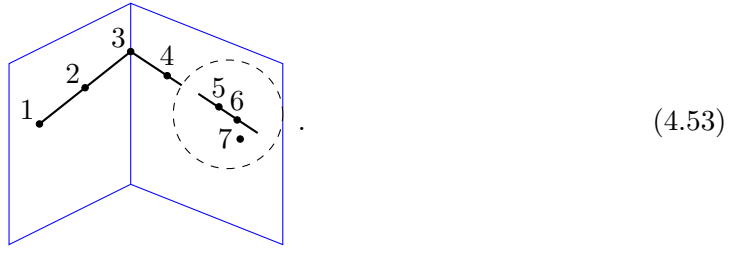
\begin{equation}
\raisebox{0cm}{
\begin{tikzpicture}[baseline={([yshift=-.5ex]current bounding box.center)},scale=0.8]
\draw[blue] (0,3) -- (2.5,2) -- (2.5,-1) -- (0,0) -- cycle;
\draw[blue] (0,3) -- (-2,2) -- (-2,-1) -- (0,0);
\draw[thick] (1.8,1) -- (0,2.2) -- (-1.5,1);
\filldraw[fill=black,draw=black] (0.6,1.8) circle (1.5pt) node[above]{$4$};
\filldraw[fill=black,draw=black] (0,2.2) circle (1.5pt) node[above left=-2pt]{$3$};
\filldraw[fill=black,draw=black] (-0.75,1.6) circle (1.5pt) node[above left=-2pt]{$2$};
\filldraw[fill=black,draw=black] (-1.5,1) circle (1.5pt) node[above left=-2pt]{$1$};
\filldraw[fill=white,dashed] (0.4+1.2,1.16) circle (0.9);
\draw[thick] (0.4+1.2-1.3*0.375,1.5) -- (0.4+1.2+1.3*0.375,1.5-1.3*0.5);
\filldraw[fill=black,draw=black] (0.4+1.2-0.4*0.375,1.5-1.3*0.5/3) circle (1.5pt) node[above]{$5$};
\filldraw[fill=black,draw=black] (0.4+1.2+0.4*0.375,1.5-1.3*0.5*2/3) circle (1.5pt) node[above]{$6$};
\filldraw[fill=black,draw=black] (1.8,0.75) circle (1.5pt) node[left=-1pt]{$7$};
\end{tikzpicture}
}\,.
\end{equation}
Here the minor $(3,4,5,6)$ vanishes at  ${\cal O}(\epsilon^{3})$ instead of ${\cal O}(\epsilon^{2})$, minors $(1,2,5,6), (1,3,5,6),$ $ (1,4,5,6), (2,3,5,6),(2,4,5,6)$ vanish at  ${\cal O}(\epsilon^{2})$ instead of ${\cal O}(\epsilon)$, minors $(i,3,5,6),(i,4,5,6)$ with $i= 8,9$ vanish at ${\cal O}(\epsilon^{2})$ instead of ${\cal O}(\epsilon)$, and minors (7,3,5,6),(7,4,5,6) vanish at  ${\cal O}(\epsilon^{3})$ instead of ${\cal O}(\epsilon^{2})$.  
Via the map, the above configuration gives the facet
\be 
{\bm S}_{[12(3)4(567)]}+ \sum\limits_{\substack{1\leq i<j\leq 4}}s_{i\,j\,5\,6}+ \sum\limits_{\substack{7\leq i \leq 9}}(s_{i\,3\,5\,6}+s_{i\,4\,5\,6})\,. 
\ee 

A combination of the above two examples  gives a third way to ``blow up'' the same black dot as shown by
 \begin{equation}
\raisebox{0cm}
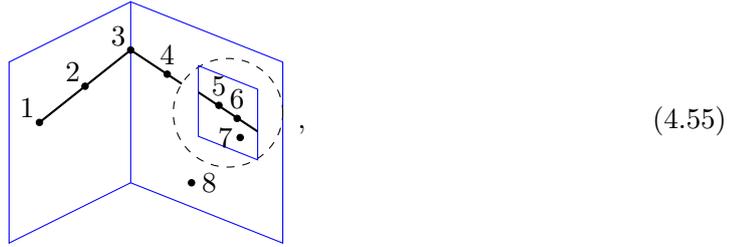
{
\begin{tikzpicture}[baseline={([yshift=-.5ex]current bounding box.center)},scale=0.8]
\draw[blue] (0,3) -- (2.5,2) -- (2.5,-1) -- (0,0) -- cycle;
\draw[blue] (0,3) -- (-2,2) -- (-2,-1) -- (0,0);
\draw[thick] (1.8,1) -- (0,2.2) -- (-1.5,1);
\filldraw[fill=black,draw=black] (0.6,1.8) circle (1.5pt) node[above]{$4$};
\filldraw[fill=black,draw=black] (0,2.2) circle (1.5pt) node[above left=-2pt]{$3$};
\filldraw[fill=black,draw=black] (-0.75,1.6) circle (1.5pt) node[above left=-2pt]{$2$};
\filldraw[fill=black,draw=black] (-1.5,1) circle (1.5pt) node[above left=-2pt]{$1$};
\filldraw[fill=white,dashed] (0.4+1.2,1.16) circle (0.9);
\draw[blue] (0.4+1.2+1.3*0.375,1.16-1.3*0.6) -- (0.4+1.2+1.3*0.375,1.16+1.3*0.3) -- (0.4+1.2-1.3*0.375,1.16+1.3*0.6) -- (0.4+1.2-1.3*0.375,1.16-1.3*0.3) -- cycle;
\draw[thick] (0.4+1.2-1.3*0.375,1.5) -- (0.4+1.2+1.3*0.375,1.5-1.3*0.5);
\filldraw[fill=black,draw=black] (0.4+1.2-0.4*0.375,1.5-1.3*0.5/3) circle (1.5pt) node[above]{$5$};
\filldraw[fill=black,draw=black] (0.4+1.2+0.4*0.375,1.5-1.3*0.5*2/3) circle (1.5pt) node[above]{$6$};
\filldraw[fill=black,draw=black] (1.8,0.75) circle (1.5pt) node[left=-1pt]{$7$};
\filldraw[fill=black,draw=black] (1,0.0) circle (1.5pt) node[right]{$8$};
\end{tikzpicture}
}\,,
\end{equation}
where 6 collides to 5 in the direction parallel to $\overline{34}$ and 7 collides to 5 along a plane parallel to $\overline{348}$.
Via the map, the above configuration gives the facet
\begin{equation}
{\bm S}_{[12(3)4(567)]}+\Bigg(s_{3567}+s_{4567}+s_{5678}+\sum\limits_{\substack{3\leq a\leq 4\\5\leq i<j\leq 7}}s_{a\,i\,j\,8}\Bigg)+\Bigg(s_{3569}+s_{4569}+\sum\limits_{1\leq i<j\leq 4}s_{i\,j\,5\,6}\Bigg).
\end{equation}

There are many more exceptional boundary configurations than normal ones in $(4,9)$, since there are many new ways to ``blow up'' black dots here. Starting from any normal configuration, one could get the first series of exceptional configurations by ``blowing up'' a black dot, and then ``blow up'' another black dot to get more exceptional ones. Repeat this process and in the end one finds all $19395$ configurations for $(4,9)$.

\section{Outlook}

In this note we study the polytopes ${\cal P}(k, n)$ which control the convergence and leading order as $\alpha'\to 0$ of natural generalizations of open-string integrals, namely the Grassmannian stringy integrals \eqref{stringyint}. The computation via Minkowski sum or SE map for such polytopes is straightforward (albeit tedious for higher $k$ and $n$), which are cut out by inequalities $F_a \geq 0$ for facets $F_a$. However, as we have emphasized, the main novelty is that we recast the linear functions $F_a$ in a manifestly gauge-invariant and cyclic manner, which are the analog of the usual planar variables for $k=2$ case. Given our results up to $n=9$, one can easily compute the canonical functions by {\it e.g.} triangulating the dual polytopes, which produce the generalized bi-adjoint amplitudes ${\cal A}_{k,n}$. This is also a straightforward computation, as no extra information is needed beyond our construction for the polytope. 

More importantly, as our notation for the facets already suggests, it is very natural to find all the boundary configurations of $G_+(k,n)/T$ which are in bijection with the facets of ${\cal P}(k,n)$ via the SE map. Such a boundary configuration is controlled by a single parameter $\epsilon \to 0$, and can be specified by given at which order in $\epsilon$ certain minors vanish. It is intriguing that these configurations have nice geometric interpretations in terms of $n$ points in $(k{-}1)$-dimensional space. As we have seen, for $k=3$, each configuration concerns which of the $n$ points are collinear or colliding, and for $k=4$, which of the $n$ points are coplanar, collinear or colliding; new configurations which require ``blowup" are also needed starting at $n=9$ for $k=3$ and $n=8$ for $k=4$. 


Our preliminary investigations have left many open questions unanswered. The most pressing question is of course how to find all facets and boundary configurations for higher $k$ and $n$. In particular, it is conceivable that all $n$ results can be obtained for $k=3$ at least, and it would be highly desirable to compare these results with the configurations of $G_+(k,n)/T$ from {\it e.g.} hypersimplex decomposition~\cite{Arkani-Hamed:2020cig}. On the other hand, there are other methods for studying these facets, configurations and the related  fan ${\cal N}({\cal P}(k,n))$, which can be derived from the tropical positive Grassmannian.  The generalized amplitudes ${\cal A}_{k,n}$ can be directly obtained from tropical positive Grassmannian, as shown in  \cite{Cachazo:2019ngv,Drummond:2019qjk,Drummond:2019cxm,Henke:2019hve}, where cluster variables are used to find a natural triangulation of the  fan ${\cal N}({\cal P}(k,n))$. Each ray of the fan, which corresponds to a vertex of the dual polytope ${\cal P}^\circ(k,n)$, is associated with a pole of ${\cal A}_{k,n}$. The rays of the fan are closely related to cluster variables \cite{fomin2002cluster,fomin2002cluster2} and it would be highly desirable to connect the cluster variables to boundary configurations of $G_+(k,n)/T$, which may help us to find the general pattern. 

Very recently, inspired by the connections between tropical Grassmannian for $k=3$ and metric arrangements of trees explained in \cite{herrmann2009draw}, Borges and Cachazo proposed~\cite{Borges:2019csl} that  planar collections of trivalent Feynman diagrams can be used to compute the generalized bi-adjoint amplitudes $\calA_{3,n}$. This idea has been generalized to planar arrays of trivalent Feynman diagrams for any $k$~\cite{Cachazo:2019xjx};  such a planar array of Feynman diagrams is believed to correspond to a cone of the fan ${\cal N}({\cal P}(k,n))$, thus a  facet of the dual polytope ${\cal P}^\circ(k,n)$, or a vertex of our ${\cal P}(k,n)$. As a strong check for both our computation and this idea, we have seen that the number of vertices of ${\cal P}(k,n)$ is equal to that of planar arrays, for $(3,6),(3,7),(3,8),(3,9)$ and $(4,8)$. Moreover, the amplitudes up to eight points also agree, and it would be interesting to compare for $k=4$, $n=9$ and even beyond. 

Based on the idea of planar arrays of Feynman diagrams, we can look at the polytope ${\cal P}(k,n)$ in a different way. Recall that for $k=2$, each vertex of ABHY associahedron,  ${\cal P}(2,n)$, corresponds to a trivalent Feynman diagrams, while higher-dimension boundaries correspond to diagrams where some cubic vertices are merged to higher-valency vertices. For higher $k$, each vertex of ${\cal P}(k,n)$ corresponds to a set of planar trivalent Feynman diagrams and one can expect that the a higher-dimension boundary of ${\cal P}(k,n)$ correspond to degenerate planar arrays where cubic vertices of Feynman diagrams get merged. In other words, the polytope  ${\cal P}(k,n)$ provides a geometry description of the relations between different (degenerated) planar arrays, and  the combinatorial study of (degenerated) Feynman diagrams in (degenerated)  planar arrays and their relations~\cite{faklfd} can also help us have a better understanding of the polytope ${\cal P}(k,n)$. Very recently more progress has been made~ \cite{Early:2019zyi,Early:2019eun} in the study of tropical positive Grassmannian using the method of matroid subdivision~\cite{herrmann2009draw}. It would be interesting to see how all these methods can help us to push further the understanding of ${\cal P}(k,n)$ and tropical positive Grassmannian. 

There are also various interesting avenues for future investigations on ${\cal P}_{k,n}$, which generalize aspects of the kinematic associahedron and string amplitudes for $k=2$. For example, it is natural to study generalized amplitudes $m_n^{(k)}(\alpha|\beta)$ with $\alpha$ and $\beta$ the orderings of the form and integration domain, respectively, which is the analog of $m_n(\alpha|\beta)$~\cite{Arkani-Hamed:2017tmz}. For $k=2$, the Parke-Taylor factor is the simplest leading singularity and all other leading singularities can be linearly expanded onto Parke-Taylor factors~\cite{Arkani-Hamed:2014bca}. For $k\geq3$, there are new kind of leading singularities that can not be expanded into Parke-Taylor factors., and it'd be interesting to find their corresponding polytopes. Relatedly, it may be worth to find the analog of polytope in~\cite{Gao:2017dek} and those in~\cite{Herderschee:2019wtl}.   Besides, one can also study $\alpha'$ expansion of Grassmannian string integrals or even the structure at finite $\alpha'$, such as the recurrence relation~\cite{Arkani-Hamed:2019mrd} and integration-by-parts reduction as in~\cite{He:2018pol,He:2019drm} for $k=2$. 

Finally, as already mentioned in~\cite{Arkani-Hamed:2019rds}, the ultimate question is how can one ask questions that are related to $G_+(4,n)/T$ and ${\cal P}(4,n)$, to which the answer are the all-loop scattering amplitudes in planar ${\cal N}=4$ SYM?

\begin{acknowledgments}
  We are grateful to N. Arkani-Hamed, A.~Guevara, F. Cachazo, N. Early, T. Lam, Z. Li, M. Spradlin and C. Zhang  for inspiring discussions. S.H. was supported in part by the Thousand Young Talents program, NSF of China under Grant No.  11947302 and 11935013.
\end{acknowledgments}

\appendix
\section{Degeneration from ${\cal P}(\calE_8)$ to $\calP(3,8)$}\label{appd1}
In this appendix we will continue our discussion of Section~\ref{sect38} on the relation between the generalized associahedron $\calP(\calE_8)$ and polytope $\calP(3,8)$. An important fact is that there are 8 disappearing facets on $\calP(3,8)$, which forms a cyclic class of length 8, with the seed
$S_{(12)3(45)6(78)}$  
which satisfy the identity,
\begin{equation}\label{fkals;f}
S_{(12)3(45)6(78)}=S_{(12)345}+S_{456(78)},\ \& \text{ 7 cyclic}.
\end{equation}
From the identity we can see that if we take $S_{(12)3(45)6(78)}=0$, the prerequisites that $S_{(12)345}\geq 0$ and $S_{456(78)}\geq 0$ would force $S_{(12)345}=S_{456(78)}=0$, making the facet $S_{(12)3(45)6(78)}$ would just ``go through" the intersection of the facets $S_{(12)345}$ and $S_{456(78)}$:

\begin{equation}
\raisebox{-2cm}{
\begin{tikzpicture}
\fill[fill=black!20] (-2,-1) -- (-1,0) -- (1,0) -- (2,-1) -- (-2,-1);
\fill[fill=black!20] (5,-1) -- (6.5,0.5) -- (8,-1);
\draw (-2,-1) node[below]{$S_{(12)345}$} -- (-0.5, 0.5);
\draw (2,-1) node[below]{$S_{456(78)}$} -- (0.5,0.5);
\draw (-2,0) -- (0,0) node[below]{$S'_{(12)3(45)6(78)}$} -- (2,0);
\draw[->] (3,0) -- node[above]{merge} (4.5,0);
\draw (5,-1)node[below]{$S_{(12)345}$} -- (6.5,0.5);
\draw (8,-1) node[below]{$S_{456(78)}$} -- (6.5,0.5);
\draw (5,0.5) -- (6.5,0.5) node[above]{$S_{(12)3(45)6(78)}$} -- (8,0.5);
\node at (0,-2) {$\calP(\calE_8)$};
\node at (6.5,-2) {$\calP(3,8)$};
\end{tikzpicture}}
\quad ,
\end{equation}
where $S'_{(12)3(45)6(78)}$ is the facet of $\calP(\calE_8)$ corresponding to $S_{(12)3(45)6(78)}$ and the dark shadow shows the region inside the polytope. We can then see the facet $S_{(12)3(45)6(78)}$ does not contribute anything to the polytope $\calP(3,8)$.

Now that those eight facets of $\calP(\calE_8)$ are true poles in the amplitudes $\calA(\calE_8)$. However, as they are no longer facets of $\calP(3,8)$, they should turn to be "\textit{spurious poles}" of the amplitudes $\calA_{3,8}$. Let us see how the change happens. We take $S_{(12)3(45)6(78)}$ as the example. By carefully checking all of the 25080 simplicial facets of the dual polytope $\dcalP(\calE_8)$, we have found that: (1) if a facet is surrounded by the vertex corresponding to $S'_{(12)3(45)6(78)}$, it \textit{must} be surrounded by another vertex corresponding to $S_{(12)345}$ \textit{or} $S_{456(78)}$; (2) those facets would come in pairs, and the only difference between the facets in each pair is just the choice between $S_{(12)345}$ or $S_{456(78)}$. Then we have the contribution of each pair to the whole amplitudes $\calA(\calE_8)$:
\begin{equation}\label{e8fskdfe}
\frac{1}{S_{(12)345}S'_{(12)3(45)6(78)}}*\frac{1}{\text{other }S}+\frac{1}{S_{456(78)}S'_{(12)3(45)6(78)}}*\frac{1}{\text{other }S}.
\end{equation}
For the degeration from $\calP(\calE_8)$ to $\calP(3,8)$, we have $S'_{(12)3(45)6(78)}\to S_{(12)3(45)6(78)}\ \& \text{ 7 cyclic}$, we have the correspondance of \eqref{e8fskdfe} in the amplitudes $\calA_{3,8}$:
\begin{equation}\label{eqcancelbound}
\begin{aligned}
 &\frac{1}{S_{(12)345}S_{(12)3(45)6(78)}}*\frac{1}{\text{other }S}+\frac{1}{S_{456(78)}S_{(12)3(45)6(78)}}*\frac{1}{\text{other }S}\\
=&\frac{S_{(12)345}+S_{456(78)}}{S_{(12)345}S_{456(78)}S_{(12)3(45)6(78)}}*\frac{1}{\text{other }S}\\
=&\frac{1}{S_{(12)345}S_{456(78)}}*\frac{1}{\text{other }S},
\end{aligned}
\end{equation}
where we have used the identity \eqref{fkals;f}. Here we can see $S_{(12)3(45)6(78)}$ is not a pole in the whole amplitudes. On the other hand, on the dual polytope $\dcalP(3,8)$, the merging manipulation would "flatten" each pair of facets into one single simplicial facet:
\begin{equation}
\begin{tikzpicture}
\draw (-4,-1) node[left]{$S_{(12)345}$} -- (-2,1) node[above]{$S'_{(12)3(45)6(78)}$} -- (0,0.5) node[right]{other $S$} -- cycle;
\draw (0,0.5) -- (0.5,-1.5) node[right]{$S_{456(78)}$} -- (-1,0);
\draw[dashed] (-1,0) -- (-2,1);
\draw[->] (1.5,-0.25) -> node[above]{merge} (3.5,-0.25);
\draw (4+1,-1.25) node[below]{$S_{(12)345}$} -- (6+1,-1.25) node[below]{other $S$} -- (6+1,0.75) node[above]{$S_{456(78)}$} -- cycle;
\draw[dashed] (6+1,-1.25) -- (5+1,-0.25) node[left]{$S_{(12)3(45)6(78)}$};
\node at (-1.2,-2) {$\dcalP(\calE_8)$};
\node at (6,-2) {$\dcalP(3,8)$};
\end{tikzpicture}
\end{equation}
from which we can see that $S_{(12)3(45)6(78)}$ would no longer correspond to a vertex. It would then stay on the edge connecting the vertices corresponding to $S_{(12)345}$ and $S_{456(78)}$. Likewise, for the other 7 disappeared facets in the cyclic class of $S_{(12)3(45)6(78)}$, also totally ``lose their role" in $\calP(3,8)$ and $\dcalP(3,8)$. They are also no longer poles in $\calA_{3,8}$.

\section{\label{fahfo}  Last two kinds of boundary configurations for $(3,9)$}
 In the main text, we show how to blow up a black dot and use a parallel line to represent that  2 is colliding to 3 in the direction determined by $4$ and 5 in eq. \eqref{newnewe}. 
More exactly to say, 2 is colliding to 3 with the angle formed by the lines $\overline{23}$ and $\overline{45}$ vanishing at order ${\cal O} (\epsilon)$. 
Similarly to  \eqref{newnewe},  we can have
\ba 
S\left[
\raisebox{-.5cm}{
 \begin{tikzpicture}[scale=.9 
 ]
\filldraw (0,0) node{\tikz{\filldraw (0,0)  circle (1pt);}}  node[below]{$4$} circle (1pt);
\filldraw (1,0) node{\tikz{\filldraw (0,0)  circle (1pt);}} node[below]{$5,6$}  circle (1pt);
\draw (-1,0) -- (1,0) (-1-.5,-.3)--(-1-.9,-.3);
\draw[dashed] (-1-.75,0) circle (0.75);
\filldraw (-1-.9,0.3) node{\tikz{\filldraw (0,0)  circle (1pt);}}  node[left]{$1$} circle (1pt);
\filldraw (-1-.5,-.3) node{\tikz{\filldraw (0,0)  circle (1pt);}}  node[above]{$3$} circle (1pt);
\filldraw (-1-.9,-.3) node{\tikz{\filldraw (0,0)  circle (1pt);}}  node[left]{$2$} circle (1pt);
\end{tikzpicture}
}
\right]=& S\left[
\raisebox{-.5cm}{
 \begin{tikzpicture}[scale=.9 
 ]
\filldraw (0,0) node{\tikz{\filldraw (0,0)  circle (1pt);}} 
node[below]{$4$} circle (1pt);
\filldraw (1,0)  node{\tikz{\filldraw (0,0)  circle (1pt);}} node[below]{$5,6$}  circle (1pt);
\draw (-1,0) -- (1,0) ;
\draw[dashed] (-1-.75,0) circle (0.75);
\filldraw (-1-.9,0.3)  node{\tikz{\filldraw (0,0)  circle (1pt);}}  node[left]{$1$} circle (1pt);
\filldraw (-1-.5,.15)  node{\tikz{\filldraw (0,0)  circle (1pt);}}  node[below]{$3$} circle (1pt);
\filldraw (-1-.9,-.3) node{\tikz{\filldraw (0,0)  circle (1pt);}}   node[left]{$2$} circle (1pt);
\end{tikzpicture}
}
\right]+ s_{2\,3\,4}+ s_{2\,3\,5}+ s_{2\,3\,6}
\,,
\ea 
 where 5 , 6 collides  and 2 collides to 3 in the direction determined by 4 and 5. This is equivalent to say $\{2, 3\}$ and $\{5,6\}$ collides respectively and at the same time 5 and 6 are approaching to the line $\overline{34}$ determined  by 3 and 4,
 \ba
S\left[
\raisebox{-.5cm}{
 \begin{tikzpicture}[scale=.9 
 ]
\filldraw (0,0)  node{\tikz{\filldraw (0,0)  circle (1pt);}} node[below]{$4$} circle (1pt);
\filldraw (1,0) node{\tikz{\filldraw (0,0)  circle (1pt);}}  node[below]{$5,6$}  circle (1pt);
\draw (-1,0) -- (1,0) (-1-.5,-.3)--(-1-.9,-.3);
\draw [dashed](-1-.75,0) circle (0.75);
\filldraw (-1-.9,0.3) node{\tikz{\filldraw (0,0)  circle (1pt);}}  node[left]{$1$} circle (1pt);
\filldraw (-1-.5,-.3)  node{\tikz{\filldraw (0,0)  circle (1pt);}} node[above]{$3$} circle (1pt);
\filldraw (-1-.9,-.3) node{\tikz{\filldraw (0,0)  circle (1pt);}}  node[left]{$2$} circle (1pt);
\end{tikzpicture}
}
\right]=& S\left[
\raisebox{-.5cm}{
 \begin{tikzpicture}[scale=.9 
 ]
\filldraw (0,0) node[below]{$4$} circle (1pt);
\draw (-1.5,0) -- (1,0) (-1-1.1,0)--(-1-.5,0);
\filldraw (1.4,0.2)  node{\tikz{\filldraw (0,0)  circle (1pt);}} node[right]{$5$} circle (1pt);
\filldraw (1.6,-.2)  node{\tikz{\filldraw (0,0)  circle (1pt);}} node[right ]{$6$} circle (1pt);
\draw [dashed](-1-.75,0) circle (0.75)  (1+.65,0) circle (0.65)  ;
\filldraw (-1-.9,0.3) node{\tikz{\filldraw (0,0)  circle (1pt);}}  node[left]{$1$} circle (1pt);
\filldraw (-1-.5,0)  node{\tikz{\filldraw (0,0)  circle (1pt);}} node[below]{$3$} circle (1pt);
\filldraw (-1-1.1,0) node{\tikz{\filldraw (0,0)  circle (1pt);}}  node[below]{$2$} circle (1pt);
\draw[blue] ($(-1-.5,0)+(0.2,0)$+) arc (0:180:.2);
\end{tikzpicture}
}
\right]
\,,
\ea 
where the distances of 5 and 6 to the line $\overline{34}$ are both at order ${\cal O} (\epsilon)$ and hence $(3,4,5),(3,4,6)$$= {\cal O} (\epsilon)$. Here  we mention that on the RHS the angle $\angle 234$ is $\pi-{\cal O} (\epsilon)$ instead of $\pi$ and hence the distance of 4, 5 and 6 to the line $\overline{23}$ are also  at order ${\cal O} (\epsilon)$.   The distance of 2 to the line $\overline{34}$ is at order ${\cal O} (\epsilon^2)$.  

However in $(3,9)$, there are such boundary configurations where 5 is approaching to the line $\overline{34}$ at order ${\cal O} (\epsilon^2)$, i.e.  the distance of 5 and 6 to the line  $\overline{34}$ are at order ${\cal O} (\epsilon^2),{\cal O} (\epsilon)$ respectively,
\be\label{newneafojwe}
 \raisebox{-.5cm}{
 \begin{tikzpicture}[scale=.9 
 ]
\filldraw (0,0)  node{\tikz{\filldraw (0,0)  circle (1pt);}} node[below]{$4$} circle (1pt);
\draw (-4.5,0) -- (4,0)-- (4.4,0) (-4-1.1,0)--(-4-.5,0);
\filldraw (4.4,0) node{\tikz{\filldraw (0,0)  circle (1pt);}}  node[above]{$5$} circle (1pt);
\filldraw (4.6,-.2)  node{\tikz{\filldraw (0,0)  circle (1pt);}} node[above right ]{$6$} circle (1pt);
\draw [dashed](-4-.75,0) circle (0.75)  (4+.65,0) circle (0.65)  ;
\filldraw (-4-.9,0.3) node{\tikz{\filldraw (0,0)  circle (1pt);}}  node[left]{$1$} circle (1pt);
\filldraw (-4-.5,0) node{\tikz{\filldraw (0,0)  circle (1pt);}}  node[below]{$3$} circle (1pt);
\filldraw (-4-1.1,0) node{\tikz{\filldraw (0,0)  circle (1pt);}}  node[below]{$2$} circle (1pt);
\draw[blue] ($(-4-.5,0)+(0.2,0)$+) arc (0:180:.2);
\end{tikzpicture}
}
\,\,,
\ee
The distance of 4,5,6 to the line $\overline{23}$ are still at order ${\cal O} (\epsilon)$. 
Hence we have the following minors at order ${\cal O} (\epsilon^2)$
\ba 
&(2,4,5), (3,4,5), \,\, (2,3,1),(2,3,4), (2,3,5), (2,3,6)={\cal O} (\epsilon^2)\,.
\ea
The remaining 7 minors are at order ${\cal O} (\epsilon)$,
\ba 
&(1,4,5),(1,4,6),(2,4,6), (3,4,6),\,\,
(1,3,4), (1,3,5), (1,3,6)={\cal O} (\epsilon)\,, 
\ea 
since their areas of the triangles are at order ${\cal O} (\epsilon)$.

  More importantly for us, considering the residue  ${\rm d}
\log R_{3,n}(\epsilon)$ for the  configuration graph \eqref{newneafojwe}, it can be  directly related to the normal configuration \eqref{hafjld} by adding several  generalized Mandelstam variables,
\ba  \label{fasjhafjld}
&S\left[
 \raisebox{-.5cm}{
 \begin{tikzpicture}[scale=.9 
 ]
\filldraw (0,0)  node{\tikz{\filldraw (0,0)  circle (1pt);}} node[below]{$4$} circle (1pt);
\draw (-1.5,0) -- (1,0)-- (1.4,0) (-1-1.1,0)--(-1-.5,0);
\filldraw node{\tikz{\filldraw (0,0)  circle (1pt);}}  (1.4,0) node[above]{$5$} circle (1pt);
\filldraw  node{\tikz{\filldraw (0,0)  circle (1pt);}} (1.6,-.2) node[above right ]{$6$} circle (1pt);
\draw[dashed] (-1-.75,0) circle (0.75)  (1+.65,0) circle (0.65)  ;
\filldraw (-1-.9,0.3) node{\tikz{\filldraw (0,0)  circle (1pt);}}  node[left]{$1$} circle (1pt);
\filldraw (-1-.5,0) node{\tikz{\filldraw (0,0)  circle (1pt);}}  node[below]{$3$} circle (1pt);
\filldraw (-1-1.1,0)  node{\tikz{\filldraw (0,0)  circle (1pt);}} node[below]{$2$} circle (1pt);
\draw[blue] ($(-1-.5,0)+(0.2,0)$+) arc (0:180:.2);
\end{tikzpicture}
}
\right]
\\
=&
S\left[
 \raisebox{-.5cm}{
 \begin{tikzpicture}[scale=.9 
 ]
\filldraw (0,0)  node{\tikz{\filldraw (0,0)  circle (1pt);}} node[below]{$4$} circle (1pt);
\filldraw (-1,0)  node{\tikz{\filldraw (0,0)  circle (2pt);}} node[below]{$1,2,3$} circle (2pt);
\draw (-1,0) -- (1,0)-- (1,0) ;
\filldraw (1,0)  node{\tikz{\filldraw (0,0)  circle (2pt);}} node[below]{$5,6$} circle (2pt);
\end{tikzpicture}
}
\right]
+( s_{2\,3\,4}+ s_{2\,3,5}+ s_{2\,3\,6}) + (s_{2\,4\,5}+ s_{3\,4\,5})
\,.
\nonumber
\ea

Now we can show the two boundary configurations for the last two seeds of facets of $(3,9)$ in  \eqref{ffadsojfjasf} in the main text,
\ba \label{fdijfslfd}
\raisebox{-1.1cm}{
\begin{tikzpicture}[scale=.8]
\node[circle,draw=black, fill=black, inner sep=0pt,minimum size=1.5pt] (v4) at (-2.15,2) {};
\node[circle,draw=black, fill=black, inner sep=0pt,minimum size=1.5pt] (v5) at (-1.5,2) {};
\node[circle,draw=black, fill=black, inner sep=0pt,minimum size=1.5pt] (v3) at (-1.9,1.625) {};
\node[circle,draw=black, fill=black, inner sep=0pt,minimum size=1.5pt] (v2) at (-1.25,0.875) {};
\node[circle,draw=black, fill=black, inner sep=0pt,minimum size=1.5pt] (v1) at (-0.75,0.125) {};
\node[circle,draw=black, fill=black, inner sep=0pt,minimum size=1.5pt] (v6) at (-0.5,2) {};
\node[circle,draw=black, fill=black, inner sep=0pt,minimum size=1.5pt] (v7) at (0.5,2) {};
\draw (v1) -- (v7);
\draw (v1) -- (-2,2);
\filldraw[fill=white,dashed] (-1.8,2) circle (0.75);
\filldraw[fill=white,dashed] (0.5,1.9) circle (0.7);
\filldraw (v4)  node{\tikz{\filldraw (0,0)  circle (1pt);}} circle (1pt) node[above] {$4$};
\filldraw (v5)  node{\tikz{\filldraw (0,0)  circle (1pt);}} circle (1pt) node[below] {$5$};
\filldraw (v3) node{\tikz{\filldraw (0,0)  circle (1pt);}}  circle (1pt) node[left] {$3$};
\filldraw (v1) node{\tikz{\filldraw (0,0)  circle (1pt);}}  circle (1pt) node[left] {$1$};
\filldraw (v2)  node{\tikz{\filldraw (0,0)  circle (1pt);}} circle (1pt) node[left] {$2$};
\filldraw (v6) node{\tikz{\filldraw (0,0)  circle (1pt);}}  circle (1pt) node[above] {$6$};
\filldraw (v7) node{\tikz{\filldraw (0,0)  circle (1pt);}}  circle (1pt) node[above] {$7$};
\draw (-2,2) -- (v7);
\draw (v3) -- (v4) -- (v5);
\node[circle,draw=black, fill=black, inner sep=0pt,minimum size=1.5pt] (v9) at (-0.25,0.875) {};
\node[circle,draw=black, fill=black, inner sep=0pt,minimum size=1.5pt] (v8) at (0.25,1.625) {};
\draw (v8)  node{\tikz{\filldraw (0,0)  circle (1pt);}} node[right]{$8$};
\draw (v9)  node{\tikz{\filldraw (0,0)  circle (1pt);}} node[right]{$9$};
\draw[blue] ($(v5)+(0.2,0)$+) arc (0:180:.2);
\end{tikzpicture}
}, \qquad 
\raisebox{-1.1cm}{
\begin{tikzpicture}[scale=.8]
\node[circle,draw=black, fill=black, inner sep=0pt,minimum size=1.5pt] (v4) at (-2,2) {};
\node[circle,draw=black, fill=black, inner sep=0pt,minimum size=1.5pt] (v5) at (-1.5,2) {};
\node[circle,draw=black, fill=black, inner sep=0pt,minimum size=1.5pt] (v2) at (-1.375,1.0625) {};
\node[circle,draw=black, fill=black, inner sep=0pt,minimum size=2.5pt] (v1) at (-0.75,0.125) {};
\node[circle,draw=black, fill=black, inner sep=0pt,minimum size=1.5pt] (v6) at (-0.5,2) {};
\node[circle,draw=black, fill=black, inner sep=0pt,minimum size=1.5pt] (v7) at (0.5,2) {};
\draw (v1) -- (v4);
\filldraw[fill=white!,dashed] (-1.75,2) circle (0.65);
\filldraw (v4) node{\tikz{\filldraw (0,0)  circle (1pt);}}   circle (1pt) node[above] {$4$};
\filldraw (v5) node{\tikz{\filldraw (0,0)  circle (1pt);}}  circle (1pt) node[below] {$5$};
\draw (v1) -- (v7);
\filldraw[fill=white!,dashed] (0.5,1.9) circle (0.7);
\filldraw (v1)  node{\tikz{\filldraw (0,0)  circle (1pt);}} circle (1pt) node[left] {$1,2$};
\filldraw (v2) node{\tikz{\filldraw (0,0)  circle (1pt);}}  circle (1pt) node[left] {$3$};
\filldraw (v6) node{\tikz{\filldraw (0,0)  circle (1pt);}}  circle (1pt) node[above] {$6$};
\filldraw (v7)  node{\tikz{\filldraw (0,0)  circle (1pt);}} circle (1pt) node[above] {$7$};
\draw (v4) -- (v7);
\node[circle,draw=black, fill=black, inner sep=0pt,minimum size=1.5pt] (v9) at (-0.25,0.875) {};
\node[circle,draw=black, fill=black, inner sep=0pt,minimum size=1.5pt] (v8) at (0.25,1.625) {};
\draw (v8) node{\tikz{\filldraw (0,0)  circle (1pt);}}  node[right]{$8$};
\draw (v9) node{\tikz{\filldraw (0,0)  circle (1pt);}}  node[right]{$9$};
\draw[blue] ($(v5)+(0.2,0)$+) arc (0:180:.2);
\end{tikzpicture}
}
\,.
\ea

 \bibliographystyle{JHEP}
\bibliography{reference}

\end{document}